\documentclass[usegraphicx,useAMS,11pt,preprint]{aastex}

\newcounter{romnum}

\newcommand{\ptl}{\partial}
\begin{document}

\title{A Magnetic $\alpha\omega$ Dynamo in Active
Galactic Nuclei Disks: II. Magnetic Field Generation,
Theories and Simulations}

\author{Vladimir I.
Pariev\altaffilmark{1}\altaffilmark{2}, Stirling A.
Colgate}
\affil{Theoretical Astrophysics Group, T-6, Los Alamos
National Laboratory, Los Alamos, NM 87545}
\altaffiltext{1}{Lebedev Physical Institute, Leninsky
Prospect 53, Moscow 119991, Russia}
\altaffiltext{2}{Currently at Physics Department,
University of Wisconsin-Madison, 1150 University Ave.,
Madison, WI 53706}
\and
\author{J. M. Finn}
\affil{Plasma Theory Group, T-15,  Los Alamos National
Laboratory, Los Alamos, NM 87545}

\begin{abstract} It is shown that a dynamo can operate in
an Active Galactic Nuclei (AGN) accretion disk due to
the  Keplerian shear and due to the helical motions of
expanding and twisting  plumes of plasma heated by many
star passages through the disk.  Each plume rotates a 
fraction of the toroidal flux into poloidal flux, always
in the same direction, through a finite angle, and
proportional to its diameter. The predicted growth rate
of poloidal magnetic flux, based upon two analytic
approaches and numerical simulations, leads to a rapid
exponentiation of a seed field, $\sim 0.1$ to $\sim
0.01$ per Keplerian period of the inner part of the disk. The
initial value of the seed field may therefore be
arbitrarily small yet reach, through dynamo gain,
saturation very early in the disk history. Because of
tidal disruption of stars  close to the black hole, the
maximum growth rate occurs at a radius of about $100$
gravitational radii from the central object. The
generated mean magnetic field, a quadrupole field, has
predominantly even  parity so that the radial component
does not reverse sign across the midplane.  The  linear
growth is predicted to be the same by each of the
following three theoretical analyses:  the flux conversion model,
the mean field approach, and numerical modeling. 
The common
feature is the conducting fluid flow, considered in
companion Paper I  \citep{pariev06} where two coherent
large scale flows occur naturally: the differential
winding of Keplerian motion and differential rotation of
expanding plumes.
\end{abstract}


\section{Introduction}\label{sec_intro}

The need for a magnetic dynamo to produce and amplify the
immense magnetic fields observed external to galaxies and
in clusters of galaxies has long been recognized. The
theory of kinematic magnetic dynamos has had a long
history and is a well developed subject by now. 
There are numerous
monographs and review articles devoted to the magnetic
dynamos in astrophysics, some of which are:
\citet{parker79}; \citet{moffatt78};
\citet{stix75}; \citet{cowling81}; \citet{roberts92};
\citet{childress90}; \citet{zeldovich83};
\citet{priest82};
\citet{busse91}; \citet{krause80}; \citet{biskamp93};
\citet{mestel99}. Hundreds of  papers on magnetic dynamos
are published each year. Three main astrophysical areas,
in which dynamos  are involved, are the generation of
magnetic fields in the convective zones of planets and
stars, in differentially rotating spiral galaxies, and
in the accretion disks around compact objects.  The
possibility of production of magnetic fields in  the
central parts of the black hole accretion disks in AGN
has been  pointed out by
\citet{chakrabarti94} and the need and possibility for a
robust dynamo by \citet{colgate97}. Dynamos have 
been also observed in the laboratory in the
Riga experiment \citep{gailitis00, gailitis01} and in
Karlsruhe experiment \citep{stieglitz01}, although these
flows only partially simulate astrophysical ones.   The
flow resulting  in a dynamo is essentially three
dimensional flow and often, especially under
astrophysical circumstances, is a chaotic  or turbulent
flow.
 
The shear in a rotating conducting fluid amplifies
the magnetic field in the direction perpendicular to the
shear and facilitates the growth of the magnetic field.
Originally, \citet{parker55} proposed to combine the
effects of kinetic helicity of the small scale motions of
the fluid with the differential rotation to generate
large scale magnetic fields in the Sun. Here we  consider
just such  a dynamo in its application to the
differentially rotating flow in the accretion disk
around  Central Massive Black Holes (CMBH) in the centers
of galaxies. The necessary and robust source of helicity
is provided by the rising and expanding plumes of the gas
heated by the star passages through the accretion disk.
This property of the rotation of expanding plumes in a
rotating frame is discussed at length in the companion
paper,
\citet{pariev06}, "A Magnetic $\alpha\omega$ Dynamo in
AGN Disks: I. The Hydrodynamics of Star-Disk Collisions
and Keplerian Flow", which is referred to as paper~I.
This natural and unique coherent flow is supported by
experimental evidence  \citep{beckley03} and is
fundamental to the origin of a robust dynamo in an AGN
accretion disk.

The magnetic dynamo in the disk is the essential part of
the whole emerging picture of the formation and
functioning of AGNs, closely related to the production of
magnetic fields within galaxies, within clusters of
galaxies, and the still greater energies and fluxes in
the inter-galactic medium. Black hole formation, 
Rossby wave torquing of
the  accretion disk
\citep{lovelace99, li00, li01b, colgate03}, jet formation
\citep{li01a} and magnetic field redistribution by
reconnection and flux conversion, and finally  particle
acceleration in the radio lobes and jets are the key
parts of this scenario \citep{colgate99, colgate01}.
Finally we  note that if almost every galaxy contains a
CMBH and that if a major fraction of the free energy of
its formation is converted into magnetic energy, then
only a small fraction of this magnetic energy,  as seen
in the giant radio lobes \citep{kronberg01}, is
sufficient to propose a possible feed back in structure
formation and in galaxy  formation.

This work is arranged as  follows: in section~\ref{sec2}
we briefly overview the ingredients of the star-disk
collisions dynamo with a brief review of the disk
conditions and star disk collisions from Paper~I.  In
section~\ref{sec3} we introduce the flux conversion
dynamo analysis with a discussion of the necessary
reconnection and turbulence driven resistivity.   In
section~\ref{sec4} the mean field theory is developed, 
in section~\ref{sec6} the  dynamo equations and
numerical method are developed, and in
section~\ref{sec7} the results of 
numerical calculations are
presented  in  support of all three approaches.  Finally,
we end with the conclusions in section~\ref{sec_concl}.  
CGS units are used throughout the paper.

\section{ The Ingredients of  the Star-disk Collisions
Dynamo }
\label{sec2}

A poloidal magnetic field can be of two types distinguished by
the reflectional symmetry in the equatorial plane:
quadrupole (or even) and dipole (or odd). Quadrupole
field has the same sign of the radial component above and
below the disk plane. The radial component of the dipole
field changes sign under the reflection in the disk
plane, it vanishes exactly at the disk plane. Rigourous
definitions and properties of the odd and even fields are
given in Appendix~A. As is evident in
Figure~\ref{fig1a}A, the quadrupole field has a large
radial component, both within and external to the disk
and furthermore maintains the same radial direction in
both spaces. On the other hand the differential shear of
a dipole field, symmetric about the midplane and
therefore with zero radial component, results in no
winding of the flux within the disk and therefore no
toroidal gain. Various higher multipoles than the
quadrupole have  an alternating radial component as a
function of radius and therefore a greater possibility of
cancellation by reconnection.  Differential winding of a
symmetric poloidal field by the Keplerian flow results
in a  uniform toroidal field having the same direction
over the disk height and within the disk,
Figure~\ref{fig1a}B. An
$\alpha$ deformation resulting in a large scale
helicity, on the scale comparable to the radius of the
disk, will transform toroidal field into poloidal field.
This transformed field, or new poloidal flux  must have
the same polarity as the original poloidal flux.  Then
the closure of the dynamo cycle demands that this
transformed flux be merged or reconnected with the
original poloidal flux in order that it is augmented and
hence produce gain. If this transformed flux alternates
in direction (as would be the case for a dipole field
across the thickness of a disk),  then  the merged flux
will be averaged to near zero.  Only in the case of the
quadrupole field is there a possibility of a coherent
addition to the original poloidal field when the
$\alpha$ deformation, as produced by star-disk
collisions, changes sign across the midplane and further
rotates only $\pi/2$ radians. One notes that star
collisions in the opposite, axial, direction equally
contribute to the quadrupole poloidal flux.  The toroidal
field produced by the shear of differential rotation from
the quadrupole field, (Figure~\ref{fig1a}B), has
opposite directions far above the surface of the  disk
from that inside the disk. The opposite direction of the
toroidal field above the disk is not shown in these
drawings, because, in addition to the dynamo, it presumes
the formation of a force-free oppositely directed helix
in the conducting half space above and below the disk. We
have, however, predicted and calculated this force-free
helix
\citep{li01a}, and furthermore, as mentioned above, we
associate partial dissipation of its free energy with
the visible structure of AGN jets. However, the magnitude
of the quadrupole field in the region closer to  the disk
surface and to the midplane should be stronger (as
computations actually prove). Therefore, the $\alpha$
deformation will primarily take the bottom portion of
the quadrupole flux and convert it into radial flux
above the disk plane directed in the same way as the
upper portion of the quadrupole field,
Figure~\ref{fig1a}D.  Therefore, in the accretion disk
dynamo, plumes from star-disk collisions entrain and
rotate toroidal flux by
$\sim \pi/2$ radians, originating primarily from within
the disk, Figure~\ref{fig1a}C.  Furthermore these plumes
terminate  close to or not far above  the  surface of the
disk, and so produce negligible rotation  of flux not so
displaced from the disk. We then expect this rotated
flux, before rotating a further
$\pi$ radians and so before self cancellation,  to
reconnect as loops of poloidal flux,
Figure~\ref{fig1a}D. These loops of flux now merge with
the initial poloidal field, Figure~\ref{fig1a}A, thereby
completing the cycle.   To proceed with the dynamo
problem we need to utilize the following results from
Paper~I:
\begin{enumerate}
\item The distribution of stars in coordinate and
velocity space  in the central star cluster of an AGN.
\item The velocity, and density  of the plasma in  the
disk and in the corona of the disk.
\item The hydrodynamics of the flow resulting from the
passage  of the star through the disk, the plumes.
\end{enumerate}

We review  briefly the properties of the plumes produced
by the star disk collisions as they relate to the
dynamo.  Then with these results we estimate the
conductivity in order to develop a flux rotation theory
of the dynamo. 

\subsection{The Untwisting or Helicity Generation by the
Plume}
\label{subsec_2.1}

Let us first introduce a term which is used frequently
below. Because of high conductivity of the plasma
considered in this paper, the magnetic field is close to 
a "frozen-in" state, when the magnetic field lines
follow the motions of the plasma. Imagine now a closed
contour attached to the particles of plasma with some
magnetic flux passing through this contour. Let us also
draw this contour such that it is close to being a plane
contour. As the result of plasma motions, this contour
can be rotated by some angle. If this rotation happens
quickly enough, so that no substantial magnetic field
crosses the contour due to diffusion, the magnetic flux
passing through this closed contour remains almost
unchanged. The component of the magnetic field normal to
the plane of the contour and averaged over the surface of
the closed contour should have rotated by the same angle
as the contour. We will name this process "flux
rotation".

We describe the number density of central stellar cluster
and the kinematics of the stellar orbits in Paper~I. The most 
important result of this consideration is that the rate at which 
stars cross the unit area of the disk surface peaks at a radial 
distance of about  $100 r_g$ to $200 r_g$ from the CMBH,
where $r_g=2GM/c^2 = 3.0 \cdot 10^{13}M_8 \,\mbox{cm}
= 9.5\cdot 10^{-6}M_8\,\mbox{pc}$ is the gravitational 
radius of the CMBH and $M_8$ is the mass $M$ of the CMBH 
expressed in units of $10^8$ solar masses: $M_8 = M/
10^8\,M_{\odot}$.  The rate of stars-disk collisions closer to the 
CMBH than $100 r_g$ is depleted because of the tidal destruction
of stars in the gravitational field of CMBH and because of 
 the grinding of the star orbits into the accretion disk plane.
 This grinding occurs because of the action of a drag, which 
 every star experiences on its passage through the accretion 
 disk. After many passages this drag causes an inclined 
 Keplerian orbit of a star to become coplanar with the disk 
 plane and this star becomes trapped inside the disk.
 
 The physics and dynamics of a star-disk collision is also 
 considered in Paper~I.  Here we briefly summarize 
 the results of Paper~I for the convenience of the reader. 
 A star collides with the disk at a 
 typical velocity of $5\cdot 10^3\,\mbox{km/s}$ to 
 $10^4\,\mbox{km/s}$. The velocity of escape from the 
 surface of a solar like star is $600\,\mbox{km/s}$. This 
 is one order of magnitude smaller than the velocity of 
 the star moving through the gas in the accretion disk. 
 Also, the sound speed in the accretion disk at a radial 
 distance of $\sim 200 r_g$ is $\sim 50\,\mbox{km/s}$
 (see Appendix~A in Paper~I for details and more accurate
 numbers).  Therefore, the gravitational field of the star 
 itself does not influence a highly supersonic flow of 
 gas onto the star. In this regard, the physics of a star-disk
 collision is radically different from the physics of the 
 classical accretion process on either the moving or the
 resting star. The classical theory of accretion of interstellar 
 gas with zero angular momentum onto stars was developed 
 in \citet{bondi44, bondi47,  bondi52, mccrea53}. Since 
 the peculiar velocities of stars in the Galaxy are much less 
 than $600\,\mbox{km/s}$
 and the sound speed in the interstellar material is also
 much less than $600\,\mbox{km/s}$, the gravitational 
 potential of a star dominates the dynamics of the 
 accretion flow in the near proximity of a star. The radius 
 of the gravitational capture of the gas is much larger 
 than the radius of the star. Captured gas falls almost 
 radially down to the star surface. The presence of a small
asymmetry or non-homogeneity of the surrounding gas
causes nonzero angular momentum, which strongly 
influence the dynamics of the accretion flow below the 
gravitational capture radius.
 
 The term "collision" rather than "accretion" is much more 
 appropriate for the description of the interaction of 
 a passing star with the accretion disk. Because of the 
 high velocity of the star, the cross section of the interaction
 of a star with the gas is equal to the geometric cross section
of a star. The high ram pressure of the incoming stream 
 with the density $\sim 10^{-8}$ to $\sim 10^{-10}\, \mbox{g}\,
 \mbox{cm}^{-3}$ strips away the outer layer of a star.
 The underlying layers with the temperature $\sim 10^6\,\mbox{K}$
 and density $\sim 10^{-5}\,\mbox{g}\,\mbox{cm}^{-3}$ 
 are exposed. This picture is completely different from 
 the physics of the mixing of the radial accretion stream
 with the stellar (solar) atmosphere in the classical 
accretion theory as described by \citet{hoyle49}.
A radiation shock is formed in front of the star 
 and the channel of the hot gas is left behind the star. This 
 channel expands sideways inside the accretion disk and 
 heats the surrounding gas.  The hot gas is subject to 
 buoyancy force acting away from the equatorial plane of the 
 disk. As a result of this force, two plumes rising from the 
 two sides of the accretion disk are formed at the location 
 of the star-disk crossing. Note, that the amount of gas in 
 the rising plumes and the size of the plumes are much 
 larger than the initial mass and size of the hot 
 channel made by the star.
 
As explained in Paper~I,  the  plume should expand to
several times its original  radius by the time it reaches
the height of the order $2H$ above the disk surface, 
where $H$ is the
semi-thickness of the disk. The corresponding increase in
the moment of inertia of the plume and  the conservation
of the angular momentum of the plume causes the plume to
rotate slower relative to the inertial frame. From the
viewpoint of the observer in the frame corotating with
the Keplerian flow at the  radius of the disk 
at the location of the
plume, this means that the plume rotates in the direction
opposite to the Keplerian rotation with an angular
velocity equal to some fraction of the local Keplerian
angular velocity depending upon the radial expansion
ratio. Since the expansion of the plume  will not be
infinite in the rise and fall time of
$\pi$ radians of  Keplerian rotation of the disk, we
expect that the average of the plume rotation will be
correspondingly less, or
$\Delta \phi < \pi$ or $\sim \pi/2$ radians. Any force or
frictional drag that resists this rotation will be
countered by the Coriolis force. Finally we note that
kinetic helicity is proportional to

\begin{equation} h = {\bf v} \cdot ({\bf  \nabla}
\times {\bf v})
\label{eqn3.28} \mbox{.}
\end{equation}

For the dynamo one requires one additional dynamic
property of the plumes. This is, that the total  rotation
angle must be finite and preferably
$\simeq
\pi/2$ radians, otherwise a larger angle or after many
turns  the vector of the entrained magnetic field would
average to a small value and consequently the  dynamo
growth rate would be correspondingly small.  This
property of finite rotation, $\Delta
\phi
\sim \pi/2$ radians,  is  a natural property of plumes
produced above a Keplerian disk.

Thus we have derived the approximate properties of an
accretion disk around a massive black hole: the high
probability of star-disk collisions, the three
necessary  properties of the resulting plumes all
necessary for a robust dynamo. What is missing from this
description is the necessary electrical properties of
the medium.

\section{The Flux Rotation Dynamo}
\label{sec3}

\subsection{The Conductivity of the Disk and the Corona}
\label{subsec_3.1}

The dynamos producing large scale magnetic fields require
a compromise between high and low conductivity. A poor
conductor or an insulator will not allow the field to be
dragged with the motion of the medium. Ohmic dissipation
will cause the magnetic field to decay, and if
sufficiently rapid, no dynamo will be possible. In the
limit of very high conductivity, kinematic exponential
growth of magnetic field has been predicted to occur in
the presence of random or chaotic three dimensional
motions of the medium (e.g., \citealt{zeldovich83};
\citealt{roberts92}). The problem with analytic three
dimensional motions is that being described as kinematic,
they are reversible in the sense that little or no
entropy is generated by the motions themselves. The field
can be unwrapped by a "non-Maxwell" demon following the
line of force.  Since the "demon"  does not have to
"throw away" any information in following the reverse
path, no entropy is generated. The chaotic behavior in
time of kinematic mathematically  reversible motions
does not create entropy because of time reversal
invariance of the equations. The plumes from star disk
collisions indeed occur randomly in time, but since the
initial state is as random as the final state, no entropy
is generated just due to the randomness in time of the
plumes themselves. By comparison, if the initial state
were a large amplitude coherent wave, then phase
scrambling would indeed alter the entropy, but by a
relatively small amount compared to a scattering process
that leads to a Maxwell distribution. This lack of a
change in entropy is then equivalent to laminar flow
(without molecular diffusion) where mixing is
reversible. By contrast, the randomness created by fluid
turbulence is irreversible, satisfying a principle of
maximizing the dissipation of the free energy of shear
flow in a fluid. The plumes, although random in time,
result in a coherent addition of poloidal flux, because
every plume translates axially, expands radially, and
rotates through nearly the same angle,
$\sim \pi/2$, for every plume.

The negative effect of turbulence on dynamo gain has been
documented in three major liquid sodium dynamo
experiments: Lyon, Cadarache ~\citep{bourgain02}, 
Maryland \citep{sisam04}, and Madison \citep{nornberg06,
spence06}, all using the similar flow configurations.
Numerous theoretical simulations of these flows, the
Dudley-James flow in a sphere~\citep{dudley89} or
similar  von K\'a{}rm\'a{}n flow in a cylinder (i.e., two
counter-rotating radially converging and axially
diverging flows) in the kinematic or laminar limit have
been performed: (Lyon, Cadarache)
\citet{bourgain02, petrelis03, marie03}, (Maryland)
\citet{peffley00, sweet01}, and (Madison)  
\cite{bayliss06, oconnell05}.  They all predict 
exponential dynamo gain at a critical magnetic Reynolds
numbers, $\mbox{Rm}_{crit}  \sim 50$.  Yet the
experiments give a null result, i.e., no exponential
dynamo gain for experimental flows where
$\mbox{Rm}_{exp} > 130 \simeq 2.5 \mbox{Rm}_{crit}$ are
achieved in the experiments. These null results are
interpreted as due to the negative effects of turbulent
diffusion \citep{bourgain04, spence06, nornberg06, 
laval06}. Our generalized interpretation of these
results is that turbulence behaves as an enhanced
diffusion of magnetic flux or an enhanced resistivity
\citep{boldyrev04, ponty05}.   In these experiments the
turbulent velocity
$v_{turb}  \simeq 0.4 <v>$ where $<v>$ is the average
shear velocity \citep{nornberg06}.  Then the turbulence
leads to a decreased conductivity or an enhanced
resistivity as described by
\citet{krause80} as:
\begin{equation}
\sigma_{turb} = \frac{\sigma_0}{1 + 4\pi \beta \sigma_0
/c^2},
\end{equation} where $\sigma_0=c^2/(4\pi\eta_0)$ is the
conductivity of the fluid, $\eta_0$ is the magnetic
diffusivity of the same fluid, and
$\sigma_{turb}$ is the effective conductivity in the
presence of turbulence.  The constant $\beta$ is derived
from mean-field electrodynamics assuming isotropic
turbulence:
\begin{equation}
\beta \simeq (\tau_{corr} / 3)  v_{turb}^2,
\end{equation} where $\tau_{corr}$ is the mean
correlation time of a turbulent fluctuation.   Since the
correlation time is an eddy turn over time, then
$\tau_{corr} = L_{corr} / v_{turb}$, where $L_{corr}$ is
an eddy size. We then identify $L_{corr} v_{turb}/3 =
\beta$ as a turbulent diffusion coefficient  and the
turbulent conductivity becomes the original conductivity
decreased by the factor
$1 + 4\pi\beta \sigma_0/c^2$.  In the limit of a large
$\beta >>
\eta_0$, the effective magnetic diffusivity then becomes
just the turbulent diffusivity, $\eta_{eff} = \beta
\simeq  L_{corr} v_{turb}/3$. It is the combination of
turbulent diffusivity with fluid resistivity or
equivalently, effective restivity that determines dynamo
gain.  In unconstrained flows $L_{corr}$ becomes the
dimension of the largest eddy that can "fit"  in the flow
or
$L_{corr} = (d(ln<v>) /dx)^{-1} \simeq L/2$ where $L$ is
the dimension of the shear flow. This larger effective
resistivity then results in a smaller effective magnetic
Reynolds number that determines dynamo exponential gain,
$\mbox{Rm}_{eff} = L <v> /\eta_{eff}$ for a given $<v>$
and $L$. Since $v_{turb}  \simeq <v>/2$, then
$\eta_{eff} \simeq (1/3)(1/4) <v> L$ and 
$\mbox{Rm}_{eff} \simeq 12$. This value  of 
$\mbox{Rm}_{eff}$ is significantly smaller than the
predicted value of $\mbox{Rm}_{crit} \sim 50$ for the
Dudley-James or von K\'a{}rm\'a{}n flows used in the
current major experiments.  It is even smaller than
$\mbox{Rm}_{crit} \approx 17$ for Ponomarenko flow used in
Riga dynamo experiment 
\citep{ponomarenko73, gailitis76}.

The critical threshold for dynamo gain,
$\mbox{Rm}_{crit}$ is determined from kinematic dynamo
calculations without enhanced turbulent resistivity.  For
bounded sheared flows, namely except for all but these
special flows mentioned above,
$\mbox{Rm}_{crit} \sim 100$.  The question is whether
there can be any  exponential dynamo gain in any
unconstrained shear flows.

By way of confirmation, in the two dynamo experiments
that have demonstrated positive exponential dynamo gain,
the Riga experiment \citep{gailitis00, gailitis01} and
the Karlsruhe experiment
\citep{stieglitz01}, turbulence was greatly constrained
by the presence of a ridged wall(s) separating the
counter flowing shear flows. It was therefore well
recognized that these experiments did not represent
astrophysical dynamos, but, on the other hand, strongly
confirmed dynamo theory. We were therefore convinced
that a natural constraint of turbulence must exist for
the dynamos of astrophysics.

There are at least five constraints of turbulence in
shear flow that may alter the magnitude of turbulence as
well as its isotropy. We list five of these constraints,
expecting others to be identified:

\begin{enumerate}
\item  viscosity.
\item a ridged wall.
\item a positive outward gradient of angular momentum.
\item a gradient of entropy in a gravitational field
(e.g., the base of the convection zone of stars).
\item delay in the onset of fully developed turbulence 
in unconstrained shear flow.
\end{enumerate} 

Viscosity may inhibit all turbulence so that in this
limit  the properties of turbulence are not relevant.  

A ridged wall  affects the magnitude of turbulence and
its isotropy (the law of the walls).

The gradients of angular momentum and entropy apply to
astrophysical circumstances, where depending upon the
presence of other instabilities, e.g. the magnetorotational 
instability or Rossby vortex instability, 
turbulence may be  a small fraction of the average  shear
flow.

Time dependence is similar to viscosity,  in the case
where the initiation time of the shear flow may be very
short compared to the development time of the turbulence
as in the case of the plumes driven by star-disk
collisions.  This limit leads to negligible levels of
turbulence compared to the  shear flow.

We chose a gradient of angular momentum
and time dependence of the shear flow, e.g., plume
flow,   as the probable mechanisms of constraint of
turbulence for the most likely circumstances for
producing an astrophysical dynamo; angular momentum as
the circumstance for the accretion in massive black hole
formation, and time dependence for the constraint of the
transient period of the rise and fall of  plumes in
astrophysical circumstances.

It is with this uncertainty of the role of turbulence in 
dissipating the magnetic flux as opposed to amplifying it
that the current work was undertaken.  Hence, when we
found  the  possibility of a combination of (a), a near
laminar shear flow, Keplerian flow, and (b), a
repeatable, transient, non-turbulent  source of helicity,
could the possibility of a robust astrophysical dynamo
become evident. Although the star-disk collisions are
random in time, the flow, to first order is repeatable
and therefore not turbulent. The flow resulting from a
superposition of many plumes may be chaotic in time, but
the superposition of many plumes, all with the same
rotation, leads to a net rotated flux in the same
direction. On the other hand, the vortices in anisotropic
turbulence make an arbitrary number of turns and so the
instantaneous mean value of rotated flux is proportional
to the square root of the  number of vortices. Thus we
characterize the plumes as semi-coherent  rather than a
truly chaotic phenomena in which the entropy would be
increased. In addition we discuss next the analysis in
which turbulence may augment or possibly limit the "fast
dynamo".

The dynamos with non-vanishing growth rate in the limit
of very high conductivity are called fast dynamos
\citep{vainshtein72}. A classical picture of the fast
dynamo mechanism is stretch-twist-fold process
\citep{sakharov82,vainshtein72}. There are strong
indications that the fast dynamo action is typical for
chaotic flows
\citep{lau93,finn92,finn91}. However, in the kinematic
stage of the dynamo, a sharp exponential decrease in some
spatial scales of the magnetic field occurs.  The
magnetic field becomes concentrated in the narrow sheets
or narrow filaments until the frozen-in picture becomes
invalid for any conductivity. In the kinematic limit the
thickness of these structures of strong magnetic field
is estimated as $\delta l \sim L\,\mbox{Rm}^{-1/2}$,
where the magnetic Reynolds number, $\mbox{Rm} = v
L/\eta$, $v$ is the velocity of the conducting fluid, $L$
is the characteristic dimension of the fluid, and $\eta$
is magnetic diffusivity due to finite resistivity. The
growth of small scale fields invalidates the kinematic
approximation beginning with the resistive scale and up
to the larger scales. The structure and the spectrum of
this, hydromagnetic, turbulent dynamo is expected to be
different from the hydrodynamic turbulence because of
the action of the magnetic forces. The size
$\delta l$ of the smallest magnetic structures depends on
the details and properties of the hydromagnetic regime of
the turbulence, which are still the subject of active
debate in the literature  \citep{iroshnikov63, 
kraichnan65,
  goldreich95, boldyrev06}, but is always much smaller
than the large scale $L$ by some positive power of
$\mbox{Rm}$. In the limit of infinite conductivity, no
flux can merge in an infinite time and hence, there can
be no multiplication of flux at a scale of the system
(large scale). The motions with nonzero helicity $h$ at a
large scale and the ability to reconnect is required to
obtain the growth of the large scale fields and magnetic
flux comparable to the growth rate of the small scale
field. When the large scale field growth is at a rate
comparable to the growth rate of the small scale field,
the characteristic growth time is of order of the
diffusion time
$t_{diff} = L^2 / \eta$.

Yet our ionized disk of thickness $H = 2.6
\times 10^{13}\,\mbox{cm}$, velocity
$v_K \approx 10^9\,\mbox{cm}\,\mbox{s}^{-1}$ and
resistivity
$\eta\simeq 10^7\,\mbox{cm}^2 \,\mbox{s}^{-1}$ at 1~eV
$\simeq 10^4\,\mbox{K}$ temperature, results in 
$\mbox{Rm} \simeq 10^{15}$. This is a number so large as
to preclude useful growth of the magnetic flux and the
large scale fields in a Hubble time, which is much
shorter than the diffusion time of the magnetic field
$t_{diff} = H^2 / \eta = 10^{20}\,
\mbox{s}$.

Only by invoking the phenomena of turbulent resistivity,
(above) can the existence  of an accretion disk dynamo
producing large scale magnetic fields be made convincing.
Turbulent resistivity within the disk is likely to be due
to the same turbulence that creates the
$\alpha$-viscosity of the Shakura--Sunyaev disk or the
Rossby vortices of the RVI disk. Within the disk we
expect turbulent diffusion of the magnetic flux to be the
same as that of angular momentum, and thus proportional
to the Shakura--Sunyaev parameter
$\alpha_{ss}$  (see paper~\uppercase{i}).

Reconnection may be occurring within the turbulence
leading to  more rapid dissipation of the magnetic flux
than the turbulent cascade alone.  However, the
force-free fields above (and below) the disk that are
produced by the winding of the dynamo-produced large
scale fields need not be dissipated until they are
projected large distances away from the disk.

Recognizing this lack of fundamental understanding, but
that laboratory and astrophysical observations lead to
the same order of magnitude for reconnection, we proceed
with the assumption that a value of
$\mbox{Rm} \simeq 200$ approximates the magnetic
diffusion within the disk. In what follows this
parameter could be several orders of magnitude larger,
but not much smaller and still  result in an effective
accretion disk  dynamo.

\subsection{Estimates of the Dynamo Growth Rate: The
Flux Rotation Dynamo}
\label{subsec_3.2}

With these values of  plume size, frequency, and
magnetic diffusivity as well as Keplerian flow, let us
make some estimates of the threshold parameters and the
growth rate of the
$\alpha\omega$ dynamo, which has been outlined in a
previous section.  The approach developed below, 
in the rest of section~\ref{sec3}, we call the flux
rotation dynamo as opposed to the mean field dynamo.
Later we will compare these approaches with emphasis
upon the difference between coherent motions versus
random averaged variables.  We consider for now the linear
growth, i.e. when the magnetic field is not strong enough,
such that one can neglect the back reaction of the
generated magnetic field on both the Keplerian flow and
on the plume flow fields.

Suppose that at some moment of time  we have an even
symmetry poloidal magnetic field,
${\bf B}_P$ (see Appendix~\ref{appendix_C} for
definitions of even and odd symmetries). The radial
component of this field within the disk defines a
poloidal flux, $F_P$, such that at a given radius $r$,
the poloidal flux through one half of the disk, either
side of the mid-plane, is
$F_P = B_r \cdot(area) = B_r \cdot H\cdot 2\pi r$, where
$H$ is a semi-thickness of the disk. This flux, within
the conducting and differentially rotating Keplerian
disk, will be wrapped up into a toroidal magnetic field
within the disk,
$B_T$. This toroidal field will be stronger or a multiple
of the initial poloidal field depending upon the number
of turns and the resistive dissipation of the currents.
Initially we consider no dissipation so that an initial
flux line of poloidal magnetic field
$B_P$ will be differentially wrapped $n$ times around the
axis leading to an enhanced
$B_T$. Let us introduce the number of differential turns,
$n$, that occurs at a radial distance, $r$, during time,
$t$, as
\begin{equation} 2\pi n = -t
\cdot r
\frac{d\Omega_K}{dr}\mbox{,}
\label{eqn_ndiff}
\end{equation} where $\displaystyle \Omega_K= \left(
GM/r^3 \right)^{1/2}$ is the Keplerian angular velocity
of the disk. We consider the toroidal flux, $F_T$, in the
azimuthal direction and within the half thickness of the
disk, $H$: $F_T = r H B_T$. Then, the increment of this
toroidal flux added to the original $F_T$ per
$dr$ and per $dn$ differential turns  becomes
\begin{equation} d F_T = d B_T \cdot H dr = r\cdot B_r
\cdot dt \frac{d\Omega_K}{dr}
\cdot Hdr =-2\pi \cdot dn \cdot B_r \cdot Hdr
\label{eqn3.29}\mbox{.}
\end{equation} Since $dF_T = dB_T \cdot H dr$, this is
equivalent to $dB_T = -2\pi dn \cdot B_r$.

If we integrate over $dn$ and integrate over $dr$ to
give the change in toroidal flux,
$\Delta F_T$, per revolution we obtain an estimate
\begin{equation}
\Delta F_T \approx -2\pi r \cdot n \cdot H\cdot B_r =
-n\cdot F_P
\label{eqn3.30}\mbox{.}
\end{equation} The poloidal flux, $F_P$, in turn is
derived from the toroidal flux by the helicity, $h$, of
the plumes driven by star-disk collisions. Each plume
lifts a loop of toroidal flux with cross section
$dA = H
\cdot R_{shk} \simeq H^2$, Fig.~\ref{fig1a}, where
$R_{shk}$ is the radius of the shock produced by a star
(see paper~I for details). The small distortion $\sim (1
+ H/r)$ from the circular cross section is neglected. When
this unit of area or flux is rotated
$\pi/2$ radians into the poloidal direction with an
efficiency of a single plume,
$\alpha_{plume}$, it creates an equal unit of poloidal
flux
$dF_{P,plume} = -\alpha_{plume} \cdot B_T \cdot H^2$
inside the disk. The top parts of the field loops created
by the plumes are rising quickly because of the low
density of plasma in the corona: even a very weak
magnetic field can overcome the gravity acting on a
rarefied gas and the expansion will be at a relativistic
Alfv\'en speed $v_A \la c$. Also, the shearing of the top
part of the loops is small, so the toroidal field
produced is also small. The rate of this removal of the
toroidal and poloidal fields from above the disk to the
magnetized jet is higher than the diffusion feeding from
the inside of the disk, if $\eta/H \ll c$. We know that
$\Omega_K H^2/\eta \sim 1$. Therefore,
$\eta/H \sim \Omega_K H$ and, indeed, we have $\Omega_K
H \ll c$. In this approximation, only the evolution of
$F_{P}$ and $F_{T}$ inside the disk defines the dynamo
and it is separate from the evolution of the magnetic
fields in the corona of the disk. Each such unit of flux
$dF_{P,plume}$ is only 2H in length and in order to 
create or to affect a flux tube of length
$r$, poloidal, or
$2\pi r$,  toroidal,  requires an aligned sum of
increments
$r/2H$ in number in the poloidal direction and
$2 \pi r/2H$ in number in the  toroidal direction. (Each
plume also creates an increment, a pair of  equal and
opposite vertical fluxes,
$dF_Z \simeq
\pm dF_{P,plume}$, which exponentially decrease  to
near zero regardless of overlapping plumes.)

\subsection{Plume Coverage}
\label{subsec_3.3}

The fractional area of one side of the disk inside radius
$r$ covered by plumes at any one time, ${\bar q}_{<r}$,
can be estimated as
${\bar q}_{<r} \approx N(<r) \cdot  H^2/r^2$ because
each of total $N(<r)$ stars (equation~2 in paper~I) with
impact radii inside a given radius $r$ crosses the disk
two times in approximately one Keplerian period
$T_K(r)=2\pi/\Omega_K (r)$. Each such crossing produces
one plume of radius $\approx H$ on each side of the disk.

Each plume exists for the time $T_K/2$ before falling
back to the disk surface. Then in the spirit of a flux
rotation explanation of the dynamo we evaluate the
fractional contribution to the total poloidal flux by
each rotated plume. The plumes occur randomly over the
area of the disk, but their contribution to the average 
flux of either poloidal of toroidal is independent of
position on the surface, because the  coherence of the
plume rotation ensures an effect proportional to the
algebraic sum of the number of plumes regardless of their
location. Therefore  we can rearrange, gedanken-wise,
the location of the plumes over the disk without
affecting the result. We therefore rearrange and align 
a fraction of the plumes,
$r/2H$ in number, to create a single, continuous poloidal
flux tube of poloidal flux,
$B_{plume}H^2$, and length $r$ where $B_{plume} =
\alpha_{plume} B_T$.  We have enough plumes,
$N(<r)$, to create
$N_{tubes} = N(<r)/(r/2H)$ such flux tubes.   These
poloidal flux tubes of width $2H$ then collectively
cover a fraction in azimuth or a sector of angular width
of
\begin{equation}
\Delta \phi = \frac{2H N_{tubes}}{2 \pi r} =
\frac{2H^2}{\pi r^2} N(<r) \label{eqn3.31}
\mbox{.}
\end{equation} and produce a poloidal flux per half
revolution of
\begin{equation}
\Delta F_P = N_{tubes} \cdot H^2 \cdot B_{plume} =
-N_{tubes} \cdot H^2 \cdot
\alpha_{plume}\cdot B_T
\label{eqn3.32}\mbox{.}
\end{equation} One can express $N_{tubes}$ from
equation~(\ref{eqn3.31}) as
\begin{equation} N_{tubes}=\frac{\pi
r}{H}\Delta\phi\label{eqn_3.32a}
\mbox{.}
\end{equation} We then note that $\Delta \phi = {\bar
q}_{<r}$ by construction,  and therefore the poloidal
flux created by $N(<r)$ plumes per half revolution
becomes
\begin{equation}
\Delta F_P = -{\bar q}_{<r}\frac{\pi r}{H} \cdot H^2
\cdot \alpha_{plume}B_T = -\pi rH {\bar q}_{<r}
\alpha_{plume} B_T = -{\bar q}_{<r} \cdot
\alpha_{plume}\pi\cdot  F_T
\label{eqn3.33}\mbox{.}
\end{equation} We define a parameter $\alpha_m$ as
$\alpha_m={\bar q}_{<r}\alpha_{plume}$ to give $\Delta
F_P = -2\pi\alpha_m \cdot  F_T$ per revolution. Here
$\alpha_m$    becomes an efficiency for the rotation of
toroidal flux into poloidal  flux per radian of
revolution  and averaged over all the plumes.

Then the time derivative considering two plumes per
Keplerian period becomes
\begin{equation}
\frac{dF_P}{dt} = -\Omega_K \cdot
\alpha_m \cdot F_T
\label{eqn3.F_T}\mbox{.}
\end{equation}

Similarly from equation~(\ref{eqn3.30}) we obtain
\begin{equation}
\frac {d F_T}{dt} = -\frac{\Omega_K}{2\pi} \cdot
\frac{3}{2} \cdot F_P
\label{eqn3.35}\mbox{,}
\end{equation} because one Keplerian revolution,
$T_K=2\pi/\Omega_K$, corresponds to $n=3/2$ differential
turns according to expression~(\ref{eqn_ndiff}).

In addition we must consider the  fractional flux
cancellation  of each of these two flux transformations.
This leads   to partial cancellation of each orthogonal
component by the other and to partial  self cancellation
as well.  Finally  second order effects,  as well as the
different dissipation rates of the two fluxes must be
considered. These effects are usually averaged in
mean-field theory, but here we consider them separately,
because we are concerned with a semi-coherent flow as opposed
to a turbulent dynamo.

\subsection{Toroidal Multiplication with Losses}
\label{subsec_3.4}

We considered above that the flux is frozen within the
disk fluid flow.   Now  we  consider the relaxation of
this condition by resistive diffusion or reconnection.

Resistive dissipation of the currents supporting these
fields limits the growth of toroidal field. Here we
consider the saturation of the toroidal multiplication
alone with a fixed poloidal field.  After  many turns,
this additional toroidal magnetic field reaches a
saturation value determined by the balance of the
multiplication rate  with resistive diffusion. The
toroidal magnetic field changes fastest in the vertical
direction on the scale $H$, therefore the dissipation
rate is estimated as
\begin{equation} dF_T/dt
\approx - \frac{\eta}{H^2}F_T = - \frac
{F_T}{\mbox{Rm}_{\Omega}}\Omega_K\frac{r^2}{H^2}
\label{eqn3.34a}\mbox{,}
\end{equation} where magnetic Reynolds number with
respect to Keplerian rotation is defined as
\begin{equation}
\mbox{Rm}_{\Omega}=\frac{\Omega_K r^2}{\eta}
\label{eqn_rmdef}\mbox{.}
\end{equation} If we add this loss to the gain of
equation~(\ref{eqn3.35}), we then have
\begin{equation}
\frac {d F_T}{dt} = \left(-\frac{3}{4\pi}F_P -
\frac{F_T}{\mbox{Rm}_{\Omega}}\frac{r^2}{H^2}\right)
\cdot
\Omega_K
\label{eqn3.36}\mbox{.}
\end{equation} Thus the toroidal field saturates after
$\mbox{Rm}_{\Omega}H^2/r^2$ turns. In view of
equation~(\ref{eqn3.36}) the limiting, steady state is
achieved when
$$
\frac{F_T}{F_P} = -\frac{3}{4\pi}\mbox{Rm}_{\Omega}
\frac{H^2}{r^2}\mbox{.}
$$ This limiting value of the ratio $F_T/F_P$ can be
measured separately in the laboratory (in $H\approx r$ 
geometry), without the motions producing the
$\alpha$-effect of the  complete dynamo, by applying an
external, fixed poloidal initial  field. The same
situation may apply to the galaxy if a small residual 
poloidal flux is left over from the initial AGN phase,
i.e., this  dynamo,  and therefore  no further dynamo in
the galactic disk  would be  required, even though one 
likely  exists
\citep{ruzmaikin88, ferriere00, kulsrud99}. Furthermore
this  ratio represents the maximum possible toroidal
multiplication that  should off set losses in the
rotation of toroidal flux back  into poloidal flux for
achieving net positive dynamo growth  rate. Thus if the
toroidal amplification is large, the efficiency of 
rotation of toroidal flux back into poloidal flux,
$\alpha_m$, can be  small, and the dynamo will still be
growing.

To this toroidal  multiplication  and resistive loss we
must add the back reaction  effects of the helicity or
flux rotation mechanism(s), of the  "$\alpha_m$-effect".

\subsection{Production of Poloidal Flux and  Losses}
\label{subsec_3.5}

In order to calculate $\alpha_m$ of 
equation~(\ref{eqn3.F_T}), we require both
${\bar q}_{<r}$ and  $\alpha_{plume}$. The coverage
factor,
${\bar q}_{<r}$, is more straight forward to estimate
from the star disk collision rate and plume size, but the
efficiency of the helical deformation by a single 
plume~(\ref{eqn3.28}) is more problematic.  The
simplest and ideal concept of poloidal flux production 
by a plume  is that a plume of radius
$\approx H/2$ rises a distance $\approx 2H$ 
above the disk with entrapped flux,
$dF_{T,plume} = B_T
\cdot H^2$, rotates  this flux exactly $\pi /2$ radians,
i.e., into the poloidal direction, falls back, merges
with the disk matter, and releases this now poloidal flux
by diffusion or reconnection so that this unit of 
poloidal flux adds in the same direction, i.e.
coherently,  to $F_P$. Of course this sequence of
rotation, rise, fall, and merging of the fluxes  will
happen episodically and only when averaged, leads to the
factor $\alpha_m$ such that
$\Delta F_P=-2\pi\alpha_m\cdot F_T$ per revolution
(equation~(\ref{eqn3.F_T})). The associated experimental
paper,
\citet{beckley03}, on laboratory measurements of plume
rotation implies that a rising and   expanding plume, in
a rotating frame, indeed rotates a finite angle $\sim
\pi/2$ radians before merging with the background fluid.
In this case the finite angle of rotation occurs for the
same reason  as expected in the accretion disk. In the
laboratory case the velocity of the plume  relative to
the velocity of rotation is chosen such that the plume is
destroyed or broken up by striking the end wall of the
apparatus in a chosen finite fraction of a period of
rotation.  In an accretion disk, as pointed out earlier,
the plume indeed rises and falls in
$\pi$  radians with the rotation angle as well as the
merging with background disk material, both  increasing
monotonically during this rise and fall time.  Hence the
ideal angle,
$\pi/2$ radians, occurs as a result of the product and
average of the three progressive deformations, but most
importantly throughout the entire sequence the
incremental addition to
$\alpha_m$  is always positive.  Also one should note
that with each plume there is an equal upward vertical
flux,
$+dF_Z$,  as downward vertical flux,
$-dF_Z$, which presumably averages to zero with flux
merging.

As described in equations~(\ref{eqn3.33})
and~(\ref{eqn3.F_T}) the number of plumes adding to the
poloidal flux is described by the filling factor,
${\bar q}_{<r}$, of the disk by plumes, 
where ${\bar q}_{<r}=N(<r) H^2/r^2$. The average number
of plumes on one side of the disk within a radius $r$ at
any given time is approximately equal to the number of
stars with impact radii inside $r$,
$N(<r)$, given by expression~(2) in paper~I.

$H$ is given by expression~(6) in paper~I, valid for the
inner zone~(a) of a standard or Shakura--Sunyaev disk.
The reader is referred to paper~I for the details,
arguments and caveats of using this model of an
accretion disk originally proposed by \citet{shakura72}
and further developed in \citet{shakura73}. 
Here we only remark that we use
the expressions for the disk parameters valid for
$r<r_{ab}$, where $r_{ab}$ is the transition radius
between zones (a), (radiation dominated)  and (b),
(particle pressure dominate) generally a few hundreds of
$r_g$. As we see from expression~(\ref{sc_eqn3.44})
below, the dynamo growth rate is maximal at about
$r_{ab}$. To obtain the number density of stars
$n(r)$ in the vicinity of CMBH we use analytic and
numerical models of the stellar dynamics to extrapolate
from the observed $n$ at the distance of $\approx
1\,\mbox{pc}$ from CMBH down to few tens of $r_g$.
Observations typically suggest number densities of the
order of
$10^5\,\mbox{pc}^{-3}$ at a distance
$1\,\mbox{pc}$. So we write $n(1\,\mbox{pc}) =n_5 \cdot
10^5\,\mbox{pc}^{-3}$. The most notable feature of this
distribution of stars is the sharp decrease of their
density for $r$ less than about
$10 r_t$, where
$r_t=2.1\cdot 10^{-4}\,\mbox{pc}\cdot M_8^{1/3} = 21 r_g
M_8^{-2/3}$ and is the tidal disruption radius for a
solar mass star by the tidal forces near the CMBH with
mass $M=M_8 \cdot 10^8 \, M_{\odot}$. This decrease is
the effect of physical collisions of stars with each
other, tidal disruptions by CMBH, and multiple passages
of the stars through the accretion disk, which grind
their orbits into the disk plane and reduce the number of
remaining stars not trapped by the accretion disk (see
paper~I for greater details).

Using the approximate analytical model (2) from 
paper~\uppercase{i} we have for 
$r<10^{-2}\,\mbox{pc}$
\begin{eqnarray} && {\bar q}_{<r} = 1.9 \cdot 10^{-4}
\cdot n_5
\left(\frac{l_E}{0.1}\right)^2
\left(\frac{\epsilon}{0.1}\right)^{-2}
\left[\frac{r}{10 r_t}-\left(\frac{r}{10 r_t}
\right)^{-2}\right]
\quad \mbox{for}
\quad 10 r_t<r<10^{-2}\,\mbox{pc}\mbox{,}
\nonumber \\ && {\bar q}_{<r} = 0 \quad
\mbox{for}
\quad r<10 r_t \quad
\mbox{(no star-disk collisions)}\mbox{,}
\label{sc_eqn3.37}
\end{eqnarray} where the factor $(1-\sqrt{3r_g/r})$
coming from the Shakura-Sunyaev model is omitted since
$r\gg r_g$. Here $l_E=L/L_{Edd}$ is the ratio of the
total luminosity of the disk
$L$ to the Eddington limit
$L_{Edd}$ for the CMBH of mass $M$ and
$\epsilon$ is the fraction of the rest mass energy of the
accreted matter, ${\dot M} c^2$, which is radiated away
by the disk,
$L=\epsilon {\dot M} c^2$. The number given by
expression~(\ref{sc_eqn3.37}) is not large, so the
probability that any given plume is overlapped with
another is small and therefore, on the average,  each
plume will be an individual, isolated  event.

With ${\bar q}_{<r}$ given by
expression~(\ref{sc_eqn3.37}) the corresponding
$\alpha_m={\bar q}_{<r}
\alpha_{plume}$ becomes
\begin{eqnarray} && \alpha_m = 1.9
\cdot 10^{-4}\, n_5 
\alpha_{plume}
\left(\frac{l_E}{0.1}\right)^2
\left(\frac{\epsilon}{0.1}\right)^{-2}
\left[\frac{r}{10 r_t}-\left(\frac{r}{10 r_t}
\right)^{-2}\right]
\quad \mbox{for}
\> 10 r_t<r<10^{-2}\,\mbox{pc}
\mbox{,} \nonumber \\ && \alpha_m = 0 \quad
\mbox{for}
\quad r<10 r_t \quad
\mbox{(no star-disk collisions)}\mbox{.}
\label{sc_eqn3.38}
\end{eqnarray}

The system of linear differential
equations~(\ref{eqn3.F_T}) and~(\ref{eqn3.35}) has a
growing solution
\begin{equation} F_T=F_{T,0}e^{\Gamma t}
\label{eqn3.40a}\mbox{,}
\end{equation} where
\begin{equation}
\Gamma=\Omega_K\sqrt{\frac{3\alpha_m}{4\pi}} =
\Omega_K\sqrt{\frac{3 {\bar q}_{<r} \cdot 
\alpha_{plume}}{4\pi}}
\label{eqn3.41a}\mbox{.}
\end{equation} Similar to the toroidal field, the
gradient of the poloidal magnetic field is greatest in
the vertical direction on the scale $H$, and therefore
the dissipation rate of the poloidal flux is estimated
analogous to the dissipation rate of the toroidal flux
(equation~(\ref{eqn3.34a})):
\begin{equation} dF_P/dt
\approx - \frac{\eta}{H^2}F_P = - \frac
{F_P}{\mbox{Rm}_{\alpha}}\frac{l^2}{H^2}\Omega_K
\label{eqnF_PDisp}\mbox{,}
\end{equation} where magnetic Reynolds number with
respect to the
$\alpha$-deformation is defined as
\begin{equation}
\mbox{Rm}_{\alpha}=\frac{\Omega_K l^2}{\eta}
\label{eqn_rmalpha}\mbox{,}
\end{equation} 
and $l\approx 3H$ is the height above the disk mid-plane 
reached by the plume before falling back to the disk.
In our approximation
$\mbox{Rm}_{\alpha}\approx\mbox{Rm}_{\Omega} l^2/r^2$,
but we keep $\mbox{Rm}_{\alpha}$ and
$\mbox{Rm}_{\Omega}$ separate to evaluate the effects of
Keplerian and plume motions separately. Adding the
resistive dissipation, equation~(\ref{eqnF_PDisp}), to
the poloidal gain, equation~(\ref{eqn3.F_T}), results in:
\begin{equation}
\frac{dF_P}{dt} = -\Omega_K \alpha_m F_T
-\frac{l^2}{H^2}\frac{\Omega_K}{\mbox{Rm}_{\alpha}} F_P
\label{eqn3.34ab}\mbox{.}
\end{equation}

The system of linear differential
equations~(\ref{eqn3.36}) and~(\ref{eqn3.34ab}) has a
growing solution of the form~(\ref{eqn3.40a}) but with
the growth rate modified as
\begin{equation}
\Gamma = \frac{\Omega_K}{2}\left[
\left (\left(\frac{r^2}{H^2 \mbox{Rm}_{\Omega}} -
\frac{l^2}{H^2\mbox{Rm}_{\alpha}}\right)^2 +
\frac{3\alpha_m}{\pi} \right)^{1/2} -
\left(\frac{r^2}{H^2 \mbox{Rm}_{\Omega}} +
\frac{l^2}{H^2\mbox{Rm}_{\alpha}}\right)\right]
\label{eqn3.43}\mbox{.}
\end{equation} We note that in the limit of small
resistivity, large magnetic Reynolds numbers, we recover
the growth rate of equation~(\ref{eqn3.41a}), otherwise
we note the surprising circumstance that the   difference
in the resistive terms adds to the growth rate  whereas,
as expected, the sum decreases the growth rate as we
expect for a purely diffusive resistivity. For positive 
growth rate, the first term, of course, must be greater
than the second. In the purely diffusive limit,  if one
uses that
$\mbox{Rm}_{\alpha}=\mbox{Rm}_{\Omega}l^2/r^2$,
expression~(\ref{eqn3.43}) simplifies to
\begin{equation}
\Gamma=\Omega_K\sqrt{\frac{3\alpha_m}{4\pi}}-
\frac{\eta}{H^2}  = \Omega_K
\left(\sqrt{\frac{3\alpha_m}{4\pi}}-
\frac{l^2}{H^2 \mbox{Rm}_{\alpha}} \right)
\label{eqn3.43a}\mbox{.}
\end{equation} To the extent that the resistive terms are
small and therefore Rm is large and the compensating
effect of the requirement for merging of newly minted
poloidal flux with old poloidal flux is neglected, then
the dynamo growth rate is large, of order $\Omega_K
\cdot \alpha_m^{1/2}$. However, there is no reason to
expect that the resistivity is purely diffusive, and we
expect that tearing mode reconnection drives the merging
of flux at some near constant   and large
$\mbox{Rm}_{\alpha} \simeq
\mbox{Rm}_{\Omega} \simeq 200$ \citep{sovinec01}.

Then from equations~(\ref{sc_eqn3.38}) 
and~(\ref{eqn3.41a}) the maximum dynamo growth rate
occurs between $r_{ab}$ and $10 r_{t}$ and becomes (for
$10^8 \, M_{\odot}$ CMBH)
\begin{eqnarray} && \Gamma  \approx 7\cdot 10^{-3}
\Omega_{Kt} \cdot n_5^{1/2}
\alpha_{plume}^{1/2}\left(\frac{l_E}{0.1}\right)
\left(\frac{\epsilon}{0.1}\right)^{-1}
\left(\frac{r}{10 r_t}\right)^{-1}\left[1-
\left(\frac{r}{10 r_t}\right)^{-3}\right]^{1/2}
\>
\mbox{for} \> r>10 r_t\mbox{,} \nonumber \\ && \Gamma
= 0 \quad \mbox{for}
\quad r<10 r_t\mbox{,}\label{sc_eqn3.44}
\end{eqnarray} where $\Omega_{Kt}=2.08\cdot
10^{-7}\,\mbox{s}^{-1}$ is the Keplerian rotation
velocity at $10 r_t$ for a $10^8\,M_{\odot}$ black hole.
The exponential rate of multiplication, in view of
equation~(\ref{sc_eqn3.44}), maximizes at
$r=13.6 r_t$, where
\begin{eqnarray} && \Gamma_{max}   \approx  1.4 \cdot
10^{-9}\,\mbox{s}^{-1}
\cdot n_5^{1/2}\alpha_{plume}^{1/2}
\left(\frac{l_E}{0.1}\right)\left(\frac{\epsilon}{0.1}
\right)^{-1} \;
      \mbox{at} \; 13.6 r_t  \quad \mbox{or}
\nonumber \\ && \Gamma_{max}   \approx 4\cdot
10^{-2}\,\mbox{yr}^{-1}
\cdot n_5^{1/2}\alpha_{plume}^{1/2}
\left(\frac{l_E}{0.1}\right)\left(\frac{\epsilon}{0.1}
\right)^{-1} \; \mbox{at} \; 13.6 r_t
\label{eqn3.45}\mbox{.}
\end{eqnarray} Since the density of stars 
does not actually drop
sharply to $0$ at $r=10 r_t$ as in our approximate 
analytical model, the estimate of
$\Gamma_{max}$ above is approximate and the actual
maximum of the growth rate is achieved at $r$ somewhat
smaller than $13.6 r_t$.

One can find the ratio of toroidal to poloidal flux in the
growing dynamo mode by substituting
expression~(\ref{eqn3.43}) for $\Gamma$ into
equation~(\ref{eqn3.36}):
\begin{equation}
\frac{F_T}{F_P}=-\frac{3/(2\pi)}{\left(\frac{r^2}{H^2\
\mbox{Rm}_\Omega}-\frac{l^2}{H^2 \mbox{Rm}_\alpha}
\right) +
\sqrt{\frac{3\alpha_m}{\pi}+\left(\frac{r^2}{H^2
\mbox{Rm}_\Omega} -\frac{l^2}{H^2 \mbox{Rm}_\alpha}
\right)^2}} \label{tpratio}\mbox{.}
\end{equation} In the purely diffusive limit, when
$\mbox{Rm}_\alpha=
\mbox{Rm}_\Omega l^2/r^2$, this ratio simplifies to
\begin{equation}
\frac{F_T}{F_P}=\frac{B_T}{2\pi B_P}=
-\frac{1}{2}\sqrt{\frac{3}{\pi\alpha_m}}
\label{tpratio1}\mbox{.}
\end{equation} As in any $\alpha\omega$ dynamo, the
averaged toroidal magnetic flux is much larger than the
averaged poloidal magnetic flux 
(recall that $\alpha_m \ll 1$).

Regardless how small $\alpha_{plume}$, which we believe
is
$\simeq 1$,  the dynamo gain is so large within the time
of formation of the CMBH,
$10^8$ years,  that saturation  will occur early in the
history of the disk-dynamo  regardless how small the
initial seed field. The origin of such a seed field,
e.g., a star, the Biermann battery from decoupling, or
primordial fields becomes moot. Nevertheless, for
completeness, we explore how this gain can be reduced by
flux rotated by the plume process such that it opposes
rather than augments either of the mean toroidal or
poloidal fluxes in the above estimates of dynamo gain.

\subsection {Flux Compensation by Plumes}
\label{subsec_3.6}

So far we have considered only the positive increment  of
poloidal flux to the dynamo gain by plume rotation of the
toroidal flux. This  same rotation  will rotate the
coexisting poloidal flux into opposition with the
primary toroidal flux.  In addition to the extent that
the plumes or any other cyclonic motion continues the
rotation beyond $\pi$ radians, a further reduction in
dynamo gain occurs because of averaging of this opposed
flux. First we consider that $B_P/B_T = - (\alpha_m
/3\pi)^{1/2}
\ll 1$ according to expression~(\ref{tpratio1}).

The deformation leading to the rotation of an increment of
toroidal flux into an increment of poloidal flux by the
rotation  of the plumes leads to a similar fraction of
poloidal flux being rotated such as to oppose the
toroidal flux. However, since only a small fraction is
rotated by plumes as opposed by the shear of rotation,
the negative effect on the toroidal flux is small, $\simeq
\alpha_m$. Similarly a rotation by
$\pi$ radians causes a decrement of both the toroidal and
poloidal fluxes to oppose themselves, so that the
fraction of flux rotated
$\pi$ radians must be small for high gain.  The fraction
of flux rotated $3 \pi/2$ radians must be even  smaller
for high  gain, because a rotation of $3 \pi/2$ radians
causes the larger  toroidal flux to oppose the much
smaller  poloidal flux even though a  small positive
effect can occur when the same rotation  causes  the
smaller poloidal flux to add to the toroidal. This
assumes that  the plumes are circular in cross section,
so that the cross  sectional areas for the radial and
toroidal fluxes are the same. The  distortion of the
plume cross section by differential rotation  in
$\pi/2$ radians of  rotation is similarly small,
$\Delta 
\phi
\simeq (1/4)(H/r)$.  

We designate these fractions by
$\tau_1$, 
$\tau_2$, $\tau_3$ for the fraction of flux rotated
$\pi/2$, 
$2\pi/2$,  $3\pi/2$ radians. In general we consider
$\tau_1  \sim 
\tau_2  \gg \tau_3$, otherwise we do not expect positive
gain.  One  can think of these coefficients as
correlation coefficients of the  decaying plume
rotation:  a plume undergoes little rotation beyond 
$\pi$ radians, when it falls back to the disk. All plumes 
are considered to behave the same so that these
coefficients remain  constant and therefore do not
describe turbulence.  These partial  flux cancellations
all reduce the dynamo gain.  The
equations~(\ref{eqn3.36}) and~(\ref{eqn3.34ab}) are then
extended  to become
\begin{eqnarray} && \frac {dF_T}{dt} = 
\Omega_K
\left(-\frac{3}{4\pi} F_P  -
\frac{F_T}{\mbox{Rm}_{\Omega}}\frac{r^2}{H^2} -\alpha_m 
\cdot\tau_1
\cdot F_P - \alpha_m \cdot  \tau_2 \cdot F_T +
\alpha_m 
\cdot
\tau_3\cdot F_P\right)
\nonumber \\ && \frac{dF_P}{dt} = 
\Omega_K
\left( -\alpha_m \cdot
\tau_1 \cdot F_T  -
\frac{F_P}{\mbox{Rm}_{\alpha}}
\frac{l^2}{H^2}
     - \alpha_m 
\cdot
\tau_2 \cdot F_P +
\alpha_m \cdot\tau_3 \cdot  F_T
\right)
\label{eqnfluxcomp}\mbox{.}
\end{eqnarray} Here we  have introduced $\alpha_m \cdot
\tau_1$  in the first term of the 
$F_P$ equation where tacitly we had assumed
$\tau_1 = 1$  before. Terms with $\tau_3$ in both 
equations~(\ref{eqnfluxcomp}) are small compared to the
terms with 
$\tau_1$ because $\tau_1 \gg \tau_3$.  Solving 
system~(\ref{eqnfluxcomp}) for  exponentially growing
solutions we  find a generalization of
expression~(\ref{eqn3.43}) for the growth  rate $\Gamma$:
\begin{eqnarray} && \Gamma = 
\frac{\Omega_K}{2}\left[
\left (\left(\frac{r^2}{H^2 
\mbox{Rm}_{\Omega}} -
\frac{l^2}{H^2\mbox{Rm}_{\alpha}}\right)^2  +
\alpha_m(\tau_1 -\tau_3) 
\left(\frac{3}{\pi}+ 4\alpha_m (\tau_1 - 
\tau_3) \right) 
\right)^{1/2}
\right. \nonumber \\ &&  \left. - 
\left(\frac{r^2}{H^2 \mbox{Rm}_{\Omega}}  +
\frac{l^2}{H^2\mbox{Rm}_{\alpha}}+
2\alpha_m\tau_2\right)\right]
\label{Gamma3}\mbox{.}
\end{eqnarray} One  can see that the effects of
considering finite 
$\tau_2$ and $\tau_3$  both act to reduce the growth 
rate of the dynamo from the one given  by 
expression~(\ref{eqn3.43}). Specifically, if $\tau_3 >
\tau_1$  (recall that $\alpha_m \ll 1$) then the  second
term in the sum under  the square root  in~(\ref{Gamma3})
becomes negative, and 
$\Gamma$  cannot have positive real part.  This means
that the dynamo is  impossible for 
$\tau_3 > \tau_1$. In fact, $\tau_3$ enters only in 
combination $(\tau_1-\tau_3)$ and effectively reduces
the value of 
$\tau_1$. This implies that the plume must terminate its 
contribution to flux rotation by
$\pi$  radians, but this is expected  on general grounds
because  by this time the plume matter will have  fallen
back to and be merged with the disk.  

The effect of finite 
$\tau_2$ on the growth  rate of the dynamo is much weaker
than the  effect of finite $\tau_3$. The leading 
positive contribution to 
$\Gamma$ comes  from the second term in the sum under the 
square  root and is 
$\propto (\alpha_m \tau_1)^{1/2}$.  The negative 
contribution of $\tau_2$ is 
$-2\alpha_m \tau_2$ term. Since 
$\alpha_m \ll 1$ we see that this negative contribution 
will be  always small compared to the  positive
contribution for any 
$\tau_2 
\sim \tau_1 \sim 1$. Thus with $\tau_3$ assumed small,
we  expect to  recover the very large growth rate,
$\Gamma
\simeq 0.04$  per  revolution, of
expression~(\ref{eqn3.45}).

\section{Mean Field  Theory for the Star Disk Collision
Driven Dynamo}
\label{sec4}

The  mean field approach to the problem of generation
of the large scale magnetic fields by the motions of the
fluid with random component was developed in
\citet{steenbeck66} and later was widely used for all
possible astrophysical and geophysical applications
\citep{moffatt78, krause80, ruzmaikin88, kulsrud99}. The
basic idea of the mean field approach is to average the
equations for the evolution of the magnetic field over
the small scale motions of the conducting liquid. Such
small scale motions can be either a collection of waves
with random phases, or turbulent pulsations, or randomly
occurring jets or plumes  with the sizes considerably
smaller than the scale of the whole system. Formal
application of the mean field theory to the star-disk
collision dynamo provides one more mean of
justification that such a dynamo is operational. 

The number of plumes produced by star-disk collisions is
large. At any given moment of time there exist $\sim
10^4$ plumes inside  $r\sim 10^{-2}\,\mbox{pc}$ (see
Paper I).
The radius of each plume is
$r_{p} \simeq  H \simeq 3.7
\cdot 10^{-3} r$  at $r \le r_{ab}$ as shown 
in section~4.2 of
paper~I and equation~(8) of paper~I. Therefore, the
distance between neighboring plumes is
$\sim 10^{-2} r$ and the radial and azimuthal sizes of
the plumes cannot  exceed $\sim 10^{-2} r$ without
overlapping each other. This condition is well satisfied
with  ${\bar q}_{<r}$ given by
expression~(\ref{sc_eqn3.37}). The magnetic field on the
scale of the  order of
$r$ will be the average over many individual plumes.  The
occurrences of plumes are  statistically independent but
each  plume can be considered nearly identical to any
other, because the  star velocities at any given radius
are about the same.  However, to  the extent that the star
sizes vary, the energy input to each plume  will vary
accordingly and therefore the size of plumes could  be
considered as a random noise process, but the spectral
range is  limited. It is attractive to apply mean field
theory for the  generation of  the large scale magnetic
field by plumes.  The averaging over the patches of the 
disk surface exceeding the  size of individual plumes is
well justified. The averaging over the  vertical
direction is more problematic, since the sign of
helicity  produced by  plumes exactly reverses above and
below the disk  midplane. In addition  the typical size 
of a plume is of the same  order as the vertical scale of
the change of helicity. Still  we explore the results of
the application of mean field theory  equations for the
excitation of the global large scale field and  attempt
to identify the departure points of mean field theory
from  the more coherent flux rotation analysis in this
section.

The random  motions induced by the star-disk collisions
are  clearly  statistically anisotropic due to the
existence of a preferred  direction  perpendicular to the
disk plane as well as a preferred  direction of rotation
on either side of the disk. Still using  isotropic
expressions for the equations of the mean field  theory
provides so  much simplifications that for our purpose of 
obtaining a proof of principle  estimate as well as a
comparison to  flux rotation theory, we will use
isotropic equations of the mean  field theory.   The mean
electromotive force is given  by
\begin{equation}
\overline{ {\bf  v}^{\prime}\times{\bf
B}^{\prime}}=\alpha {\bar {\bf  B}}-\beta
\nabla\times{\bar {\bf  B}}\label{eqn3.47}\mbox{,}
\end{equation} where ${\bf v}^{\prime}$  are velocities
of small scale motions,
${\bf B}^{\prime}$ is a small  scale field and the bar
means averaging  over small scales (the  distances 
between individual plumes and sizes of the plumes in 
our case). The  expressions
for the coefficients
$\alpha$ and $\beta$  are
\begin{eqnarray}  &&
\alpha=-\frac{\tau}{3}\left<{\bf
v}^{\prime}\cdot(\nabla\times{\bf  v}^{\prime})
\right>\label{eqn3.48}\mbox{,} \\  &&
\beta=\frac{\tau}{3}\left<{\bf  v}^{\prime
2}\right>\label{eqn3.49}\mbox{.}
\end{eqnarray} Here 
$\tau$ is the time of the decorrelation of the Lagrangian
velocities,  i.e. the time of the ``memory'' of a fluid
particle about the past  history of its velocity. The
$<>$ brackets denote averaging over the  statistical
ensemble and in practice can   usually be replaced by the 
averaging over the volume larger than the typical scale
of  the  random flow, ${\bf v}^{\prime}$, but smaller
than the scale of the  change of the  statistical
properties of ${\bf v}^{\prime}$ and  smaller than any
large scale of  the variability of the mean flow and 
mean magnetic field.

If the mean large scale flow and large scale  magnetic
fields are axisymmetric, then one needs to solve  the
following system of equations for the evolution of  mean
axisymmetric magnetic field in cylindrical coordinates
$r$, 
$\phi$, $z$ (corresponding unit vectors are
${\bf e}_r$, ${\bf  e}_{\phi}$, ${\bf e}_z$)
\citep{roberts92}
\begin{eqnarray} && \frac{\partial  A}{\partial
t}+\frac{1}{r}{\bf  v}_P\cdot\nabla(rA)=
(\beta+\eta)\left(\nabla^2
A-\frac{1}{r^2}A\right)+\alpha  B_{\phi}
\label{eqn3.50}\mbox{,} \\ && 
\frac{\partial B_{\phi}}{\partial t}+r{\bf 
v}_P\cdot\nabla\left(
\frac{1}{r}B_{\phi}\right)=r{\bf  B}_P\cdot\nabla\Omega
+  (\beta+\eta)\left(\nabla^2
B_{\phi}-\frac{1}{r^2}B_{\phi}\right) 
\nonumber\\  && -\alpha\left(\nabla^2
A-\frac{1}{r^2}A\right)-\frac{1}{r}\nabla\alpha\cdot
\nabla(rA)
\label{eqn3.51}\mbox{.}
\end{eqnarray} Here 
$A$ is related to the poloidal magnetic flux $F_P$ as
$F_P=2\pi rA$, 
$B_{\phi}$ is the toroidal magnetic field,
${\bf B}_P$ is the  poloidal magnetic field,
${\bf v}_P$ is the  poloidal velocity field,  and
$\Omega=\Omega(r,z)$ is the angular velocity of
differential  rotation. The quantity $A$ is also a
$\phi$-component of a vector  potential of the mean
magnetic field and
${\bf B}_P=\nabla\times  (A{\bf e}_{\phi})$, where
${\bf e}_{\phi}$ is a unit vector in  toroidal direction.

Averaging over the statistical ensemble $<>$  in
equations~(\ref{eqn3.48}) and~(\ref{eqn3.49}) is
replaced by  averaging  over many neighboring plumes. The
correlation time $\tau$  is approximately half of the
Keplerian period,
$\tau=T_K/2$.

In  fact, $\alpha$ and $\beta$ are tensors because of the
statistical  anisotropy of the plumes. The
generalisation  of expression~(\ref{eqn3.47}) for the
mean electromotive force,  including effects of 
anisotropy, is (e.g., 
\citealt{moffatt78}):
$\overline{ {\bf  v}^{\prime}\times{\bf B}^{\prime}}_i
=\alpha_{ik} {\bar  B}_k  -\beta_{ijk}
\partial {\bar B}_j/\partial x_k,$ where there is a 
summation over repeated indices. In the limit of the
$\alpha\omega$  dynamo, when
$\mbox{Rm}_{\alpha} \ll \mbox{Rm}_{\Omega}$, the most 
important term in the expression for the  mean
electromotive force  is
$\alpha_{\phi\phi} {\bar B}_{\phi}$. This term describes
the  conversion of the toroidal to poloidal magnetic flux
using the  language of the mean field theory. It is
analogous to the term on the  right hand side of
equation~(\ref{eqn3.F_T}) describing the  production of
the poloidal flux in the language of the flux rotation 
dynamo. It is this term, which determines the growth
rate of the 
$\alpha\omega$ dynamo. The generation of the mean field
in  anisotropic random medium is possible for the mean
kinetic helicity, 
$\left<{\bf v}^{\prime}\cdot(\nabla\times{\bf
v}^{\prime})
\right>$,  equal to zero, but for non-vanishing
components of the tensor $\left<  v^{\prime}_i (\nabla
\times {\bf v}^{\prime})_k 
\right>$
\citep{krause80, molchanov83}.
\citet{ferriere93a,  ferriere93b, ferriere98} performed 
detailed calculations of  the
$\alpha$ and $\beta$ tensors resulting from the
plume-like  motions of gas in differentially rotating
Galactic disk caused by  randomly placed supernovae
explosions. These motions have some  limited similarity 
to the plumes considered in the present work in  that
they also result in the conversion of the toroidal to  the
poloidal magnetic flux and are anisotropic due to the
vertical  density gradient in the Galaxy.  Subsequently,
these results were  used by \citet{ferriere00} to
calculate kinematic anisotropic 
$\alpha\omega$ dynamos. In the present work,
$\alpha$ and $\beta$  effects are assumed to be isotropic.

We now estimate the magnitude  of the coefficients
$\alpha$ and
$\beta$ in equation~(\ref{eqn3.51}).  The half thickness
of the slab with the  helicity produced by plumes  is
about the vertical extent of a plume,
$l$.  We assume the  dependence of
$\alpha$ on
$z$ as $\alpha=\alpha_0 z/l$, where 
$\alpha_0$ is a characteristic value of  helicity which
can vary  with the radius
$r$. This assumption for $\alpha$  satisfies symmetry 
requirement that $\alpha(-z)=-\alpha(z)$ while exact 
knowledge of  the dependence of $\alpha$ on $z$ is beyond
our accuracy.  We assume  that $l>H$ and that
$\alpha=\alpha_0 z/l$ in the whole region 
$-l<z<l$, i.e. we neglect the fact, that helicity is
almost zero  inside the disk for
$-H<z<H$. We also assume the turbulent magnetic 
diffusivity, $\beta$, to be uniform over
$-l<z<l$. The fact that  the maximum  height of the plume
is $l$ means that the characteristic  vertical velocity 
of the plasma in the plume is $v_z^{\prime}\approx  v_K
l/r$. We assume that the  characteristic velocity of the
sideways  expansion of the plume is
$v_s^{\prime}\approx v_z^{\prime}/2$. Then,  by the time
$T_K/2$ the plume expands to $\approx l/2$ in horizontal 
dimension (we neglect the fact that the shape of the
plume becomes  elliptical). We estimate
$\nabla\times{\bf v}^{\prime}\approx  -2\Omega_K {\bf
e}_z$, and therefore
${\bf  v}^{\prime}
\cdot(\nabla\times{\bf  v}^{\prime})\approx
-2v_z^{\prime}\Omega_K=-2l\Omega_K^2$. Similarly 
${\bf v}^{\prime 2}=2v_s^{\prime 2}+v_z^{\prime
2}\approx (3/2) v_K^2  (l^2/r^2)$ for the plume. Let us
introduce the filling factor 
$q=q(r)$ equal to the fraction of the surface of the one
side of the  disk covered by plumes. Then averaging,
$<>$, is  reduced to the  multiplication of the values
for one plume by $q$.  From expression~(\ref{eqn3.48})
and the above estimate of
${\bf  v}^{\prime} \cdot(\nabla\times{\bf v}^{\prime})$
we  have
\begin{equation}
\alpha_0=\frac{2\pi}{3}\cdot l \cdot \Omega_K 
\cdot q
\label{eqn3.52}\mbox{,}
\end{equation} and from  expression~(\ref{eqn3.49}) and
the above  estimate of ${\bf  v}^{\prime 2}$ we have
\begin{equation}
\beta=\frac{\pi}{2}\cdot 
\Omega_K \cdot l^2
\cdot q\label{eqn3.53}\mbox{.}
\end{equation} Our  estimate of $\beta$ coincides with
the estimate of the characteristic  value of $\beta$ for
an ensemble of supernovae explosions occurring  at the
midplane of the Galaxy considered by \citet{ferriere93b} 
(formula [35] in that work). The numerical coefficient
in  our estimate of $\beta$ is slightly different  from
\citet{ferriere93b}.

The dynamo activity is present inside the  thin layer
with thickness $l\ll r$. This situation is the same as
for  the traditional model of the $\alpha \omega$
Galactic dynamo. We can  use the extensive theory of the
$\alpha \omega$ dynamo  in thin disks  developed in the
connection with the Galactic dynamo. An  extensive
treatment of $\alpha \omega$ Galactic dynamo can be found 
in  \cite{stix75}, \cite{zeldovich83}, and
\cite{ruzmaikin88}.  One looks for the solution of 
equations~(\ref{eqn3.50}) and~(\ref{eqn3.51}) in the
$\alpha \omega$  limit when
$\mbox{Rm}_{\alpha}\ll \mbox{Rm}_{\Omega}$. Since  the
thickness of the disk, $2H$,  is small, one can neglect
radial  derivatives of the magnetic field compared to the
$z$-derivatives. In  this way the problem becomes local
with the eigenfrequency  of the  dynamo determined by
solving the one dimensional eigenvalue problem  in
$z$-direction. This local approximation is similar to
the local  approximation used in Appendix~A to derive the
vertical structure  of the  accretion disk. We will use
results from
\cite{ruzmaikin88}  and replace their  parameters with
ours. The important parameter is  the dynamo number
\begin{equation}  D=r\frac{d\Omega_K}{dr}\frac{\alpha_0
l^3}{(\beta+\eta)^2}=-\frac{\pi\Omega_K^2  q l^4}{
\left(\eta+\frac{\pi}{2}\Omega_K l^2  q\right)^2}
\label{eqn3.54}\mbox{.}
\end{equation} The $D$ is  negative for anticyclonic
vortices and $d\Omega_K/dr<0$.

The density  of particles in equilibrium non-magnetized
disk falls off with $z$  precipitously: $\propto \exp
(-z^2/H^2)$ when  the gas pressure  dominates and even
steeper when radiation pressure  dominates
\citep{shakura73}. This means that even a small magnetic 
field will have a significant influence on the dynamics
of the disk  corona. Thus, the  kinematic dynamo
approximation does not work in  the disk corona. There
the force-free approximation
$\nabla\times{\bf  B}=\lambda {\bf B}$ describes the
magnetic field evolution at 
$|z|>l$. In the particular case
$\lambda=0$ the force-free magnetic  field satisfies the
vacuum equation
$\nabla\times{\bf B}=0$. 
\cite{reyes99}  investigated the
$\alpha \omega$ turbulent dynamo in  accretion disks
with  linear force-free coronae. They  match
axisymmetric solutions of the dynamo
equations~(\ref{eqn3.50})  and~(\ref{eqn3.51}) inside the
disk to the  solutions with constant 
$\lambda$ of a force-free equation $\nabla\times{\bf
B}=\lambda{\bf B}$ outside  the disk. They find that the
results for the dynamo eigenvalues and  dynamo eigenmodes
do not change significantly with the value of 
$\lambda$.  The
$\alpha$-quenched saturated mode also depends weekly  on
$\lambda$. Thus, in order to obtain estimates for the
star-disk  collisions driven dynamo we can assume that
$\lambda=0$ and the  magnetic fields obey the vacuum
condition $\nabla\times{\bf B}=0$  outside the disk.
Note, however, that some of the poloidal magnetic  field
lines  obtained in \cite{reyes99} have inclination
angles to  the surface of the accretion  disk less than
$60^{\circ}$.  This  means that MHD outflow should start
along these poloidal magnetic  field lines
\citep{blandford82}. The presence of the MHD  outflow
would make the  force-free approximation invalid.
However,  these field lines, although radial initially,
after many turns become  wrapped up into a force-free
helix where the radial magnetic field  becomes smaller
than either the external poloidal or toroidal fields. 
Both these external fields, in turn are smaller than the
toroidal  field inside the disk
\citep{li01a}. Since the magnetic field inside  the disk
is much stronger than outside the disk, the  boundary
condition at the top of the plume zone, $z=\pm l$, can be 
approximated as on the boundary with the vacuum:
$B_{\phi}=0$ and 
$B_r=0$.

The eigenvalue problem for the $\alpha \omega$ dynamo in 
the thin slab $-l(r)<z<l(r)$ with the vacuum outside the
slab  (\cite{ruzmaikin88}) can be reduced to solving a
one-dimensional  eigenvalue problem in the
$z$-coordinate. In this way, the local  growth rate of
the dynamo $\Gamma(r)$ is obtained. The growth  rate  of
the global mode $\Gamma$ is very close to the maximum
value of 
$\Gamma_m=
\Gamma(r_m)$  over the disk radius. The  corresponding
eigenmode is localized in the ring of the  disk  near
radius $r_m$. The characteristic radial width of the
eigenmode  for the  dynamo numbers, that do not much
exceed the threshold  limit, is $\sim (lr_m)^{1/2}$
\citep{ruzmaikin88}. The most easily  excited mode of
the dynamo has quadrupole  symmetry and is steady.  The
excitation condition of this most easily excited mode  is
$D<-\pi^4/16$ for  the vertical dependence of  the
$\alpha$-coefficient $\alpha=\alpha_0 z/l$
\citep{ruzmaikin88}.  The excitation  condition varies
somewhat depending on the choice of  the profile of the
$\alpha$-coefficient but is of the same order as  for the
linear profile of $\alpha$. The growth rate of the most 
easily excited steady state quadrupole mode not far from
the  excitation threshold  is
\begin{equation}
\Gamma=\frac{\beta+\eta}{l^2}
\left(-\frac{\pi^2}{4}+\sqrt{|D|}\right)
=\frac{\pi}{2}\Omega_K 
\cdot q \left(-\frac{\pi^2}{4}+
\frac{2}{\sqrt{\pi q}}\right)  -
\frac{\pi^2}{4}\frac{\eta}{l^2}
\label{eqn3.55}\mbox{.}
\end{equation} The  growth rate for large dynamo numbers,
$|D|\gg \pi^4/16$, or for small 
$\eta$ is
\begin{equation}
\Gamma=  0.3\frac{\beta+\eta}{l^2}\sqrt{\pi|D|} =
0.3\cdot 
\Omega_K
\pi\sqrt{q}
\label{eqn3.56}\mbox{.}\end{equation}

This  differs from Eq.~(\ref{eqn3.41a}) by  a negligible
factor, $\sim  0.35$, for
$\alpha_{plume} =1$, in view of the many  approximations.
We therefore conclude that mean field dynamo theory 
results in a similar growth rate to that predicted by
the  flux  rotation analysis.  In either case the growth
is so rapid in view of  Eq.~(\ref{eqn3.45}) that nearly
the entire history of the accretion  disk dynamo will be
dominated by the near steady state saturated 
conditions.  Unfortunately this steady state is beyond
the scope of  the present paper where instead we feel
satisfied in demonstrating an  understanding of the
dynamo gain using a flux rotation model, a mean  field
theory, and numerical simulations.

We see that the filling  factor $q(r)$ is crucial for the
mean field dynamo. Let us estimate 
$q(r)$. The cross section area of the plume is $\pi r_p^2
\approx \pi  H^2$, the number of plumes present at any
moment of time on one  side of the disk is $2\cdot nv/4
\cdot T_K/2$. Therefore, one has
$$  q=\frac{nv}{4}2\frac{T_K}{2}\pi H^2 \mbox{.} $$
Using expression~(16)  of paper~I for the flux of stars,
$nv/4$, and expression~(A5) of  paper~I for the disk
half-thickness, one obtains
\begin{eqnarray} &&  q=1.52\cdot 10^{-3} \cdot n_5 
\left(\frac{r}{10^{-2}\,
\mbox{pc}}\right)\left(\frac{l_E}{0.1}\right)^2
\left(\frac{\epsilon}{0.1}\right)^{-2}
\quad 
\mbox{for} \quad 10 r_t < r < 10^{-2}\,\mbox{pc}
\mbox{,} \nonumber 
\\ && q=0 \quad \mbox{for} \quad r < 10 r_t
\label{q_differ} 
\mbox{.}
\end{eqnarray}

The ratio of the toroidal to the poloidal  or radial
magnetic field in the growing mode and  inside the volume 
occupied by plumes is
$$
\frac{B_T}{B_P}\approx  |D|^{1/2}=
\frac{2}{\sqrt{\pi q}}
\mbox{.}
$$ Using  expression~(\ref{q_differ}) for the value of
$q$ one  has
$$
\frac{B_T}{B_P}\approx 63\,n_5^{-1/2} \cdot
\left(\frac{r}{10  r_t}\right)^{-1/2}
\left(\frac{l_E}{0.1}\right)^{-1}
\left(\frac{\epsilon}{0.1}\right)
\mbox{.}
$$

As  in all $\alpha\omega$ dynamos, the generated toroidal
field is larger  than the poloidal field. However, the
toroidal field in the vacuum  outside  the region of
dynamo activity vanishes, because the  normal component
of the current at the vacuum boundary must be zero.  If
there is conductivity, as we expect,   and therefore
force-free  magnetic field  above the plume region, then
the toroidal magnetic  field generated by the  dynamo
penetrates into this  region
\citep{reyes99}. However, due to the  quadrupole
symmetry of  the poloidal magnetic field, the toroidal
field in the  force-free  corona has the opposite
direction from the toroidal field inside the  disk.  The
axial component of the magnetic field, $B_z$, is  much
smaller than the  radial component inside the slab
occupied by  plumes, $B_z\approx (l/r) B_r$.  However,
the radial component of the  magnetic field decreases
down to the  value comparable to $B_z$ at 
$|z|=l$. The quadrupole poloidal field in the  corona is
weaker than  the poloidal magnetic field inside the disk
by the factor $l/r$. The  structure of the force-free
corona above the dynamo  generation  region cannot be
determined without further knowledge about boundary 
conditions at the outer boundaries of the force-free
region  or physical  processes, which limit the
applicability of force-free  ideal MHD approximation in
the corona (i.e., fast reconnection of  magnetic fields).
If one requires that the magnetic field in the 
force-free region vanishes for $|z|\gg l$, as
\citet{reyes99} assume,  then, the toroidal magnetic
field is comparable to  the poloidal  field in the
corona. In this case, the toroidal magnetic field  in  the
force-free corona is much smaller than the toroidal
magnetic  field  inside the disk, and so we neglect it in
the simulations. In  the actual case of the black hole
accretion disk dynamo, we expect  the coronal field to be
force-free and to progressively remove the  flux and
magnetic energy  generated by the dynamo in a force-free 
helix as described in \citet{li01a} where the field
strength, as  discussed above, is of the order of the
poloidal  field.

\section{The Dynamo Equations and Numerical  Method}
\label{sec6}

Because of limited numerical resolution and  limited
computing time we cannot attempt to directly simulate the 
dynamo problem for the real astrophysical parameters.
Three  dimensional simulations of just one star passage
through the  accretion disk is already quite a challenge
for computational gas  dynamics. Even if we assume that
we know the velocity field for a  single star-disk
collision and treat only the kinematic dynamo  problem,
the existence of $\sim 10^4$ plumes, the necessity  of
good  resolution in the space between and above the
plumes, and long  evolution times required by the dynamo
problem make the  direct  computations very difficult
and demanding of major computer  resources. Numerical
modeling done in this work illustrates and  proves
essential features of the star-disk collisions  dynamo
described above. We simulate the kinematic dynamo with
only a  few plumes  present and adopt a simplified flow
model for  individual plumes.  Then, we compare the
numerical growth rate and  magnetic field structure to
the predictions of flux rotation and mean  field theories
extrapolated to a small number of plumes. Qualitative 
agreement between all three approaches in the limit of
only a  few  plumes is observed.

\subsection{Basic  Equations}
\label{subsec6.1}

We have computed  order of magnitude  estimates of the
growth rate and  threshold   for the dynamo by  direct
numerical simulations. For that purpose we have written a 
3D kinematic dynamo code evolving the vector potential
${\bf A}$ of  the magnetic field in a given velocity field
${\bf v}$ and with  resistive  diffusion. The code is
written in cylindrical geometry. We  start with the
equations describing the  evolution of fields in 
nonrelativistic quasineutral plasmas.
\begin{eqnarray} && \nabla\cdot  {\bf B}=0\label{eqn7},
\\  && \frac{1}{c}\frac{\partial {\bf B}}{\ptl  t}=
-\nabla\times{\bf E}\label{eqn8}, \\ &&
\nabla\times {\bf  B}=\frac{4\pi}{c}{\bf j}\label{eqn9},
\\ && {\bf j}=\sigma\left({\bf  E}+\frac{1}{c}{\bf v}
\times{\bf  B}\right),\label{eqn10}
\end{eqnarray}  where $\sigma$ is the  conductivity of
the plasma. Because we are considering the kinematic 
dynamo,
$\bf v$ is specified and the momentum equation is 
ignored. Substituting the expression for the current
${\bf j}$ from  the equation~(\ref{eqn9}) into Ohm's law,
equation~(\ref{eqn10}), and  introducing a coefficient of
magnetic diffusivity $\eta$  as
$\displaystyle\eta=\frac{c^2}{4\pi\sigma}$ we obtain
Ohm's law in  the form
\begin{equation} {\bf E}+\frac{1}{c}{\bf  v}\times{\bf
B}=\frac{\eta}{c}\nabla\times{\bf  B}
\label{eqn11}\mbox{.}
\end{equation} subject to the  constraint (\ref{eqn7}).

The conventional and widely accepted way of  writing and
solving the kinematic MHD equations (MHD without the 
hydrodynamical part) is to obtain a single equation for
the evolution  of the magnetic field. Substitution of the
electric field
${\bf E}$  from the equation~(\ref{eqn11}) into Faraday's
law, equation  (\ref{eqn8}), results in
\begin{equation}
\frac{\ptl {\bf  B}}{\ptl
t}=-\nabla\times\left(\eta\nabla\times{\bf  B}\right)
+\nabla\times({\bf v}\times{\bf  B})\label{eqn12}\mbox{.}
\end{equation}  Introducing the vector  potential $\bf
A$ with
$$ {\bf B}=\nabla\times{\bf A}\mbox{,}
$$  equation (\ref{eqn12}) takes the form
\begin{equation}
\frac{\ptl{\bf  A}}{\ptl
t}+\eta\nabla\times\nabla\times{\bf A}  -{\bf
v}\times(\nabla\times{\bf
A})+c\nabla\varphi=0\label{eqn13}\mbox{,}
\end{equation} where 
$\varphi$ is the scalar potential; no gauge has been
chosen. Any  solution of equation (\ref{eqn13})
satisfying the boundary and  initial conditions for the
magnetic field should give a physical  result for the
evolution of the magnetic field. The  equation
(\ref{eqn13}) has  the same second order in  space
derivatives as equation (\ref{eqn12}) for the evolution
of the  magnetic field.

The gauge freedom can be used to simplify the  procedure
for solving equation (\ref{eqn13}). The scalar  potential
$\varphi$ may be chosen to be an arbitrary function by an 
appropriate choice of gauge transformation. For
instance, one can  choose to  set
$\varphi=0$, in which case the remaining equation  for
${\bf A}$ takes the form
\begin{equation}
\frac{\ptl {\bf  A}}{\ptl
t}+\eta\nabla\times(\nabla\times{\bf A})-  {\bf
v}\times(\nabla\times{\bf  A})=0\label{eqn19}\mbox{.}
\end{equation} The boundary conditions  for
${\bf A}$ should be consistent with the gauge chosen. In 
principle, equation~(\ref{eqn19}) requires three
separate  boundary conditions for the components of ${\bf
A}$. This number is  the  same as the number of boundary
conditions required to solve the  equation for the
evolution of the magnetic field~(\ref{eqn12}).  Note,
however, that there is still a freedom to add
$\nabla\chi$ to 
${\bf A}$ and therefore to the boundary conditions for
${\bf A}$,  where $\chi$ is an arbitrary time independent
function, and still  preserve the gauge condition
$\varphi=0$. Although any arbitrary  initialization of
${\bf A}$ satisfying the boundary conditions can be 
allowed, many initializations would result in the same
magnetic field 
${\bf B}$. Initializing eq.~(\ref{eqn12})
$\nabla\cdot{\bf B}=0$ is formally  required. We have
the following requirements for the  boundary and  initial
conditions for ${\bf A}$:
\begin{enumerate}
\item There must  be boundary and initial conditions on
all three components of ${\bf  A}$.
\item Boundary and initial conditions should be
consistent with  the gauge used.
\item The physical boundary conditions and the  initial
conditions for the magnetic  and electric fields (or  any
other quantities) specific to a particular problem must
be  satisfied.
\end{enumerate} The last requirement means that  the
physical boundary conditions must be derivable from the
boundary  conditions equations imposed on ${\bf A}$. The
reverse is not  necessarily true, i.e. for one specific
physical boundary conditions  there may be many possible
boundary conditions for ${\bf A}$. The  situation with
the boundary conditions for ${\bf A}$  is analogous to 
the situation with the initial conditions for ${\bf A}$.
With this  specification of initial and boundary
conditions, the curl of the  solution to
equation~(\ref{eqn13}) will be equal to the solution of 
equation~(\ref{eqn12}).

If one chooses to evolve the magnetic field  directly,
then   in addition to the equation  of
evolution~(\ref{eqn12}) the magnetic  field must obey 
the constraint
$\nabla\cdot {\bf B}=0$, which should be  specified  as
an initial condition. Although it follows from
(\ref{eqn12}) that,  once initialized to zero,
$\nabla\cdot {\bf B}$ will be kept equal  to zero, the
numerical methods used to solve (\ref{eqn12}) introduce 
discretization errors, which after a sufficient time can
accumulate  so that
$\nabla\cdot {\bf B}$ is no longer zero (e.g., 
\citealt{lau93}). Special procedures are employed in
the  codes to  deal with this problem such as
"divergence cleaning".  However, in  the case of the
evolution of the vector potential there are  three
equations~(\ref{eqn13}) to solve, while there are four
dynamic  variables in them (i.e. three components of
${\bf A}$ and one scalar  function
$\varphi$). Therefore, one can utilize this one extra
degree  of freedom in choosing
$\varphi$  for a suitable gauge constraint  without
actually imposing any constraints on three components  of
${\bf A}$. This will allow us to have freedom to choose
the gauge  and at the same time will not introduce the
necessity of taking  special measures in order to ensure
that the gauge will be kept  correctly throughout the
computation. The magnetic field is than  obtained  by
taking curl of ${\bf A}$. This way
$\nabla\cdot {\bf B}$  vanishes automatically within the
discretization error associated  with approximating the
curl by finite differencing.

In the  simulations presented in this work we used the
following  gauge
\begin{equation} c\varphi-{\bf  v}\cdot{\bf
A}+\eta\nabla\cdot{\bf A}=0\label{eqn20}
\end{equation}  One can show that for this gauge the
basic equation (\ref{eqn13})  reduces to
\begin{equation}
\frac{\ptl{\bf A}}{\ptl  t}=-A^k
\frac{\ptl v^k}{\ptl x^i}-({\bf v}\cdot \nabla)  {\bf
A}+\eta\nabla^2{\bf A}+(\nabla\cdot {\bf  A})
\nabla\eta\label{eqn21}\mbox{.}
\end{equation}  We choose the  gauge (\ref{eqn20})
because the resulting equation for
${\bf A}$ has  similarity with the equation  for the
advection of a vector quantity.  It has the familiar
advection term
$({\bf v} \cdot \nabla){\bf A}$  and diffusion term
$\eta\nabla^2{\bf A}$. The term $-A^k
\frac{\ptl  v^k}{\ptl x^i}$ corresponds to a stretching
term
$({\bf B}\cdot 
\nabla){\bf v}$ in the equation for the advection of the
magnetic  field. Finally,
$(\nabla\cdot {\bf A})\nabla\eta$ term is  associated
with the nonuniformity of electric conductivity. In  this
work we will consider the case of
$\eta=\mbox{constant}$ only  and concentrate on the
effects of the plasma flow producing the  dynamo. Thus
this  term drops out of the equations. Note, that  the
equation~(\ref{eqn21}) is valid both for incompressible
and  compressible flows.

Finally, we present equations (\ref{eqn21})  written out
in cylindrical  coordinate system
$r$, $\phi$,
$z$  (corresponding unit vectors are
${\bf e}_r$, ${\bf e}_{\phi}$, ${\bf  e}_z$)
\begin{eqnarray} && \frac{\ptl A^r}{\ptl
t}=-\left(v^r\frac{\ptl  A^r}{\ptl r}+
\frac{1}{r}v^{\phi}
\frac{\ptl A^r}{\ptl 
\phi}+v^z\frac{\ptl  A^r}{\ptl z}-\frac{1}{r}v_{\phi}
A_{\phi}\right)-\left(A^r\frac{\ptl  v^r}{\ptl
r}+A^{\phi}\frac{\ptl v^{\phi}} {\ptl  r}+\right.
\nonumber\\  && \left. A^z\frac{\ptl v^z}{\ptl r}\right)+
\eta\left(\frac{1}{r}\frac{\ptl}{\ptl
r}\left(r\frac{\ptl A^r}{\ptl  r}\right)+
\frac{1}{r^2}\frac{\ptl^2 A^r}{\ptl
\phi^2}+\frac{\ptl^2  A^r}{\ptl z^2}-
\frac{A^r}{r^2}-\frac{2}{r^2}\frac{\ptl  A^{\phi}}{\ptl
\phi}\right)+
\frac{\ptl\eta}{\ptl r}(\nabla\cdot{\bf  A})
\label{eqn22}\mbox{,} \\ && \frac{\ptl  A^{\phi}}{\ptl
t}=-\left(v^r\frac{\ptl A^{\phi}}{\ptl  r}+
\frac{v^{\phi}}{r}\frac{\ptl A^{\phi}}{\ptl
\phi}+v^z\frac{\ptl  A^{\phi}} {\ptl
z}+\frac{1}{r}v^{\phi}A^r\right)-
\nonumber\\  &&
\left(A^r\frac{1}{r}\frac{\ptl  v^r}{\ptl
\phi}+A^{\phi}\frac{1}{r}
\frac{\ptl v^{\phi}}{\ptl 
\phi}+ A^z\frac{1}{r}\frac{\ptl  v^z}{\ptl
\phi}+
\frac{1}{r}A^{\phi}v^r-\frac{1}{r}A^r  v^{\phi}\right)+
\label{eqn23} \\  &&
\eta\left(\frac{1}{r}\frac{\ptl}{\ptl
r}\left(r\frac{\ptl  A^{\phi}} {\ptl
r}\right)+\frac{1}{r^2}\frac{\ptl^2  A^{\phi}}{\ptl
\phi^2}+
\frac{\ptl^2  A^{\phi}}{\ptl
z^2}-\frac{A^{\phi}}{r^2}+\frac{2}{r^2}
\frac{\ptl  A^r}{\ptl 
\phi}\right)+
\frac{1}{r}\frac{\ptl\eta}{\ptl\phi}(\nabla\cdot{\bf
A})\mbox{,} 
\nonumber\\ && \frac{\ptl A^z}{\ptl
t}=-\left(v^r\frac{\ptl A^z}{\ptl  r}+
\frac{1}{r}v^{\phi}
\frac{\ptl A^z}{\ptl \phi}+v^z\frac{\ptl  A^z}{\ptl
z}\right)-
\left(A^r\frac{\ptl v^r}{\ptl  z}+A^{\phi}\frac{\ptl
v^{\phi}} {\ptl z}+\right. \nonumber\\  && 
\left. A^z\frac{\ptl v^z}{\ptl z}\right)+
\eta\left(\frac{1}{r}\frac{\ptl}{\ptl
r}\left(r\frac{\ptl A^z}{\ptl  r}\right)+
\frac{1}{r^2}\frac{\ptl^2 A^z}{\ptl
\phi^2}+\frac{\ptl^2  A^z}{\ptl z^2}
\right)+\frac{\ptl\eta}{\ptl  z}(\nabla\cdot{\bf
A})\label{eqn24}\mbox{,}
\end{eqnarray} where 
$\displaystyle
\nabla\cdot{\bf A}=\frac{1}{r}\frac{\ptl}{\ptl  r}(rA^r)
+\frac{1}{r}\frac{\ptl A^{\phi}}{\ptl
\phi}+\frac{\ptl  A^z}{\ptl z}$. The gauge condition
(\ref{eqn20}) takes the  form
\begin{equation} c\varphi=v^r  A^r+v^{\phi}A^{\phi}+v^z
A^z-\eta\left(\frac{1}{r}
\frac{\ptl}{\ptl  r}(rA^r)+\frac{1}{r}\frac{\ptl
A^{\phi}}{\ptl \phi}+
\frac{\ptl  A^z}{\ptl z}\right)\label{eqn25}\mbox{.}
\end{equation} Also  expressions for the magnetic field
components in cylindrical  coordinates are
\begin{eqnarray} && B^r=\frac{1}{r}\frac{\ptl  A^z}{\ptl
\phi}-\frac{\ptl A^{\phi}}{\ptl  z}\mbox{,}\quad
B^{\phi}=\frac{\ptl A^r}{\ptl z}-\frac{\ptl  A^z}{\ptl
r}\mbox{,}\quad  B^z=\frac{1}{r}\frac{\ptl}{\ptl
r}(rA^{\phi})-
\frac{1}{r}\frac{\ptl  A^r}{\ptl
\phi}\mbox{.}\label{eqn26}
\end{eqnarray}

\subsection{Boundary  and Initial Conditions}
\label{subsec6.2}

Although the use of the  vector potential eliminates the
problem with the divergence cleaning,  the boundary
conditions in terms of the vector potential may  be
somewhat more complicated and not so obvious from
intuitive  physical standpoint than the boundary
conditions for magnetic fields.  In this work we used
perfectly conducting boundary conditions at  all
boundaries of the cylinder. There is no general agreement
on what  boundary conditions are most physically
appropriate for a thick  accretion disk dynamo
simulations. For example,
\citet{stepinski88}  used vacuum boundary conditions
outside some given spherical domain  for solving the mean
field dynamo equations in axial  symmetry.  Khanna~\&
Camenzind (1996a, 1996b) also considered an axisymmetric 
mean field dynamo in the disk and in the corona
surrounding the disk  on the Kerr background
gravitational field of a rotating black hole.  They used
an artificial boundary condition that the magnetic field
is  normal to the rectangular boundary of their
computational domain and  the  poloidal component of the
current density vanishes near the  boundary. However, the
main goal of these investigations was to  demonstrate
that   certain types of helicity distributions inside the 
disk produce a dynamo. As soon as the boundary of the
numerical  domain is extended far enough from the region
of large helicity and  large differential rotation, the
influence of the boundary conditions  on the process of
the generation of the  magnetic fields far inside  from
the boundary should be small.  Since both the Keplerian
profile  of the angular rotational velocity and  the
frequency of star-disk  collisions have increasing
values toward  the central black hole,  the approximation
of a distant boundary can be applicable to the case  of
our simulations.  Therefore we have  chosen a perfectly
conducting  rotating cylindrical boundary as a simple
boundary condition  prescription. We checked that the
results of our simulations do not  strongly depend on the
position of the outer boundary.

The magnetic  field near the rotation axis is strongly
influenced by  the presence  of the black hole as well as
the general relativistic effects  associated  with the
black hole. Magnetic field lines in the region  close to
the rotation axis have their foot-points on the black
hole  horizon or  in the region between the black hole
and the inner edge  of the accretion  disk. Therefore,
one should expect that this region  of the
magnetosphere  will be also strongly influenced  by
relativistic effects of the black hole. The subject of
the  influence of the central black hole on the  magnetic
fields produced  by the dynamo is a part of the so-called
``black hole electrodynamics  '' theory  (e.g., see the 
chapter ``Electrodynamics of Black Holes''  in
\citet{frolov98}). Since the number density of stars
should  decrease near  the black hole due to their
capture by the black hole  and due to tidal disruption,
one should not expect the star-disk  collision dynamo to
operate effectively in this region, where  strong
relativistic effects are important. Therefore, for the
purpose  of this work we replace the region close to the
axis of symmetry by  imposing an inner cylindrical
boundary (also perfectly conducting).  This may be
adequate to the real astrophysical situation  in  the
coronae of the accretion disks, since there is highly
conducting  plasma there.

We choose as an initial condition a purely poloidal 
magnetic field with  even symmetry with respect to the
plane of the  disk (see Appendix~\ref{appendix_C} for
definitions and  properties of odd and even magnetic
fields). The field is contained  within the computational
boundaries such that the normal component  of the 
magnetic field is zero on all boundaries.

Let us consider  the  perfectly conducting rotating
boundaries.  There is no magnetic  flux penetrating the
boundaries. This means that the normal component  of  the
magnetic field must always remain zero on the boundary. 
If  the velocity of the boundary is
${\bf v}_{\rm b}$,  then the  tangential component of
electric field in the rest frame of the  moving boundary
$\displaystyle {\bf E}+
\frac{1}{c}{\bf v}_{\rm  b}\times {\bf B}$ is also zero
on the boundary. If ${\bf v}_{\rm b}$  and ${\bf B}$ are
both tangential at the boundary, then this implies  that
the tangential component of ${\bf E}$ is also zero
there.  This  then implies that we can chose the
$\varphi$ and the tangential  components of ${\bf A}$ to
be zero  on the boundary.  Then from 
expression~(\ref{eqn20}) and the vanishing of the normal
component of 
${\bf v}$  on the boundary, we conclude that we must
have  ${\bf 
\nabla
\cdot A} = 0$  there. Specifically we  have
\begin{equation}
\frac{1}{r}\frac{\ptl}{\ptl  r}(rA^r)=0\mbox{,}\quad
A^{\phi}=0\mbox{,}\quad A^z=0\quad
\mbox{on  the
$r=\mbox{constant}$ boundary}
\label{eqn31}
\end{equation}  and
\begin{equation}  A^r=0\mbox{,}\quad
A^{\phi}=0\mbox{,}\quad
\frac{\ptl A^z}{\ptl  z}=0\quad \mbox{on the
$z=\mbox{constant}$  boundary.}
\label{eqn32}
\end{equation} This forms a complete set of  three
boundary  conditions for three components of
${\bf A}$ on each  boundary, which are compatible both
with the physical requirements  for fields on a perfectly
conducting boundary and the gauge  condition
(\ref{eqn20}). One can also see that the  equations
(\ref{eqn31}--\ref{eqn32}) are consistent in the corners
of  the computational domain, i.e., at the intersections
of the planes 
$z=\mbox{constant}$ and cylinders $r=\mbox{constant}$.

\subsection{The Numerical Scheme}
\label{subsec6.3}

We use the  finite differences predictor-corrector scheme
to solve equations  (\ref{eqn22}--\ref{eqn24}) in
cylindrical coordinates.  For  approximating advection
and stretching terms we use central  differencing, which
gives second order accuracy in the coordinates.  The
diffusion term is approximated by the usual 7 point
stencil.  Since the numerical method is explicit, it
requires the stability  condition to be satisfied. Let us
denote discretization intervals in  coordinates and time
as
$\Delta r$, $\Delta \phi$, $\Delta z$,  and
$\Delta t$ and define the quantities
$\displaystyle  s_r=\frac{\eta \Delta t}{
\Delta r^2}$, $\displaystyle  s_{\phi}=\frac{\eta
\Delta t}{r^2 \Delta\phi^2}$,
$\displaystyle  s_z=\frac{\eta \Delta t}{
\Delta z^2}$ and $\displaystyle  C_r=\frac{v_r
\Delta t}{\Delta r}$,
$\displaystyle  C_{\phi}=\frac{v_\phi
\Delta t}{r \Delta\phi}$, $\displaystyle  C_z=\frac{v_z
\Delta t}{\Delta z}$. Then, the stability  conditions
that we used in our  simulations  are
\begin{equation}
s_r+s_{\phi}+s_z<\frac{1}{2}\mbox{,}\quad
(C_r+C_{\phi}+C_z)^2<2(s_r+
s_{\phi}+s_z)\label{eqn33}\mbox{.}
\end{equation}  One can show that these conditions follow
from the local linear  stability analysis of the  dynamo
equations~(\ref{eqn22}--\ref{eqn24}). Before doing each
new  cycle of predictor-corrector calculations we set up 
the value of the  time step
$\Delta t$. First, we choose some reasonable value  of
$\Delta t$ dictated by the accuracy requirements or  how
frequent  we want to get an output measurements from our
simulations. Then, we  decrease the value of
$\Delta t$ until the first of the conditions  in
equation~(\ref{eqn33}) is satisfied. After that we check
the  second  condition in (\ref{eqn33}) and see, if it is
satisfied. If  not, than we decrease
$\Delta t$ further. One can see, that the  second
condition in  equation (\ref{eqn33}) will be always
satisfied  at some value of
$\Delta t$ since the right hand side depends  on
$\Delta t$ linearly while the left hand side depends
on 
$\Delta t$ quadratically. The first stability criterion is the
usual one  for the diffusion equation and means that the
diffusion per single  time step propagates no further
than through only a single grid cell.  The second
condition is specific for central differences in the 
advection term and means that the distance the magnetic
field is  advected during one time step
$\Delta t$ is less than the distance  through which the
field diffuses per single time step
$\Delta t$  (e.g., \citealt{fletcher92}). In practice, we
ensure stability by  using a safety coefficient of $0.9$
in the inequalities  (\ref{eqn33}).

When coding the boundary  conditions
(\ref{eqn31}--\ref{eqn32}) we used a second order  one
sided difference scheme for approximating derivatives.
The  resulting expressions have been solved for the
unknown value of the  component of
${\bf A}$ at the point on the boundary.  Boundary
conditions  have been updated after both predictor  and
corrector steps. In the
$\phi$ direction seamless periodic  boundary conditions
have been used, i.e.  we make the first and the  last
grid points in the
$\phi$ direction identical and corresponding  to
$\phi=0$ and $\phi=2\pi$ and use the same difference
scheme as for  other values of
$\phi$ to update these points. Also we used the  same
seamless treatment of lines
$\phi=0$ and $\phi=2\pi$ at the  radial cylindrical
boundaries and at the top and bottom  boundaries.

The code is  able to treat both the domains with  an
inner radial boundary and the domains including the
symmetry axis.  In the latter case, there is a
singularity of the grid at
$r=0$,  namely, all grid points having
$r=0$ and all values of $\phi$ from 
$0$ to
$2\pi$ coincide.  One needs a special treatment of the
grid  points at $r=0$ ensuring the regularity of
Cartesian  components
$A^x$, $A^y$, $A^z$ of ${\bf A}$ and the  correct
asymptotes for $A^r$,
$A^{\phi}$ and $A^z$.  If the  values  of the Cartesian
components at $r\to 0$ are
$A^x_0$,
$A^y_0$, 
$A^z_0$, then the asymptotic behavior of the polar
components is $A^r 
\to A^x_0\cos\phi + A^y_0\sin\phi$,
$A^{\phi} \to - A^x_0\sin\phi+A^y_0\cos\phi$,
$A^z \to A^z_0$. To impose these  asymptotic conditions
we first interpolate
$A^x_0$,
$A^y_0$, and 
$A^z_0$ by calculating the average over
$\phi$ of the Cartesian  components of the vector
potential at  grid points situated on a ring  with radius
$\Delta r$. We take this  average for
$A^x_0$,
$A^y_0$,  and  $A^z_0$. Then, we assign the values of the
components of ${\bf  A}$ in the cylindrical coordinate
system at $r=0$ according  to
$A^r(\phi)= A^x_0\cos\phi +  A^y_0\sin\phi$,
$A^{\phi}(\phi)=-A^x_0\sin\phi+A^y_0\cos\phi$,
$A^z(\phi)=  A^z_0$. This finalizes the prescription for
the boundary condition  at
$r=0$.  When the symmetry axis $r=0$ is included in  the
computational region,  the code slows  considerably
because of  the small ($\Delta\phi\Delta r$) distance
between grid points in the 
$\phi$ direction and, therefore, more restrictive
limitations on the  time step imposed by the first of the
conditions  (\ref{eqn33}).

\subsection{Code Testing}
\label{subsec6.4}

In the  process of writing the code we performed tests
for separate  parts of  the code and, then, for the
complete code itself. The diffusion part  of the code has
been tested by reproducing the analytic solution  for
eigenmodes of the diffusion equation
$\displaystyle
\frac{\ptl  {\bf A}} {\ptl t}=\eta\nabla^2{\bf A}$ with
${\bf A}=0$  boundary  conditions. A variety of different
eigennumbers have been tested and  decay rates are found
to be in excellent agreement with analytic  expressions.
The code preserves the shape of eigenmodes with  very
high  accuracy even for a very moderate number of  nodes.
Coupling between equation (\ref{eqn22}) for
$A^r$ and equation  (\ref{eqn23}) for
$A^{\phi}$ has been tested by evolving  nonaxisymmetric
eigenmodes.

The advection part  of the code has been  tested by
computing the advection by the uniform  flow of  the
magnetic field of the type ${\bf B}=B{\bf n}$, where
${\bf n}$ is  a fixed  vector of unit length (we made a
few runs with different  directions of
${\bf n}$), and the magnitude of the magnetic field
$B$  has the constant  gradient vector
$\nabla B=\mbox{constant}$  perpendicular to
${\bf n}$.  The current density corresponding to  such a
magnetic field is uniform, and therefore, the magnetic
field  does not diffuse. The boundary condition for this
test was set to  time-dependent explicit values computed
from the known purely  advective behavior of the field.
We observed  good agreement with the  picture of the pure
advection of  flow.

We also compared the results  for dynamo simulations with
the two dimensional flow given by our code to  the simulations
produced by two other 2D kinematic dynamo codes,  one
evolving vector potential and another evolving magnetic
field.  The latter 2D code has a divergence cleaning
procedure for 
$\nabla
\cdot {\bf B}$. The flow was an axisymmetric Beltrami
flow  with
$\nabla\times {\bf v}=\lambda {\bf v}$. For the interior
of the  domain
$0<r<R_{o}$ and $0<z<L$ one can obtain the following
analytic  solution for the Beltrami flow:
\begin{eqnarray}  &&
v^r=J_1\left(j_{11}\frac{r}{R_o}\right)\frac{\pi}{L}\sin\frac{\pi
z}{L}
\mbox{,}\nonumber\\  &&
v^z=\frac{j_{11}}{R_o}J_0\left(j_{11}\frac{r}{R_o}\right)\cos\frac{\pi
z}  {L} \mbox{,}\nonumber\\ &&  v^{\phi}=\lambda_B
J_1\left(j_{11}\frac{r}{R_o}\right)\cos\frac{\pi  z}{L}
\mbox{,}\nonumber
\end{eqnarray} where $J_0(x)$ and
$J_1(x)$  are the Bessel functions,
$j_{11}$ is  the first root  of
$J_1(x)=0$,
$\displaystyle \lambda_B^2=\frac{j_{11}^2}{R_o^2}+
\frac{\pi^2}{L^2}$.  The solution can also be written in
terms of the flux  function
$\Psi(r,z)$:
\begin{eqnarray}  &&
\Psi=rJ_1\left(j_{11}\frac{r}{R_o}\right)\cos\frac{\pi
z}{L}
\mbox{,}\nonumber 
\\ && v^r=-\frac{1}{r}\frac{\ptl
\Psi}{\ptl z}\mbox{,}\quad  v^z=\frac{1}{r}\frac{\ptl
\Psi}{\ptl r}\mbox{,}\quad  v^{\phi}=\frac{\lambda_B
\Psi}{r}\mbox{.} \nonumber
\end{eqnarray}  The 3D kinematic code picks up the
fastest growing mode of the  dynamo. In the  case of
axisymmetric flows the nonaxisymmetric modes  of the
field ($\propto e^{im\phi-i\omega t}$) with  different
azimuthal wavenumber $m$ evolves separately. The  fastest
growing mode in our simulations was with $m=1$. The
growth  rate and the structure of the $m=1$ modes
obtained with 3D and 2D  codes agrees remarkably well. We
also studied the convergence with  respect to the grid
resolution and  found that for a magnetic  Reynolds
number
$R_m$ (defined as the product of maximum velocity  and
minimum of
$L$ and $R_o$) of about $200$ the simulations  converge
for the grid resolution of about 41x61x41 in
$r$,$\phi$, and 
$z$ directions respectively.

\section{Results of Numerical  Simulations}
\label{sec7}

\subsection{Model of the Flow  Field}
\label{subsec7.1}

We now approximate the flow for our  kinematic code  from
the analysis  of the simplified model of plumes  in
Section~\ref{sec3}. When describing the results of our 
numerical simulations, we will use  
dimensionless units with the
unit  of length equal to the radius at which the star-disk
collisions  occur, and the unit of velocity equal to the
Keplerian  velocity at  that radius. Then, one turn of
the disk at unit radius takes
$2\pi$  dimensionless units of time. The disk is assumed
to have constant  thickness. Its top boundary is at
$z=z_{top}$ and bottom boundary is  at
$z=z_{bot}$. We usually put the disk at
$z=0$, in the middle of  computational cylindrical
domain, and then,
$z_{bot}=-z_{top}$.  However, we will preserve separate
notations for top and bottom  boundaries. For simplicity
we assume that all  star-disk collisions  happen at unit
radius, but are randomly distributed in azimuthal  angle
along
$r=1$. Also, a remarkable feature of star-disk
collisions  is that the numbers of stars crossing the disk
in both directions are  equal on average. We consider two
models for the position of  star-disk collisions
addressing this property. In the first model we  assume
that collisions happen in pairs: at each time there are
two  collisions at
$r=1$, one with the star going up through the disk  and
the other at the opposite point on  the circle
$r=1$ with the  star going down through the disk. Thus,
at any moment of time the  flow is symmetric with respect
to the inversion relative to the  central point of the
disk. The second model considers random  directions of
plumes as well as random distribution of plumes  over
the  circle
$r=1$. In the next section we describe the  results
obtained with both models.

The plume flow is superimposed  onto a background of
Keplerian differential rotation  occupying the  whole
computational domain
$\displaystyle {\bf v}_K  =\frac{1}{r^{1/2}}{\bf
e}_{\phi}$. A star-disk collision is simulated  by a
vertically progressing cylinder of radius
$r_p$ in the  corotation frame.  The cylinder  starts at
the bottom of the disk  located at
$z=z_{bot}$, penetrates the disk,  and rises to a height 
of
$h$ above the disk. At the same time the cylinder rotates
about  its axis opposite to the local  Keplerian frame
such that the  cylinder does not rotate about its axis if
viewed in laboratory frame  (an inertial frame where the
central black hole is at rest), but the  axis corotates
with the local Keplerian frame. By the time the plume 
reaches its highest point,
$\pi/2$ radians of Keplerian rotation,  the  axis
corotates  with the Keplerian flow also  by
$\pi/2$ radians  on average. Since the cylinder  does not
rotate about its axis,  the  relative rotation between
the cylinder and Keplerian flow corresponds  to an 
untwisting of
$\pi/2$ radians, when the local frame  rotates
$\pi/2$ radians as measured at the radius of the axis of
the  jet.  The length of the cylinder is progressive with
time and its  velocity, $v_{pz}
\approx v_K$.  The vertical velocity of the gas  inside
the  cylinder is constant and is equal to
$v_{pz}$. After the  time the plume rotates by
$\pi/2$ it is stopped and the velocity  field is restored
to be pure Keplerian differential rotation  everywhere.
This very simplified flow field captures the basic 
features of actual complicated flow produced by randomly
distributed  star-disk collisions. We also feel that
elaborating on some of the  details of the flow  field
like taking a more realistic distribution  of star-disk
collision points in
$r$, and  introducing a weak and  distributed downflow,
is not warranted at the present initial stage  of
simulations in view of the fact that we do not know  other
important features of the flow (no actual  hydrodynamic
calculations have yet been performed ). Our model  flow
and simplified assumptions about star-disk  collisions,
frequency, and distribution capture qualitative features 
important for the excitation and symmetry properties of
the dynamo.  We feel that all elaborations mentioned
above as well as accurate  simulations of star-disk
collision hydrodynamics would  not qualitatively change
our conclusion about the possibility of such  a dynamo.

Since equations (\ref{eqn22}--\ref{eqn24}) require
spatial  derivatives  of the velocities, we apply
smoothing of discontinuities  in the flow field described
above. Also we introduce smooth switching  on and off  of
the plumes in time. For all three components of  velocity
$v_k$ we use the same interpolation rule for two  plumes
\begin{equation}
v^k=v^k_{in1}s_1+v^k_{in2}s_2+(1-s_1-s_2)v^k_{out}\mbox{.}\label{eqn34}
\end{equation}  Here $s_1(r,\phi,z,t)$ and
$s_2(r,\phi,z,t)$  are smoothing functions  for plume
$1$ and $2$ correspondingly. Each function
$s$ is close to 
$1$ in the region of space and time occupied by the plume
and is  close to $0$ in the  rest of space and during
times when the plume is  off. Transition from
$1$ to $0$ happens in the narrow layer at the  boundary
of the plume and during the interval of time short
compared  to the characteristic time  of the plume rise.
$v^k_{in1}$ and 
$v^k_{in2}$ are velocities of the  flow of plumes $1$ and 
$2$,
$v^k_{out}$ is the velocity of the flow outside the
regions  occupied by the plumes.  For spatial derivatives
of the velocity  components, one has from (\ref{eqn34})
\begin{equation}
\frac{\ptl  v^k}{\ptl x^i}=\frac{\ptl s_1}{\ptl
x^i}\left(v^k_{in1}-  v^k_{out}\right)+\frac{\ptl
s_2}{\ptl  x^i}\left(v^k_{in2}-v^k_{out}
\right)+s_1\frac{\ptl v^k_{in1}}{\ptl  x^i}+s_2\frac{\ptl
v^k_{in2}}{\ptl x^i}  +(1-s_1-s_2)\frac{\ptl
v^k_{out}}{\ptl  x^i}\mbox{.}\label{eqn35}
\end{equation}  It is easy to generalize  this approach
for an arbitrary number of plumes.

Let us assume that  the position of the axis of a 
cylindrical jet launched  upward (in
the positive direction of  the
$z$ axis) is  at
$r=r_0$ and $\phi=\phi_0$. We keep
$r_0=1$ for all plumes and the  initial
$\phi_0$ is randomly taken between $0$ and
$2\pi$. Let us  denote this plume as number
$1$ and the symmetric  plume going down  from the
equatorial plane as number
$2$. Then, after time
$(t-t_p)$  from the starting moment of the  plume
$t=t_p$, its position  is
\begin{equation}
\phi_1=\phi_0+(t-t_p)r_0\Omega_{K0}\mbox{,}\label{eqn36}
\end{equation}  where
$\Omega_{K0}=\Omega_K(r_0)$ is the Keplerian angular
rotational  velocity at
$r=r_0$ and in the simulations presented in this  work,
$\Omega_{K0}=1$.   The position of the axis of the
symmetric  plume  is
\begin{equation}
\phi_2=\phi_1+\pi\mbox{.}\label{eqn37}
\end{equation}  The radii of both plumes are
$r_p$. The bottom surface of the  plume
$1$ is  at
$z=z_{bot}$, the top surface of the plume
$1$ is  at
$z_1=z_{bot}+v_{pz} (t-t_p)$, the top surface of the
plume $2$ is  at
$z=z_{top}$, the bottom surface  of the plume
$2$ is  at
$z_2=z_{top}-v_{pz}(t-t_p)$.  Due to symmetry,
$z_2=-z_1$. The  velocity field inside the upward jet is
\begin{eqnarray} &&  v^r_1=r_0\Omega_{K0}\sin
(\phi-\phi_1)\mbox{,}\label{eqn38} \\  &&
v^{\phi}_1=r_0\Omega_{K0}\cos
(\phi-\phi_1)\mbox{,}\label{eqn39} 
\\ && v^z_1=v_{pz}\mbox{.}\label{eqn40}
\end{eqnarray} The velocity  field inside the downward
jet is
\begin{eqnarray} &&  v^r_2=r_0\Omega_{K0}\sin
(\phi-\phi_2)\mbox{,}\label{eqn41} \\  &&
v^{\phi}_2=r_0\Omega_{K0}\cos
(\phi-\phi_2)\mbox{,}\label{eqn42} 
\\ && v^z_2=-v_{pz}\mbox{.}\label{eqn43}
\end{eqnarray} We choose the  following  interpolation
functions
\begin{equation}
s_1=\left(\frac{1}{2}+\frac{1}{\pi}\arctan\frac{r_p^2-{r^{\prime}_1}^2}
{2r_p\Delta}\right)
\left(\frac{1}{2}+\frac{1}{\pi}\arctan\frac{(z-z_{bot})(z_1-z)}{
\Delta\sqrt{(z_1-z_{bot})^2+\Delta^2}}\right)S(t)\label{eqn44}
\end{equation}  and
\begin{equation}
s_2=\left(\frac{1}{2}+\frac{1}{\pi}\arctan\frac{r_p^2-{r^{\prime}_2}^2}
{2r_p\Delta}\right)
\left(\frac{1}{2}+\frac{1}{\pi}\arctan\frac{(z-z_{top})(z_2-z)}{
\Delta\sqrt{(z_{top}-z_2)^2+\Delta^2}}\right)S(t)\mbox{.}\label{eqn45}
\end{equation}  Here
${r^{\prime}_1}^2=r_0^2+r^2-2r_0r\cos(\phi-\phi_1)$ is
the  distance from the axis of the  plume
$1$,
${r^{\prime}_2}^2=r_0^2+r^2-2r_0r\cos(\phi-\phi_2)$ is
the  distance from the axis of the plume
$2$, $\Delta$ is the thickness of  the transition layer
of the functions $s_1$ and
$s_2$ from their  value $1$ inside the plume to $0$
outside the plume, $\Delta
\ll  r_p$.  Square root expressions in the
$z$-parts of
$s_1$ and $s_2$  ensure that the thickness of the
transition layer in the
$z$  direction is never less than
$\Delta$, even just after the plumes are  started, when
the differences
$(z_1-z_{bot})$ and $(z_{top}-z_2)$ are  zero. We choose
$\Delta=0.01$.

The function $S(t)$ ensures smooth  ``turning on'' and
``turning off'' of the plumes at prescribed  moments of
time. If the plumes are to be started at
$t=t_p$ and to be  turned off at $t=t_d$ ($t_d>t_p$),
then we adopt the following form  of the function
$S(t)$
$$
\left\{
\begin{array}{ll} S(t)=0\mbox{,} 
\quad &
\mbox{for}
\quad t<t_p-\delta t/2\mbox{,} 
\\
S(t)=\frac{1}{2}+\frac{1}{2}\sin\left(\pi\frac{t-t_p}{\delta
t}\right)
\mbox{,}
\quad  & \mbox{for} \quad t_p-\delta t/2 < t < t_p +
\delta t/2 \mbox{,} \\  S(t)=1\mbox{,}
\quad &
\mbox{for} \quad t_p+ \delta t/2 < t <  t_d-\delta
t/2\mbox{,}\\
S(t)=\frac{1}{2}-\frac{1}{2}\sin\left(\pi\frac{t-t_d}{\delta
t}\right)\mbox{,}
\quad  & \mbox{for} \quad t_d-\delta t/2 < t < t_d +
\delta t/2 \mbox{,} \\  S(t)=0\mbox{,}
\quad &
\mbox{for} \quad t>t_d+\delta  t/2\mbox{.}
\end{array}
\right.
$$ where $\delta t$ is the length of  the transition
period. $S=0$ corresponds to the flow without  plumes, and
$S=1$ corresponds to the flow with plumes. One needs  to
ensure that
$\delta t < t_d-t_p$. We took
$\delta t =  (t_d-t_p)/5$. The cycles with the
cylindrical jets present are  interchanged periodically
with the cycles with the pure Keplerian  rotation only.
The time between two consequent launchings of the  plumes
is
$\Delta t_p$ and we always have
$\Delta t_p > t_d-t_p$,  such that at any time only one
pair of plumes is present. This  eliminates the
occurrences of overlapping jets. Note, that during the 
time
$t_d-t_p$ the disk makes only about a quarter of the
turn.

Our  second model of random directions of the plumes
introduces obvious  changes into the expressions above.
Namely, we set
$s_2=0$ in  equations~(\ref{eqn34}) and~(\ref{eqn35}),
and we intermittently use  either
expressions~(\ref{eqn38}--\ref{eqn40}) for the
velocity,  when  the jet is directed upward,  or
expressions~(\ref{eqn41}--\ref{eqn43}), when the jet is
directed  downward. We also use the same ``switch''
function
$S(t)$ for both  models.

Finally, let us list the parameters, which are important
for  the growth of  the magnetic field in our model:  the magnetic
diffusivity 
$\eta$,  (or magnetic Reynolds  number
$\displaystyle
\mbox{Rm}_{\Omega}=\frac{r^2\Omega_K(r)}{\eta}$), the 
radius of the plumes $r_p$, the frequency of star-disk
collisions, 
$\Delta t_p$, the vertical velocity of the plume
$v_{pz}$, and  the duration of the plumes $t_d-t_p$.

\subsection{Analytic Solution  in the Asymptotic Region}
\label{subsec7.2}

Because the equations  for the evolution of the magnetic
field ${\bf B}$ and  vector  potential
${\bf A}$ are parabolic, the  boundary conditions  will
always influence the solutions inside the computational
region.  However, the distribution of the frequency of
star-disk collisions is  concentrated towards the center
meaning that most of dynamo activity  happens in a
limited region of space (around
$r=1$  in  our  dimensionless units). If one is willing
to disregard a relatively  small $\alpha$-effect for
$r\gg 1$, then the solutions of the field  equations in
the region
$r\gg 1$  can be obtained analytically because 
the flow is
just the differential rotation with the Keplerian
angular  velocity.

The equations for the evolution of the  axisymmetric
magnetic field in the presence of only the  differential
rotation are analogous to  equations~(\ref{eqn3.50})
and~(\ref{eqn3.51}) for the  evolution of  the
axisymmetric mean field.  We can obtain the necessary
equations  when replacing  the mean field by the actual
field and using the same  functions $A$ and $B_{\phi}$
for the poloidal magnetic flux and  toroidal magnetic
field. If one sets $\alpha=0$,
$\beta=0$, ${\bf  v}_P=0$, and $\Omega=\Omega_K(r)$ in
equations~(\ref{eqn3.50})  and~(\ref{eqn3.51}), the
resulting equations for axisymmetric  magnetic field in a
purely rotating flow are
\begin{eqnarray} && 
\frac{\partial A}{\partial  t}=\eta\left(\nabla^2
A-\frac{1}{r^2}A\right)
\label{asym_eqn1}\mbox{,} 
\\ && \frac{\partial B_{\phi}}{\partial  t}=
r\frac{d\Omega_K}{dr}B_r+
\eta\left(\nabla^2  B_{\phi}-\frac{1}{r^2}B_{\phi}\right)
\label{asym_eqn2}\mbox{.}
\end{eqnarray} Equation~(\ref{asym_eqn1})  is a
diffusion equation for the poloidal magnetic field
without  sources. Its solutions are determined by boundary
conditions imposed  on the poloidal magnetic field.
Equation~(\ref{asym_eqn2}) is a  diffusion equation for
the toroidal magnetic field but with  the source term due
to the $\Omega$-effect. We see that the evolution  of
poloidal magnetic field is decoupled from the evolution
of  the toroidal magnetic field (unless boundary
conditions mix them  together). After one knows the
solution for the poloidal field, one  can solve
equation~(\ref{asym_eqn2}) to find the toroidal magnetic 
field. If one looks for stationary solutions of 
equations~(\ref{asym_eqn1}) and~(\ref{asym_eqn2}) then
the outer  boundary condition is very important to
determine the solution.  However, in the case of a dynamo
the magnetic field in the dynamo  domain $r\approx 1$
grows exponentially. This growing field diffuses  into
the surrounding conducting medium according to 
equations~(\ref{asym_eqn1}) and~(\ref{asym_eqn2}). The
phenomenon is  analogous to the skin layer in plasma. The
growing magnetic field  decreases exponentially outward
from the generation  region. Therefore, if the growth
rate is sufficiently high such that  the skin depth is
smaller than the distance to the ideally conducting 
boundary, the boundary conditions at the boundary do not
influence  the dynamo process.

We computed an analytic solution  of
equation~(\ref{asym_eqn1}) in the region $r>1$ when the
magnetic  field grows exponentially. This solution is
presented in  Appendix~\ref{appendix_B}. We have checked
with numerical simulations  of the dynamo that the
magnetic field in the zone outside of dynamo  activity but
inside the outer radius of our computational domain is 
very closely approximated by
expressions~(\ref{asym_eqn8}) resulting  from our
analytic solution. We also varied the outer sizes of 
the outer ideally conducting boundaries in our
3-dimensional  simulations to verify that the growth
rate and the structure of the  growing magnetic field are
insensitive to the placement of the  boundaries. It is
necessary to stress that the simulations are 
insensitive to the boundary conditions only when the
magnetic  field is exponentially growing: the simulations
in the cases of  decaying or steady fields do depend on
how far the ideally conducting  boundaries are placed.

\subsection{Simulations of the Dynamo  Growth}
\label{subsec7.3}

The simulation is shown in a sequence of  stages. We use
dimensionless units described  in
section~\ref{subsec7.1}. Our computational domain is the
space  between two cylinders with the inner radius
$R_1=0.2$ and the outer  radius $R_2=4$, filled with a
media having uniform magnetic  diffusivity $\eta$. The
computational space is limited from below by  the surface
$z=-4$ and from above by the surface $z=4$. The total 
length of the cylindrical volume comprised  between
surfaces $z=-4$  and $z=4$ is $8$. All boundaries of the
computational volume  are ideally conducting. There is no
magnetic field penetrating the  boundaries, and the
boundary conditions~(\ref{eqn31})  and~(\ref{eqn32}) are
applied.

An initial quadrupole like field  establishes a primarily
radial field within the midplane of the  cylindrical
volume, $|z|<1/3$. The initial field is purely  poloidal,
concentrated toward the inner parts of the disk,  and  is
shown in Fig.~\ref{f_num1a} by arrows. The accretion
disk is indicated at
$z_{top}=1/3$,
$z_{bot}=-1/3$.

Keplerian  differential rotation is initiated and
generates toroidal field. At  the same time poloidal
field diffuses toward the outer boundary and  becomes
distributed over the volume more uniformly. The  magnetic
diffusivity is $\eta=0.01$, and the magnetic Reynolds
number  for rotation at $r=1$  is
$\displaystyle
\mbox{Rm}_{\Omega}=\frac{r^2\Omega_K(r)}{\eta}=100$.
With no source term present in
equation~(\ref{asym_eqn1}), the  poloidal magnetic field
will decay away in a purely toroidal flow.  The toroidal
magnetic field
$B_{\phi}$ will first grow because of the  source term
$\displaystyle r\frac{d\Omega_K}{dr}B_r$  in
equation~(\ref{asym_eqn2}), then reach a saturation value
$\approx  B_P \mbox{Rm}_{\Omega}/ (2 \pi)$ determined by
the balance between  the source term
$\displaystyle r\frac{d\Omega_K}{dr}B_r$ and  the
diffusion term
$\eta(\nabla^2B_{\phi}-r^{-2}B_{\phi})$ in 
equation~(\ref{asym_eqn2}), and finally decay as the
poloidal  magnetic field $B_r$ decays and so the source
term for $B_{\phi}$  also decays (Cowling's theorem).

Fig.~\ref{f_num1b} illustrates  the poloidal magnetic
field obtained after several revolutions  at
$r=1$, Fig.~\ref{f_num1c} shows the contours of toroidal
field  at the same moment of time as on
Fig.~\ref{f_num1b}. The time  evolution of the fluxes of
magnetic field is shown in  Fig.~\ref{f_num3}.  We also
show the process of winding up the dipole  like (odd)
field in Figs.~\ref{f_num2a} and~\ref{f_num2b} (poloidal 
and toroidal fields) and Fig.~\ref{f_num4} (the
evolution of fluxes).  Note, that the toroidal field
produced from the initial quadrupole  field (and any even
symmetry field) has the same sign throughout  the disk
thickness as well as in the space above and below the
disk.  In contrast  the toroidal field produced from an 
initial dipole  field (and any odd  symmetry field) is
zero at the equatorial plane  and has opposite signs in
the upper and lower halfs of the disk  thickness.

We now examine how the simulated star-disk  collisions
(approximated by a flow model described in 
section~\ref{subsec7.1}) deform the  wound up, toroidal
magnetic  field and create poloidal field from the
toroidal field.  Figs.~\ref{f_num5a} and~\ref{f_num5b}
illustrate the action of the  rising plume on the
poloidal magnetic field in a fluid which is at  rest. The
initial magnetic field here is a quadrupole like field 
shown in  Fig.~\ref{f_num1a}. The radius of the inner
cylinder  is
$0.2$ and the radius of the  plume is
$0.2$. The velocity of the  plume is equal to the
Keplerian  velocity at
$r=1$ and the plume  moves
$\pi/4$ radians  in the
$\phi$-direction before it  disappears.
Fig.~\ref{f_num5a} is a side view on the plume. 
Fig.~\ref{f_num5b} is a view on the plume from the top.
One can  clearly see the lifting of the field lines of
the quadrupole field  from the midplane of the disk by
the plume flow. Because the plume  flow is strongly
compressible near the  head of the plume it forms  a
narrow layer of enhanced magnetic field  near the top
boundary of  the plume. Magnetic field diffuses inside
the  plume from this layer.  On the top view one can see
the twisting of  magnetic field lines by  the  unwinding
of the flow in the plume. It creates toroidal field  from
the poloidal field. More importantly,
Figs.~\ref{f_num6a}, 
\ref{f_num6b}, and~\ref{f_num6c} illustrate the action
of the same  plume on the primarily toroidal magnetic
field wound up from the  initial quadrupole field (as in
Fig.~\ref{f_num1c}). The plume rises  through the
differentially rotating fluid with the Keplerian profile 
of angular velocity. Fig.~\ref{f_num6a} is a side view 
from
$r$-direction, Fig.~\ref{f_num6b} is a top view
from $z$-direction, and 
Fig.~\ref{f_num6c} is a side view from 
$\phi$-direction. Shown  by arrows is the flow velocity
in the reference frame corotating with  the base of the
plume with the angular velocity at the point of the 
location of the plume, i.e. the value
${\bf v}^{\prime}={\bf  v}-\Omega_{K0} r{\bf e}_{\phi}$.
As with Fig.~\ref{f_num5a}, the side  view from the
$r$-direction on Fig.~\ref{f_num6a} shows the lifting  up
of the toroidal field by the rising plume. One can see
from the  projection viewed from
$\phi$-direction that the magnetic field is  entrained into
the forming a loop of poloidal field. The top  view clearly
shows the twisting of  toroidal magnetic field and the 
creation of poloidal field from the toroidal field, i.e.
the
$\alpha$  effect. The resulting loop of flux translated
and rotated from the  toroidal plane is shown at the
time of maximum jet extension. After  that time the jet
velocities are smoothly set to zero.

By close  examination of the positions of field lines in
Figs.~\ref{f_num6a}, 
\ref{f_num6b}, and~\ref{f_num6c} one can discover the
presence of  another, more subtle effect: as the bundle
of magnetic field lines is  rotated and bent by the
plume, magnetic field lines twist around each  other in
this bundle. The direction of this twist can be observed
to  be opposite to the direction of the helical twisting
associated with  the lifting and bending of the bundle as
a whole. The bundle of  magnetic field lines behaves like
a ribbon when it is bended and  curved. The reason for
the additional opposite twist of the magnetic  field
lines in this ribbon is the conservation of magnetic  helicity
\citep{blackman03}. This small scale twist does not 
influence our flux rotation and mean field estimates of
the  kinematic stage of the dynamo.

The problem is continued with the  jets or plumes
repeated. The model of the flow described in 
section~\ref{subsec7.1} is applied. Below we present the
results for  a representative case for the model with the
plumes randomly  distributed along the circle $r=1$ and
launched in periodic intervals  in random directions up
and down through the disk. The parameters are  the
following (in dimensionless units introduced in 
section~\ref{subsec7.1}): $R_1=0.2$, $R_2=4$,
$\eta=0.01$, $r_p=0.3$, 
$\Delta t_p=\pi/2+0.4$, $t_d-t_p=\pi/2$, $v_{pz}= 1$,
$z_{bot}=-1/3$, 
$z_{top}=1/3$ and the centers of plumes are located on
the circle 
$r=1$. The run is started with the initial field being
purely  poloidal. The initial poloidal field is the linear
superposition of  odd and even magnetic fields shown in
Fig.~\ref{f_num2a} and  Fig.~\ref{f_num1b} respectively.
The exact meaning of odd (dipole  like) and even
(quadrupole like) parity fields is described in 
Appendix~\ref{appendix_C}. Here we only note, that the
total energy  of the magnetic field is equal to the sum
of energies of odd and even  components. Odd component
contributes 5\% of the total energy of the  initial field.
The remaining 95\% of the total energy is the energy  of
the even field. The first plume is launched at the moment
$t=0.2$  after the beginning of the simulation, and the
subsequent plumes are  launched in periodic moments of
time with the period $\Delta t_p$.  This rate of plume
launches corresponds to an average $2\pi/\Delta  t_p =
3.2$ plume launches per revolution at $r=1$. The
simulation is  continued until time $t=640$. By that time
the magnetic field grows  by $\sim 10$ orders of
magnitude. The resolution of our  typical
dynamo simulation is 41x81x41 nodes in radial, azimuthal
and  vertical directions respectively. Although this
resolution seems to  be quite modest to resolve the
plumes (there are typically only about  6x6 nodes to
resolve the cross section of a plume) we checked the 
convergence of our simulations by performing trial runs
with  61x121x61 resolution. The growth rate of the
dynamo and the structure  of the growing magnetic fields
do not change with the increased  resolution. We also
performed trial runs with the larger size of the 
computational domain: $-6<z<6$ and $0.2<r<6$ with
61x121x61  resolution. We did not observe significant
changes of the  growth rates and magnetic field structure
of the dynamo when  increasing the size of computational
domain. The reasons for  insensitivity to the boundary
conditions are described above in 
section~\ref{subsec7.2}.

The time evolution of the total energy of  the magnetic
field integrated over the computational volume is 
presented in Fig.~\ref{f_num7} as well as the time
evolution of  the fractions of total energy of odd and
even components of the  magnetic field. An arbitrary
value of the initial magnitude of the  magnetic field is
used. The initial rapid growth of the energy is due  to
rapid build up of the toroidal magnetic field. After a
couple of  revolutions at $r=1$ the dynamo effect
overcomes the linear growth of  the toroidal magnetic
field and the growth of the magnetic energy  becomes
exponential. The magnetic field experiences oscillations
with  the period equal to
$\Delta t_p$ due to the repeated actions of  single
plumes. More significant oscillations of odd and even 
components of the field occur on the time scale of the
diffusion over  the region of dynamo activity $\approx
100$. Despite the significant  variation of the fraction
of the odd field, which can become up to  30\%, even
(quadrupole) field dominates. Since the flow does not
have  symmetry with respect to reflections $z\to -z$,
the odd and even  components of magnetic field are
coupled to each other and grow with  the same exponential
rate.

The time evolution of fluxes of three  components of
the magnetic field is shown in Fig.~\ref{f_num8}. We 
calculate the fluxes of magnetic field through the
following three  surfaces: the flux of $B_r$ through the
part of cylindrical surface 
$r=1/2$ limited by lines $z=0$, $z=4$, $\phi=0$, and 
$\phi=\pi/2$; the flux of $B_{\phi}$ through the
rectangle in the  plane $\phi=0$ limited by lines $z=0$,
$z=4$, $r=R_1$, and $r=R_2$;  the flux of $B_z$ through
the half of the ring in the plane $z=-2$  limited by
lines $r=R_1$, $r=R_2$,
$\phi=0$, and $\phi=\pi$. Then, we  divide each of the
three fluxes by the areas of the corresponding  surfaces.
In this way, the values of the magnetic field averaged
over  the surfaces, $<B_r>$, $<B_{\phi}>$, and $<B_z>$,
are obtained. The  time evolution of the logarithms of
absolute values of these averaged  values of the magnetic
field is presented in Fig.~\ref{f_num8}. All  three
fluxes grow exponentially (if averaged over
fluctuations)  with the same growth rate $\Gamma=0.026$.
The growth rate of the mean  square of the magnetic field
plotted in Fig.~\ref{f_num7} is equal to 
$2\Gamma$ which is consistent with the growth rate of
fluxes. The  value of $<B_{\phi}>$ is larger than the
values of poloidal  fluxes meaning that the toroidal
field is predominant in the dynamo,  which is also in the
agreement with the conclusion from the mean  field
theory. While radial and toroidal fluxes grow
monotonically,  the flux of the axial magnetic field
experiences oscillations with  exponentially growing
amplitude. The $z$-flux remains zero on  average. This is
due to the fact that both dipole and quadrupole  growing
magnetic fields have zero $z$-flux through the surface 
described above. However, the $z$-flux experiences
oscillations due  to individual plumes creating
nonaxisymmetric magnetic field.

The  behavior of dynamo magnetic fields immediately
outside of  the generation region is especially
interesting in connection to  the magnetic fields in the
jets (magnetic helices) and observed  magnetic field in
galactic disks. In Fig.~\ref{f_num9} we plotted the 
fraction of energy of the magnetic field, which resides
outside of  the region of dynamo activity. In particular,
we divided the whole  computational domain into two: the
inner domain is the region 
$-2<z<2$ and $r<2$, the outer domain is the rest of the
computational  domain with $|z|>2$ or
$r>2$. Initially, the fraction of the outer  energy grows
because of the diffusion of the initial magnetic field 
outside the central region (compare the poloidal field on 
Fig.~\ref{f_num1a} and on Fig.~\ref{f_num1b}). However,
after the  dynamo action sets in, the skin effect
described  in section~\ref{subsec7.2} and
Appendix~\ref{appendix_B} occurs. The 
skin depth of the steady growing magnetic field given  by
equation~(\ref{asym_eqn10}) for $\eta=0.01$ and
$\Gamma=0.026$  is
$l_s=0.6$. Thus, the outer domain is in the zone of pure
diffusion  of the magnetic field, where the variations
due to individual plumes  are smoothed out. The average
value of the outer fraction of the  magnetic energy is
$\approx 0.06$ of the total magnetic energy. This  is
roughly consistent with the estimate one can obtain from
the skin  depth analysis of Appendix~\ref{appendix_B},
$\sim (0.6/e)^2 \approx 0.05$.  The field in the outer region
is predominantly even as well as in the  inner region.
The time dependence of the fraction of even field in 
Fig.~\ref{f_num9} follows closely the time dependence of
the fraction  of the even field in Fig.~\ref{f_num7}.
Note, however, that the  curves in Fig.~\ref{f_num9} are
more smooth than in  Fig.~\ref{f_num7}. Rapid
oscillations of the field caused  by individual plumes
are smoothed out in the diffusion process of  the
magnetic field into the outer region as the exponential
decay  scale
$l_s$ becomes shorter for higher oscillatory frequencies 
$\omega'$ (Appendix~\ref{appendix_B}). Only slow
variations with the  time scale about or longer than the
diffusive time scale remain  present in the outer domain.

Another diagnostic of our simulation is  to calculate the
time behavior of the magnetic fluxes through the 
surfaces in the outer part of computational domain. By
looking at the  time evolution of these fluxes we can
learn about the time evolution  of the magnetic field in
the asymptotic diffusion region. We  calculate magnetic
fluxes of radial magnetic field, or equivalently, 
$<B_r>$ through the following cylindrical surfaces:
radial flux~1  through the part of the surface
$r=2$ limited by lines $\phi=0$, 
$\phi=\pi/2$, $z=-1/3$, and $z=1/3$; radial flux~2
through the part  of the surface $r=3$ limited by lines
$\phi=0$, $\phi=\pi/2$, 
$z=-1/3$, and $z=1/3$; radial flux~3 through the part of
the surface 
$r=3$ limited by lines
$\phi=0$, $\phi=\pi/2$, $z=2$, and 
$z=4$; radial flux~4 through the part of the surface
$r=3$ limited by  lines
$\phi=0$, $\phi=\pi/2$, $z=-4$, and $z=-2$. The first two 
radial fluxes describe the evolution of the magnetic
field close to  the equatorial plane. The third and
fourth fluxes describe the  evolution of the magnetic
field in the outer corners of the  computational domain.
We plot these four radial fluxes in  Fig.~\ref{f_num10}.
We calculate three fluxes of the toroidal  magnetic
field, or equivalently,
$<B_{\phi}>$ through the following  rectangular areas of
the plane $\phi=0$: toroidal flux~1 through the 
rectangle limited by lines $r=2$, $r=4$,
$z=-1/3$, and 
$z=1/3$; toroidal flux~2 through the rectangle limited by
lines 
$r=2$, $r=4$,
$z=3$, and $z=4$; toroidal flux~3 through the rectangle 
limited by lines $r=2$, $r=4$,
$z=-4$, and $z=-3$. We plot these  three toroidal fluxes
in Fig.~\ref{f_num11}. We calculate two fluxes  of the
axial magnetic field, or equivalently,
$<B_z>$ through the  following ring-shaped surfaces:
axial flux~1 through the quarter of  the ring in the
plane $z=2$ limited by the lines $\phi=0$, 
$\phi=\pi/2$, $r=3$, and $r=4$; axial flux~2 through the
quarter of  the ring in the plane $z=-2$ limited by the
lines $\phi=0$, 
$\phi=\pi/2$, $r=3$, and $r=4$. We plot these two axial
fluxes in  Fig.~\ref{f_num12}. One can see that all
radial, toroidal and axial  fluxes do not change sign
during the exponential growth of the dynamo  (after the
time
$t\approx 100$). Therefore, the star-disk collisions 
dynamo produces steadily growing non-oscillating
magnetic fields. The  signs of the fluxes (not shown in 
Figs.~\ref{f_num10}--\ref{f_num12}) are consistent with
the  quadrupole geometry of the magnetic field in the
outer region of the  dynamo.

In Fig.~\ref{f_num13} we plotted two vector plots of the 
poloidal magnetic field at the plane $\phi=0$ at the
final moment of  the simulation $t=640$: on the top plot
the length of arrows is  proportional to the magnitude of
the poloidal magnetic field, on the  bottom plot all
arrows have unit length and the direction of the  arrows
indicate the direction of the same magnetic field as on
the top plot. The concentration of the magnetic field
toward the central region with the plumes is clearly
visible on the top plot. The imaging with arrows picks up
only the region of the strong field while the arrows
outside this region are so short that they cannot be
pictured at all. The bottom plot illustrates the
structure of the poloidal field in the asymptotic outer
region. This structure can be described as a ``shifted
quadrupole'' implying the presence of a significant
dipole component. The toroidal field is $\sim 20$ times
stronger than the poloidal. The direction of the toroidal
field agrees well with the direction of the field
produced by the stretching of the poloidal field by the
Keplerian differential rotation. The structure of the
field at different $\phi$ positions is similar to that
at $\phi=0$. The nonaxisymmetric variations of the field
are most significant at the location of the plumes at
$r\approx 1$ and quickly decay outwards. Each individual
plume perturbs the magnetic field significantly. This is
also reflected in the oscillations of fluxes in
Fig.~\ref{f_num8}. The three dimensional plot of the
dynamo magnetic field is presented in
Fig.~\ref{f_num_col}. Here we plotted only the poloidal
component of the magnetic field at the two meridional
slices, $\phi=\pi/2$ and $\phi=3\pi/2$, in the
computational domain. In order to smooth out the strong
contrast between magnitudes of the magnetic field in the
inner and outer regions of the computational domain, we
plotted a vector field
${\bf B}_P/|B_P|^{2/3}$. The dominance of the quadrupole
magnetic field in the outer asymptotic region is obvious
from Fig.~\ref{f_num_col}. In the central region for
$r\approx 1$, the field is strongly perturbed by
individual plumes, and the nonaxisymmetric field caused
by the action of each single plume is visible. Toroidal
magnetic field is also strongest in the central part of
the computational domain.  

Finally, let us compare the
predictions of the flux rotation and the mean field theories with
the results of our numerical simulation. All three predict
that the growing magnetic field will be quadrupole. The
simulation formally corresponds to $\displaystyle
q\approx {\bar q}_{<r}
=\frac{r_p^2}{r^2}\frac{t_d-t_p}{2\Delta t_p} = 0.036$,
$H=-z_{bot}=1/3$, $l=z_{bot}+v_{pz}(t_d-t_p)=1.24$ in
dimensionless units of simulation. Using these
parameters and $\alpha_{plume}=1$ in the expression for
the growth rate in the flux rotation theory,
equation~(\ref{eqn3.43a}), one obtains $\Gamma = 0.084$.
For the mean field theory the expression~(\ref{eqn3.53})
gives $\beta=0.09$, the dynamo number
(equation~(\ref{eqn3.54})) is $D=-28$, and both
expressions~(\ref{eqn3.55}) and~(\ref{eqn3.56}) give
$\Gamma\approx 0.18$. This is to be compared to
numerical growth rate $\Gamma=0.026$. Both the flux
rotation and especially the mean field theory growth
rates are higher, but all three are within one order of
magnitude from each other. Such a result is satisfactory
because of the far reaching extrapolations of the
applicability of both flux rotation and mean field
theories.

\section{Conclusions}
\label{sec_concl}

We believe that by theory and calculation we have
demonstrated that a robust $\alpha \omega$ dynamo is
likely to occur in conducting accretion disks with a 
robust source of helicity. The growth rates as large as
$\Gamma \simeq 0.1
\;  \mbox {to} \; 0.01 \, \Omega_K$ are expected. We
have discussed in depth one such source of helicity in the
accretion disk forming the central massive black holes
of most galaxies.  This is the almost inevitable
star-disk collisions that should occur in the dense
stellar  populations   at the center of the galaxy. We
estimate that this source of helicity is far larger than
necessary  for the dynamo fields to reach saturation in
less than the formation time of the black hole. Star-disk
collisions should also be the most robust source of
helicity because the resulting plumes are driven several 
scale heights above the surface of the disk as compared
to turbulence were the vertical motions are limited to a
fraction of a scale height.  The advantage of the $\alpha
\omega$ dynamo is that because it produces a large scale 
coherent field outside the disk, the poloidal field, the
differential winding of this poloidal field leads to a
large scale force-free helix that transports the
magnetic energy away from the disk and 
from the dynamo.  The back
reaction of this force-free  field (force-free except
at  the disk surface boundary) only acts as a torque on
the Keplerian flow and thus the field energy  of the
force-free helix can grow at the expense of the free
energy of formation of the black hole. The back  reaction
of this force-free field, being much smaller  than the
toroidal field,  does not affect the plume formation by
star-disk collisions. Only the much larger  toroidal
field affects the plumes and this  in turn must be less
than the pressure inside the disk. Thus the star disk
collisions produce a robust dynamo where the back
reaction does not quench the dynamo action at low  values
of field.  The  resulting exponential gain  of this
dynamo is an instability  converting kinetic to magnetic
energy. Since the gain is large, the dynamo fields should
rapidly grow to saturation or the back reaction limit.
This  limit  we conjecture  is the torque  corresponding
to the accretion flow of  angular momentum  away from the
black hole. Hence, the dynamo  should convert a large
fraction of the free energy of the black hole formation 
to magnetic  energy.

\acknowledgements

VP is pleased to thank Richard Lovelace and Eric Blackman
for helpful discussions and Benjamin Bromley for support
with computer simulations. Eric Blackman is thanked
again for his support during  the late stages of this
work. The facilities and interactions of Aspen Center for
Physics are gratefully acknowledged, and particular
support has been given by Hui Li through the support of
the Director Funded Research, "Active Galaxies".  We are
particularly pleased to acknowledge the careful reading
of the text  by the anonymous referee and furthermore the
significant improvement of readability 
and putting our work in more perspective due to
the referee's efforts. This work has been supported by the
U.S. Department of Energy through the  LDRD program at
Los Alamos National Laboratory. VP also acknowledges
partial support by DOE grant DE-FG02-00ER54600 and
by the Center for Magnetic Self-Organization in Laboratory
and Astrophysical Plasmas at University of Wisconsin-Madison . 
The Cray supercomputer used in this research was provided 
through funding from the NASA Offices of Space Sciences,
Aeronautics, and Mission to Planet Earth.
 
\appendix

\section{On the Parity of Magnetic Fields}
\label{appendix_C}

Any arbitrary vector field ${\bf C}={\bf C}(r,\phi,z)$
can be decomposed  into the sum of parts even and odd
with respect to the reflection $z\to -z$,
${\bf C}={\bf C}^e+{\bf C}^o$. The following
symmetry rules are valid for an even field:
\begin{equation} C^e_r(-z)=C^e_r(z)\mbox{,}
\quad C^e_{\phi}(-z)=C^e_{\phi}(z)\mbox{,}
\quad C^e_z(-z)=-C^e_z(z)\mbox{,}\label{mf_eqn11}
\end{equation} and for an odd field:
\begin{equation} C^o_r(-z)=-C^o_r(z)\mbox{,}
\quad C^o_{\phi}(-z)=-C^o_{\phi}(z)\mbox{,}
\quad C^o_z(-z)=C^o_z(z)\mbox{.}\label{mf_eqn12}
\end{equation} Often even fields are called quadrupole
type fields and odd fields are called dipole type fields.
The last terminology reflects on the largest scale modes
possible within each symmetry class and allows one to
visualize fields of each symmetry type easily. The even
and odd decomposition of an arbitrary field
${\bf C}$ can be performed as follows:
\begin{mathletters}
\begin{eqnarray} &&
C^e_r(r,\phi,z)=\frac{1}{2}(C(r,\phi,z)+C(r,\phi,-z))\mbox{,}
\\ &&
C^e_{\phi}(r,\phi,z)=\frac{1}{2}(C(r,\phi,z)+C(r,\phi,-z))\mbox{,}
\label{mf_eqn13}
\\ &&
C^e_z(r,\phi,z)=\frac{1}{2}(C(r,\phi,z)-C(r,\phi,-z))\mbox{,}
\\ &&
C^o_r(r,\phi,z)=\frac{1}{2}(C(r,\phi,z)-C(r,\phi,-z))\mbox{,}
\\ &&
C^o_{\phi}(r,\phi,z)=\frac{1}{2}(C(r,\phi,z)-C(r,\phi,-z))
\mbox{,}
\\ &&
C^o_z(r,\phi,z)=\frac{1}{2}(C(r,\phi,z)+C(r,\phi,-z))\mbox{.}
\end{eqnarray}
\end{mathletters} One can check that for any volume $V$
symmetric with respect to the plane
$z=0$
\begin{equation}
\int_V {\bf C}^2\,dV=\int_V ({\bf C}^e)^2 \,dV + \int_V
({\bf C}^o)^2 \,dV
\label{mf_eqn15}\mbox{.}
\end{equation} This implies that if ${\bf C}={\bf B}$ is
a magnetic field, then the energy  of the magnetic field
is equal to the sum of the energies of its even and odd
components. The even and odd components of solutions of
equations~(\ref{eqn3.50}) and~(\ref{eqn3.51}) decouples
if the mean  velocity field is even,
$v_{Pr}(-z)=v_{Pr}(z)$, $v_{Pz}(-z)=-v_{Pz}(z)$,
$\Omega(-z)=\Omega(z)$, the coefficient $\alpha$ is
antisymmetric with  respect to reflection $z\to -z$, and
the coefficient $\beta$ is symmetric  with respect to
reflection $z\to -z$. Thus, even (quadrupole) and odd
(dipole) modes will have different growth rates. The
axisymmetric magnetic field is even if $A(-z)=-A(z)$,
$B_{\phi}(z)=B_{\phi}(-z)$ and is odd if $A(-z)=A(z)$,
$B_{\phi}(-z)=-B_{\phi}(z)$.

\section{Skin Effect for the Magnetic Dynamo}
\label{appendix_B}

Let us consider equation~(\ref{asym_eqn1}) written in
spherical coordinates $\varrho$, $\theta$, and
$\phi$ such that $\theta=0$ and $\theta=\pi$ corresponds
to the symmetry axis of the system. In the case of
time-dependent flow described in section~\ref{subsec7.1}
there are no eigenmodes with a fixed frequency. Instead,
the magnetic field can be represented as an integral over
frequencies in the Fourier transformation. However, in
the case of a growing (and possibly oscillating) magnetic
field, there is a characteristic  growth rate $\Gamma$
of the dynamo averaged over plume pulses. In addition, the
magnetic field will possess oscillating Fourier
components associated with the period of the emergence of
plumes and, possibly, some intrinsic oscillatory
behavior of the dynamo.  We consider the behavior of one
such Fourier component assuming the dependence
$A\propto \exp(-i\omega t)$, where the complex
$\omega$ is the sum of the real and imaginary parts as
$\omega=\omega^{\prime}+ i\Gamma$. The $\Gamma$ is the
average growth rate of the dynamo, while
$\omega^{\prime}$ can take on a whole range of values,
including the frequency of plumes, the Keplerian period,
all its harmonics, etc. We impose the boundary condition
for $A$ on some sphere of radius $\varrho_{in}$ such that
$\varrho_{in}>1$ but still $\varrho_{in}$ is of the
order of $1$. We assume that the value of
$A$ at $\varrho=\varrho_{in}$ is dictated by the dynamo
process inside
$\varrho_{in}$. Then, for one Fourier component
equation~(\ref{asym_eqn1}) becomes
\begin{equation} -i\omega\frac{A}{\eta}=
\frac{1}{\varrho^2}\frac{\partial} {\partial \varrho}
\left(\varrho^2 \frac{\partial A} {\partial
\varrho}\right) +
\frac{1}{\varrho^2} {\hat L}A\label{asym_eqn3}\mbox{,}
\end{equation} where
$$ {\hat L}=\frac{1}{\sin\theta}
\frac{\partial}{\partial\theta}
\left(\sin\theta
\frac{\partial}{\partial\theta}\right)
-\frac{1}{\sin^2\theta}
$$ is the angular operator acting on $A$. In spherical
geometry, equation~(\ref{asym_eqn3}) has separable
variables
$\varrho$ and $\theta$. Thus, we look for solutions in
the form
$A=R_l(\varrho)Q_l(\theta)\exp(-i\omega t)$.

The operator
${\hat L}$ commonly occurs in problems with axisymmetric
flows, when solving the equation for the stream function.
Since the magnetic field should be finite on the axis
$\theta=0$, the quantity
$$
\frac{1}{\sin\theta}
\frac{\partial}{\partial\theta}(\sin\theta A)
$$ must be finite at $\theta=0$ and at $\theta=\pi$,
because ${\bf B}_P=\nabla
\times(A {\bf e}_{\phi})$. The eigenvalues and
eigenfunctions ${\hat L}Q_l=
\lambda_l Q_l$ satisfying these boundary conditions are
\begin{equation}
\lambda_l=-l(l+1)\mbox{,} \quad Q_l=\sin\theta
P_l^{\prime}(\cos\theta)
\label{asym_eqn4}\mbox{,}
\end{equation} where prime denotes the differentiation of
the Legendre polynomial $P_l(x)$ with respect to $x$ and
$l=1,2,3,\ldots$ Besides these eigenvalues, $\lambda=0$
is also an eigenvalue with the eigenfunction
$Q_0=(1-\cos\theta)/\sin\theta$. The first three
eigenfunctions given by formula~(\ref{asym_eqn4}) are
\begin{equation} Q_1=\sin\theta\mbox{,} \quad
Q_2=\sin\theta\cos\theta\mbox{,} \quad
Q_3=\sin\theta\left(
\cos^2\theta-\frac{1}{5}\right)
\label{asym_eqn5}
\mbox{.}
\end{equation} The angular dependence $Q_l(\theta)$
determines the symmetry of the solutions. The mode
proportional to $Q_0$ describes the radially directed
magnetic field with nonzero total flux through the
sphere from $\theta=0$ to $\theta=\pi$. All terms with
$l\geq 1$ corresponds to the magnetic field with
vanishing total flux through the sphere from
$\theta=0$ to $\theta=\pi$. The $Q_0$ term cannot be
excited by the dynamo operating inside
$\varrho_{in}$ because of $\nabla\cdot{\bf B}=0$
condition. This is also clear from the fact that
$Q_0\to \infty$ when $\theta\to \pi$, which means that
the vector potential cannot be well defined for a
magnetic field with
$\nabla\cdot{\bf B}\neq 0$. The terms with $l\geq 1$
represent multipole expansion of the magnetic field in
the far zone of the generation region.
$R_1(\varrho)Q_1(\theta)$ is a dipole term,
$R_2(\varrho)Q_2(\theta)$ is a quadrupole term, and so
on.

For the radial part of the solution we obtain the equation
\begin{equation}
\frac{d^2 R_l}{d\varrho^2}+
\frac{2}{\varrho}\frac{dR_l}{d\varrho}-
\frac{l(l+1)}{\varrho^2}R_l-
\frac{\Gamma-i\omega^{\prime}}{\eta}R_l=0
\label{asym_eqn6}\mbox{.}
\end{equation} We introduce a new variable
$z=\varrho/\chi$ where
\begin{equation}
\chi^2=\frac{\eta(\Gamma+i\omega^{\prime})}
{\Gamma^2+\omega^{\prime 2}}
\label{asym_eqn7}\mbox{.}
\end{equation} Then, equation~(\ref{asym_eqn6}) reduces
to the Bessel equation of imaginary argument. Solutions
of this equation which vanishes at $\varrho\to \infty$
are given in terms of modified Bessel function
$K_{\nu}(z)$  as
$$ R_l=\sqrt{\frac{\pi}{2z}}K_{l+1/2}(z)\mbox{.}
$$ The Bessel functions of half-integer order can be
expressed through elementary functions (e.g.,
\cite{abramowitz72}). Thus, we obtain for the dipole and
quadrupole terms
$$ R_1(z)=\frac{\pi}{2z}e^{-z}
\left(1+\frac{1}{z}\right)\mbox{,} \quad
R_2(z)=\frac{\pi}{2z}e^{-z}
\left(1+\frac{3}{z}+\frac{3}{z^2}\right)
\mbox{.}
$$ Finally, collecting all the terms together and
retaining only the leading dipole and quadrupole terms,
we obtain the following solution for $A$
\begin{eqnarray} &&
A=a_1\sin\theta\frac{\pi\chi}{2\varrho}
e^{-\varrho/\chi}\left(1+
\frac{\chi}{\varrho}\right) e^{-i\omega t} + \nonumber \\
&& a_2\sin\theta\cos\theta
\frac{\pi\chi}{2\varrho}e^{-\varrho/\chi}
\left(1+\frac{3\chi}{\varrho}
+\frac{3\chi^2}{\varrho^2}\right) e^{-i\omega
t}\label{asym_eqn8}\mbox{,}
\end{eqnarray} where the coefficients $a_1$ and $a_2$
should be determined by the condition of the continuity
of harmonics of $A$ at the surface $\varrho=\varrho_{in}$.
The values of $a_1$ and $a_2$ are determined by the
dynamo action inside the radius $\varrho_{in}$. We see
that both dipole and quadrupole components (and all
higher multipole components) decay as $\propto
e^{-\varrho/\chi}$. Using the
expression~(\ref{asym_eqn7}) for $\chi$ one obtains
\begin{equation} e^{-\varrho/\chi}=\exp\left(
-\frac{\varrho}{\sqrt{2\eta}}
\sqrt{\sqrt{\Gamma^2+\omega^{\prime 2}} +\Gamma} + i
\frac{\varrho}{\sqrt{2\eta}}\sqrt{
\sqrt{\Gamma^2+\omega^{\prime 2}}
-\Gamma}\right)\label{asym_eqn9}\mbox{,}
\end{equation} where we assumed $\Gamma>0$ and
$\omega^{\prime}>0$. The thickness of the skin layer is
determined by the real part of the expression under the
exponent in equation~(\ref{asym_eqn9}). The larger the
growth rate $\Gamma$, the faster the magnetic field
decays with the radius. Also, oscillating modes with
$\omega^{\prime}>0$ decay faster with the radius than
the steady modes with $\omega^{\prime}=0$. Thus, far
from the dynamo source one should expect the magnetic
field to be growing in time, steadily, without
oscillations.

The characteristic length of the exponential decay of the
field, $l_s$, is found from equation~(\ref{asym_eqn9}) as
\begin{equation} l_s=\sqrt{\frac{2\eta}{\sqrt{\Gamma^2+
\omega^{\prime 2}}+\Gamma}}
\label{asym_eqn10}\mbox{.}
\end{equation} For a steady magnetic field
$l_s=\sqrt{\eta/\Gamma}$. When $\varrho$ is approaching
the radius of the outer boundary $R_2$, the
solution~(\ref{asym_eqn8}) starts to ``feel'' the
boundary condition as an ideally conducting boundary and
the numerical results at $\varrho \geq R_2$ are not
approximated by formula~(\ref{asym_eqn8}).

\newpage

\begin{figure}
\epsscale{0.5}
\plotone{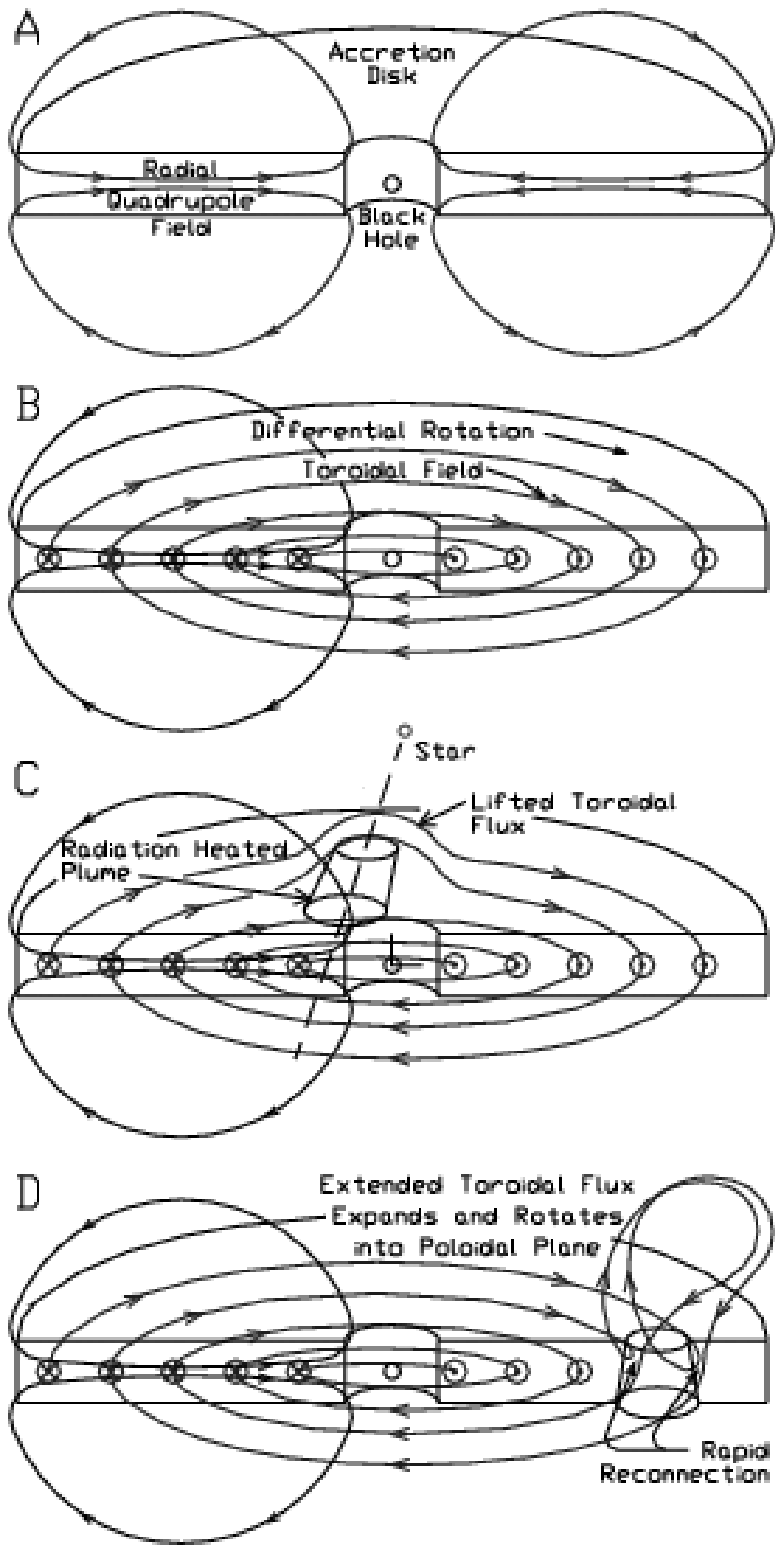}
\caption {The $\alpha-\Omega$ dynamo in a galactic black
hole accretion disk. The radial component of the
poloidal quadrupole field within the disk (A) is sheared
by the  differential rotation within the disk, developing
a stronger toroidal  component (B).  As a star passes
through the disk it heats by shock  and by radiation a
fraction of the matter of the disk, which expands
vertically and lifts a fraction of the toroidal flux
within an expanding plume (C). Due to the conservation of
angular momentum, the expanding plume and embedded flux
rotate
$\sim \pi/2$ radians before the matter in the plume and
embedded flux  falls back to the disk (D).  Reconnection
allows the new poloidal flux to merge with and augment
the original poloidal flux (D).}
\label{fig1a}
\end{figure}

\newpage

\begin{figure}
\epsscale{1.0}
\plotone{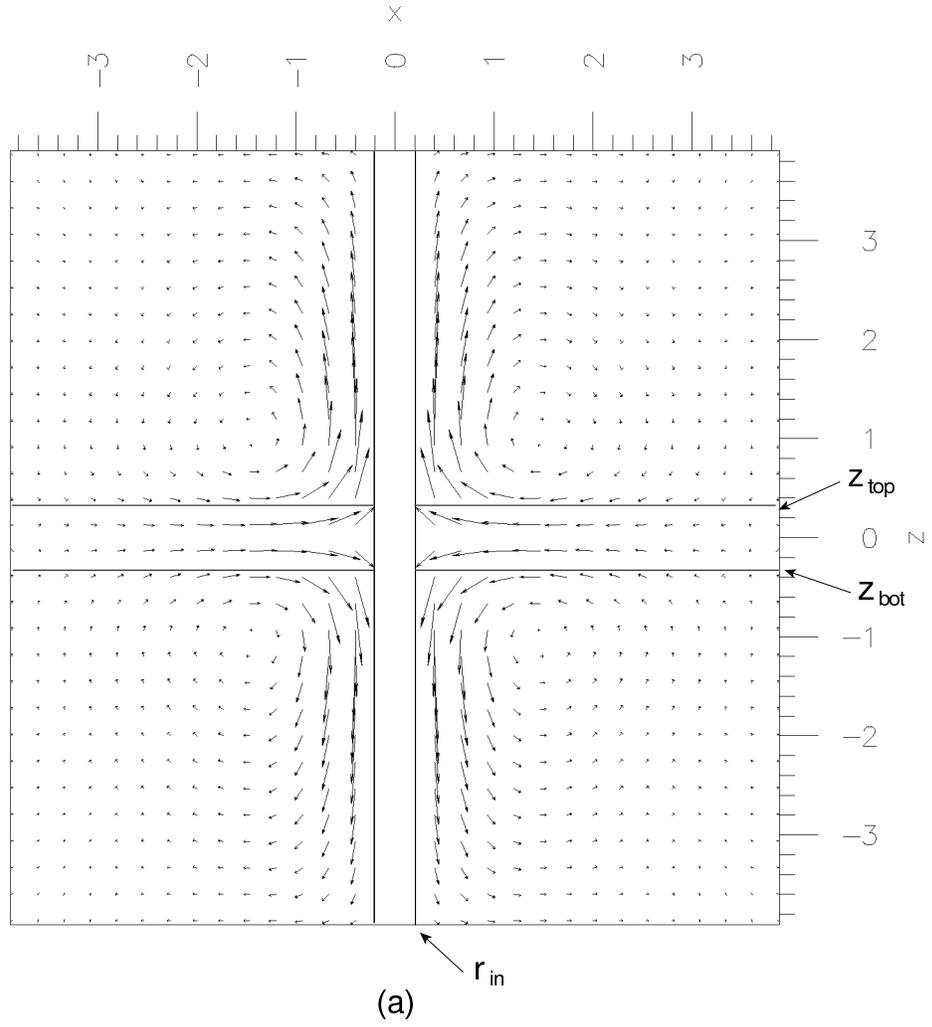}
\caption{Initial even (quadrupole) axisymmetric poloidal
magnetic field in the simulation box. \label{f_num1a} }
\end{figure}
\clearpage

\begin{figure}
\epsscale{1.0}
\plotone{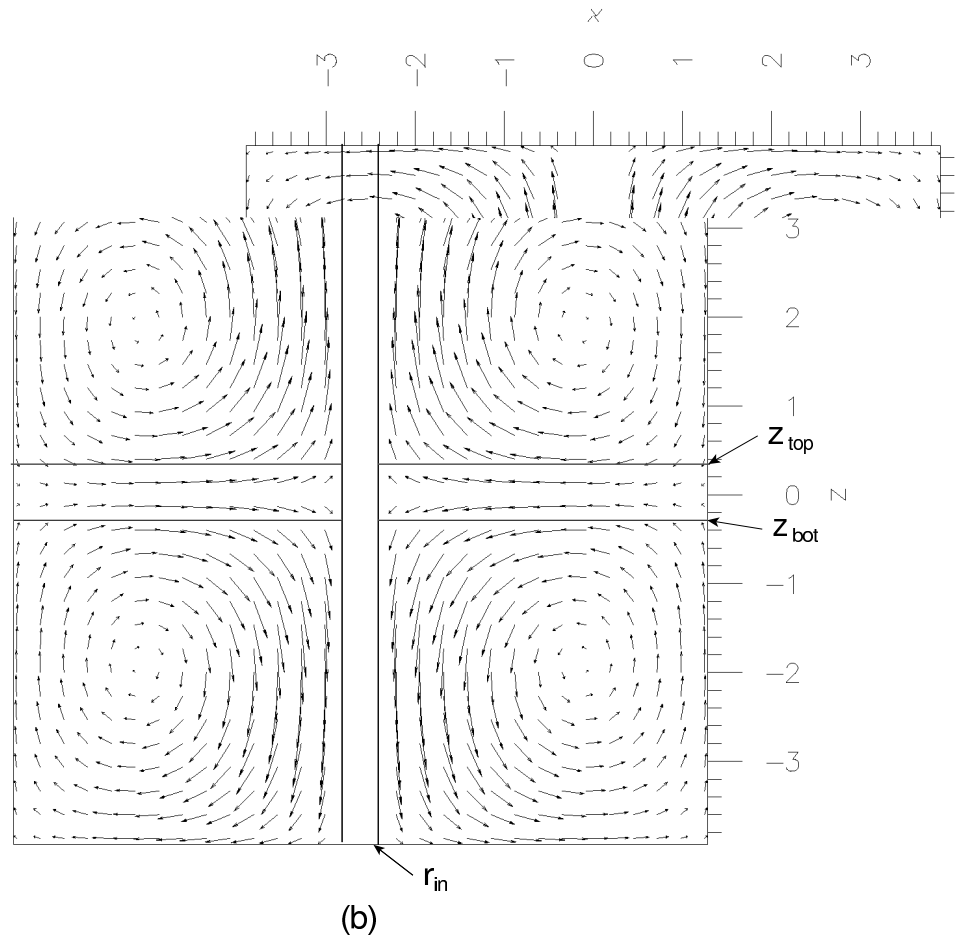}
\caption{The diffusion of the initial poloidal magnetic
field shown in Fig.~\protect\ref{f_num1a}. The
axisymmetric poloidal field at the time $t=140$ is shown.
\label{f_num1b} }
\end{figure}
\clearpage

\begin{figure}
\epsscale{0.93}
\plotone{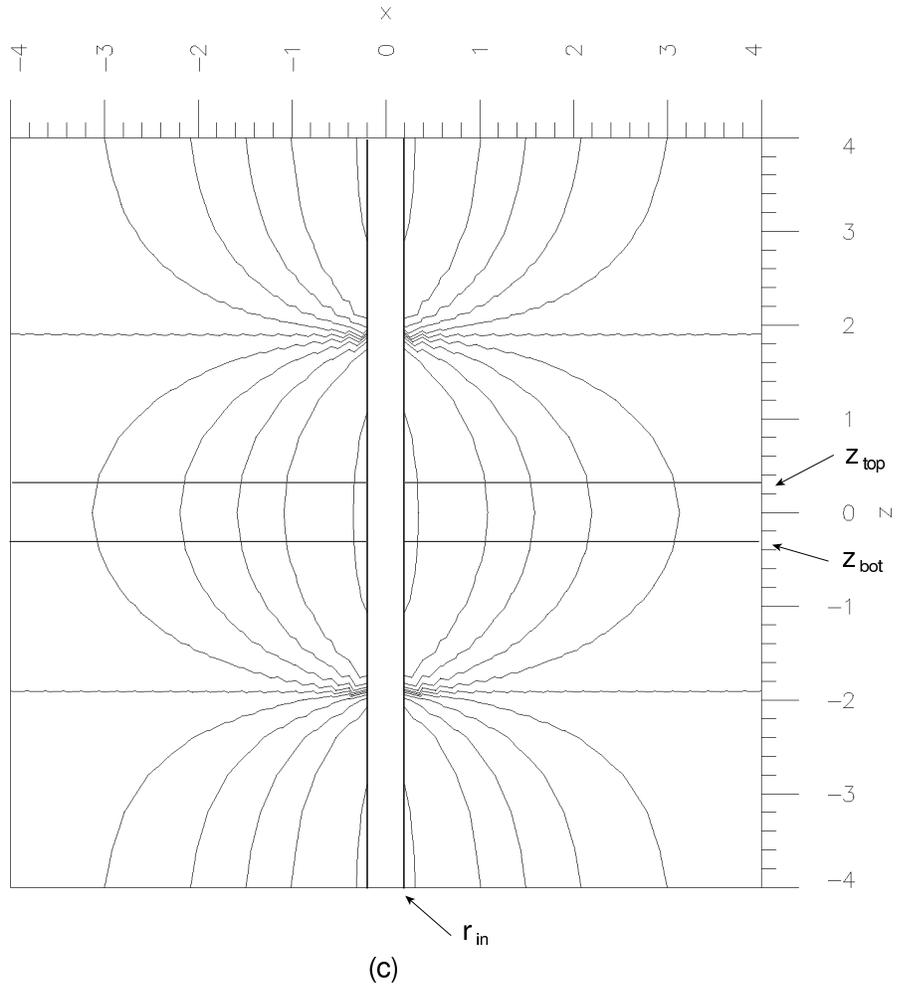}
\caption{Toroidal magnetic field produced by
differential rotation with the Keplerian angular velocity
$\Omega_K=r^{-3/2}$ starting 
from the initial quadrupole poloidal
magnetic field shown in Fig.~\protect\ref{f_num1a}. The
contours of equal magnitude of the axisymmetric toroidal
magnetic field at the time $t=140$ are shown.
\label{f_num1c} }
\end{figure}
\clearpage

\begin{figure}
\epsscale{0.95}
\plotone{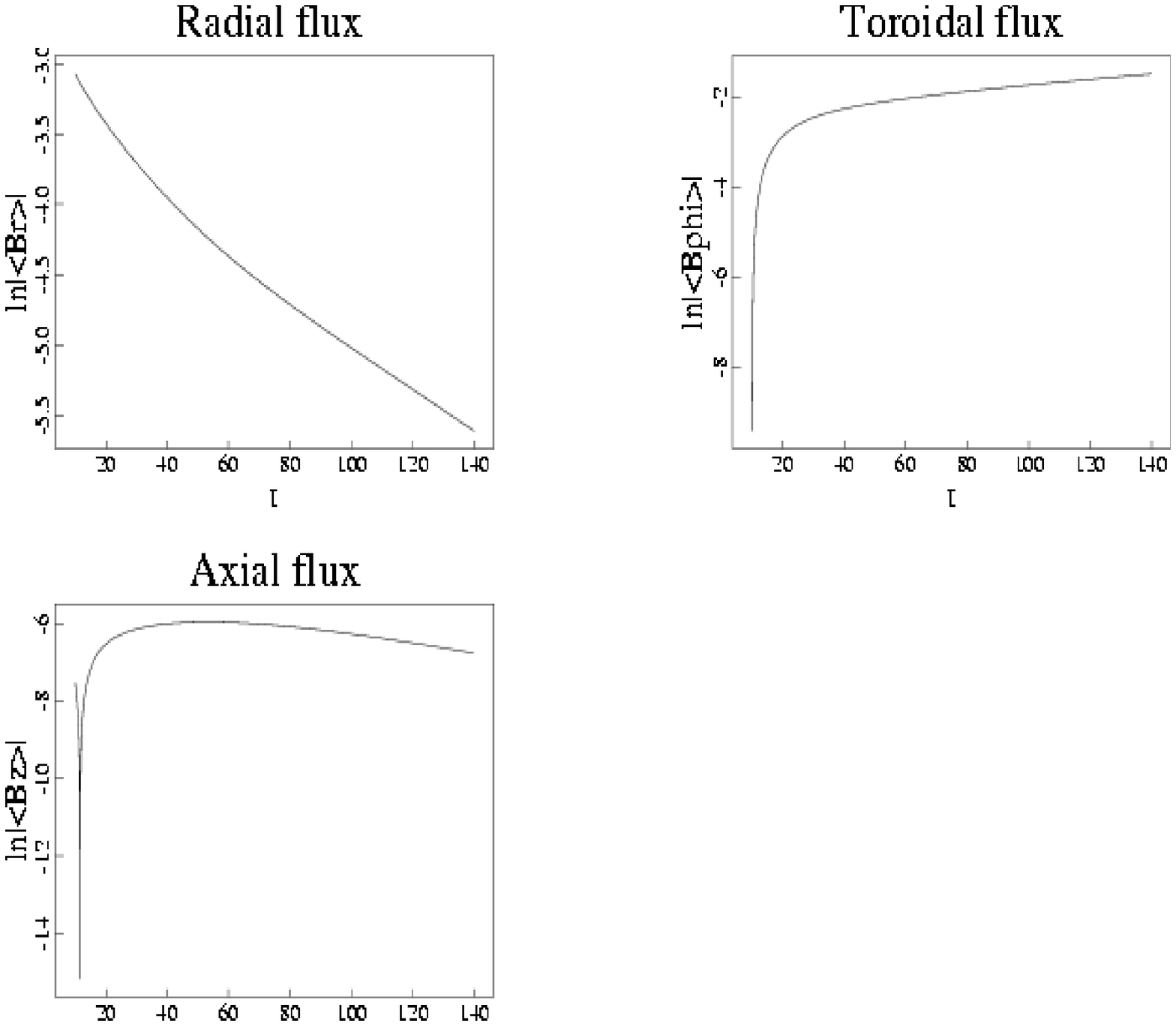}
\caption{Evolution of fluxes of three components of
magnetic field for the simulation presented in
Figs.~\protect\ref{f_num1a}--\protect\ref{f_num1c}.
Natural logarithms of the absolute value of fluxes are
plotted vs. time. Flux of $B_r$ is calculated through the
part of the cylindrical surface $r=2$ limited by lines $\phi=0$,
$\phi=\pi/2$, $z=0$, and $z=4$. Flux of $B_z$ is
calculated through the surface $z=3$ limited by lines $r=R_1$,
$r=R_2$, $\phi=0$, and $\phi=\pi$. Flux of $B_{\phi}$ is
calculated through the surface $\phi=0$ 
limited by lines $r=R_1$,
$r=R_2$, $z=0$, and $z=4$.
\label{f_num3} }
\end{figure}
\clearpage

\begin{figure}
\epsscale{1.0}
\plotone{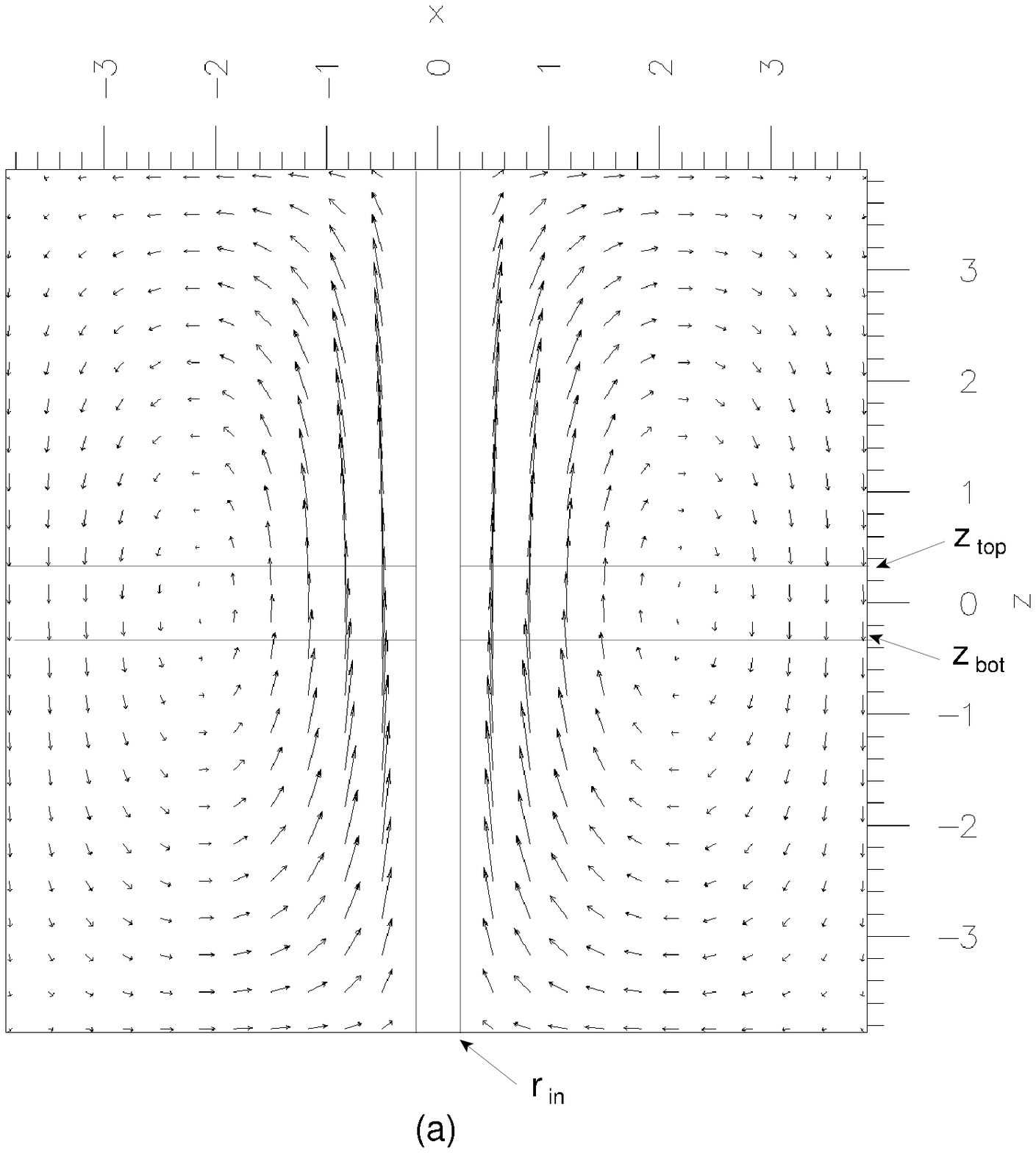}
\caption{Evolution of  the initial odd (dipole like)
poloidal magnetic field  in differentially rotating
plasma with the Keplerian angular  velocity
$\Omega_K=r^{-3/2}$. The poloidal axisymmetric magnetic 
field at the  time $t=210$ is shown.
\label{f_num2a}  }
\end{figure}
\clearpage

\begin{figure}
\epsscale{1.0}
\plotone{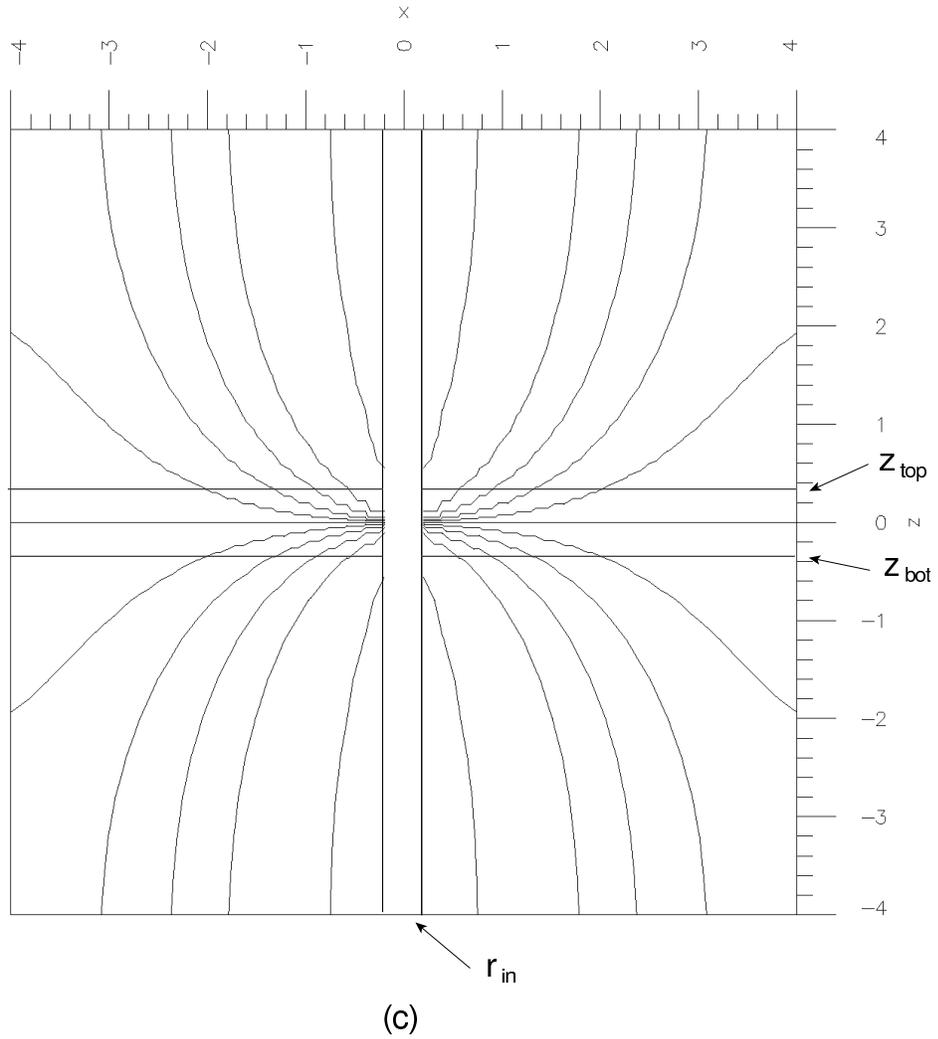}
\caption{Evolution  of the initial odd (dipole like)
poloidal magnetic field in a differentially rotating
plasma with the Keplerian angular velocity
$\Omega_K=r^{-3/2}$. The contours of toroidal
axisymmetric magnetic field at the time $t=210$ are shown.
\label{f_num2b} }
\end{figure}
\clearpage

\begin{figure}
\epsscale{1.0}
\plotone{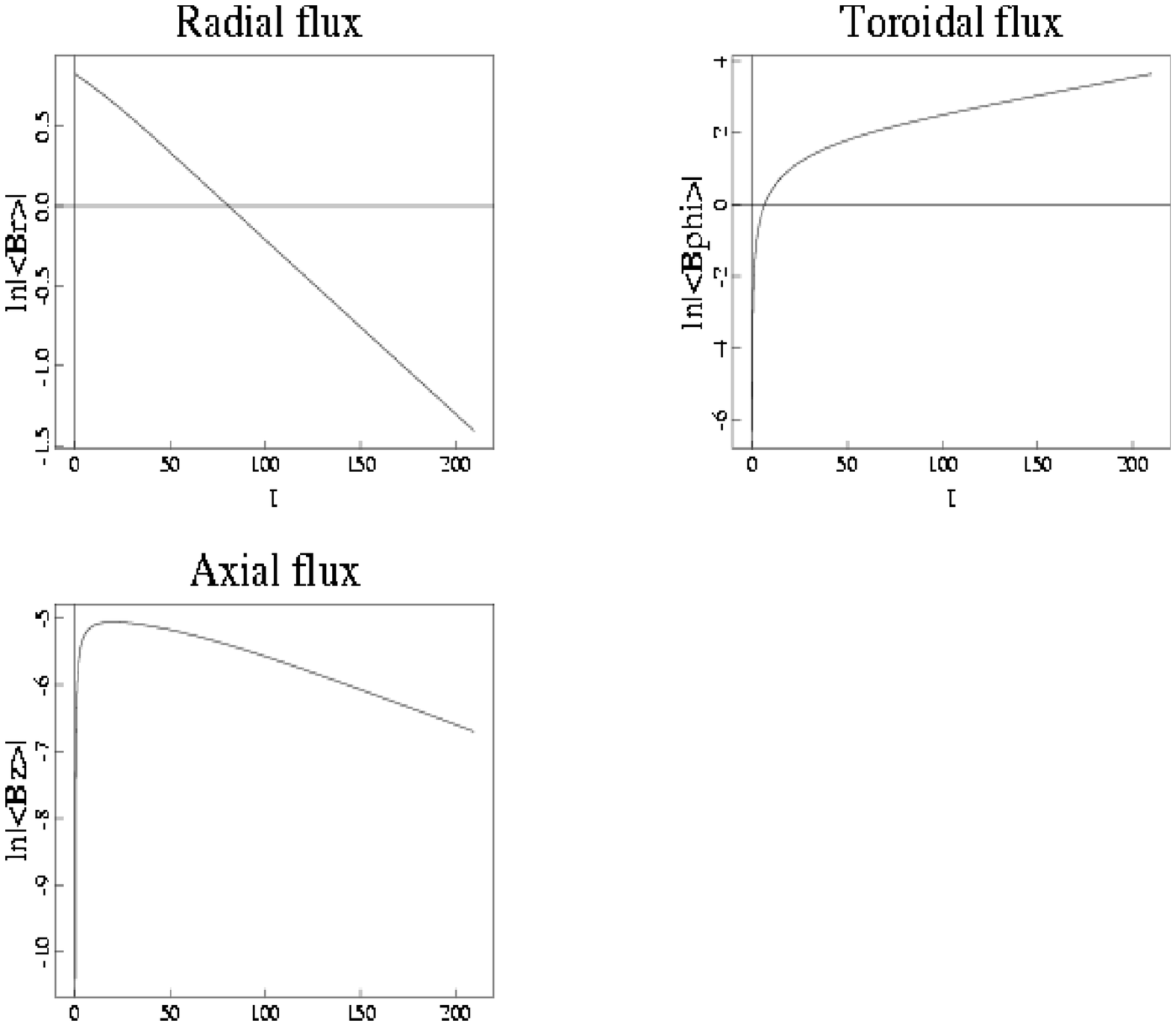}
\caption{Evolution of fluxes of three components of
magnetic field for the simulation presented in
Figs.~\protect\ref{f_num2a}--\protect\ref{f_num2b}.
Natural logarithms of the absolute value of the flux are
plotted versus time. The flux of $B_r$ is calculated
through the part of the cylindrical surface 
$r=2$ limited by lines $\phi=0$,
$\phi=\pi/2$, $z=0$, and $z=4$. The flux of $B_z$ is
calculated through the surface $z=3$ 
limited by lines $r=R_1$,
$r=R_2$, $\phi=0$, and $\phi=\pi$. The flux of
$B_{\phi}$ is calculated through 
the surface $\phi=0$ limited by
lines $r=R_1$, $r=R_2$, $z=0$, and $z=4$.
\label{f_num4} }
\end{figure}
\clearpage

\begin{figure}
\epsscale{0.6}
\plotone{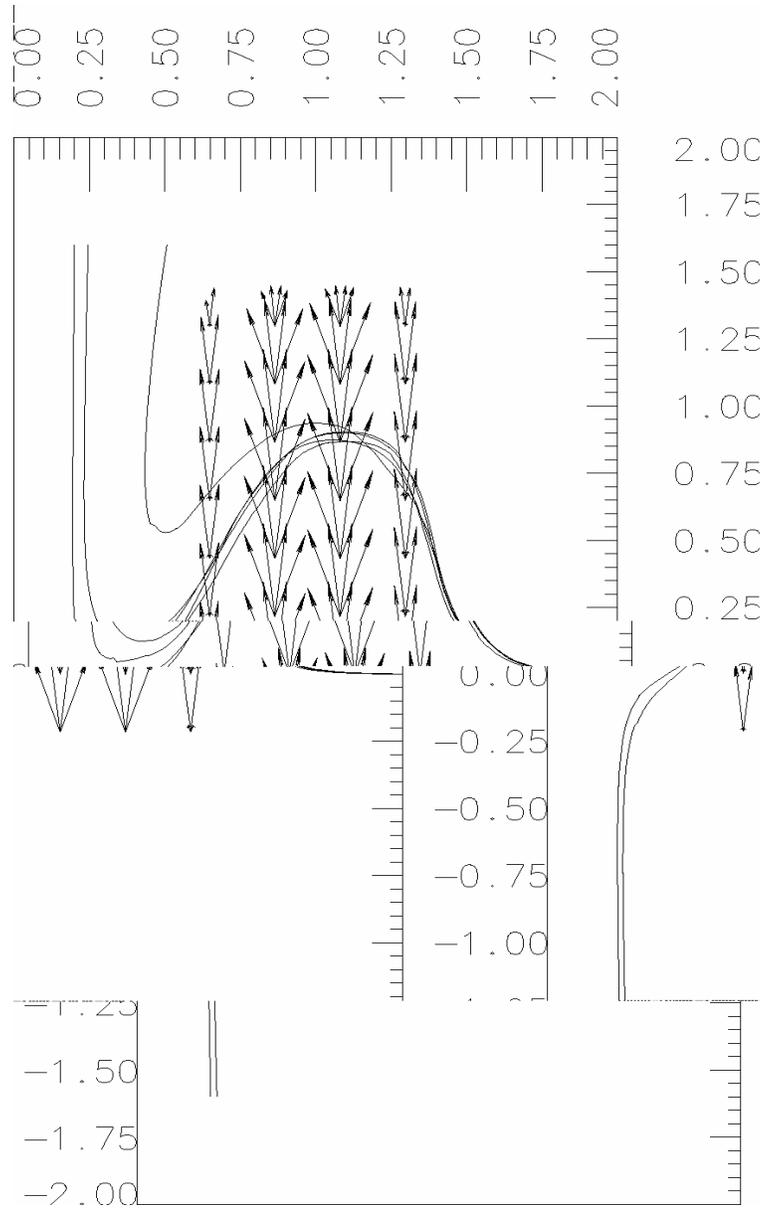}
\caption{Distortion of magnetic field lines of a
quadrupole field by a single cylindrical plume rising in
a fluid, which is at rest. Side view from
$\phi$-direction. The picture shows the lifting of a
bundle of field lines. Arrows indicate the velocities of
the flow at different depths in the plume. The corkscrew
motion of the plume is clearly seen on the side view.
\label{f_num5a}  }
\end{figure}
\clearpage

\begin{figure}
\epsscale{0.85}
\plotone{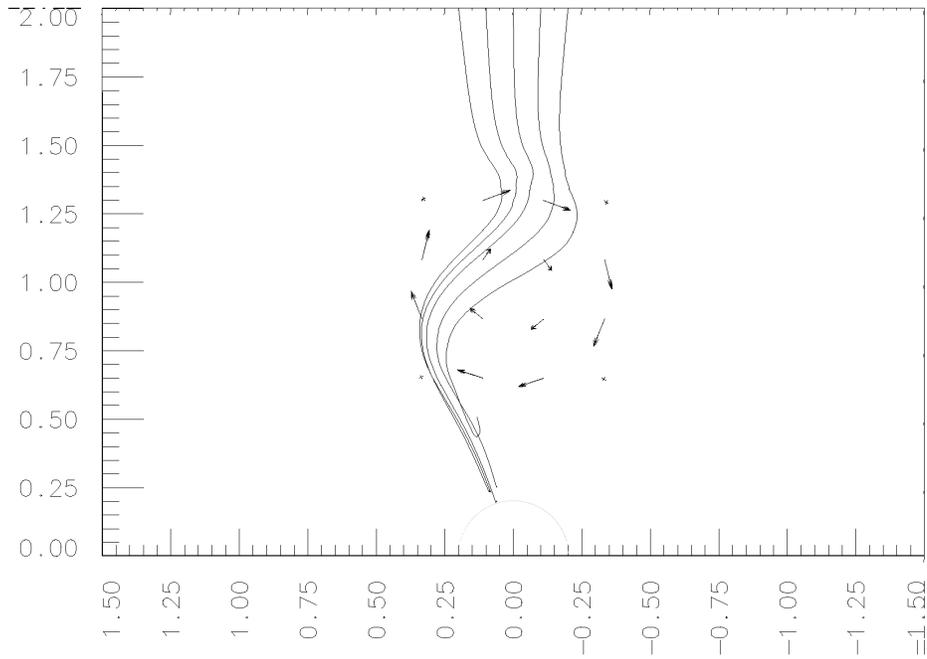}
\caption{Same  field configuration as in
Fig.~\protect\ref{f_num5a} viewed from the  top
$z$-direction. The picture shows the twisting of the
bundle of  field lines. Arrows indicate velocities of the
flow. The time is the  same as on
Fig.~\protect\ref{f_num5a}.
\label{f_num5b}  }
\end{figure}
\clearpage

\begin{figure}
\epsscale{1.0}
\plotone{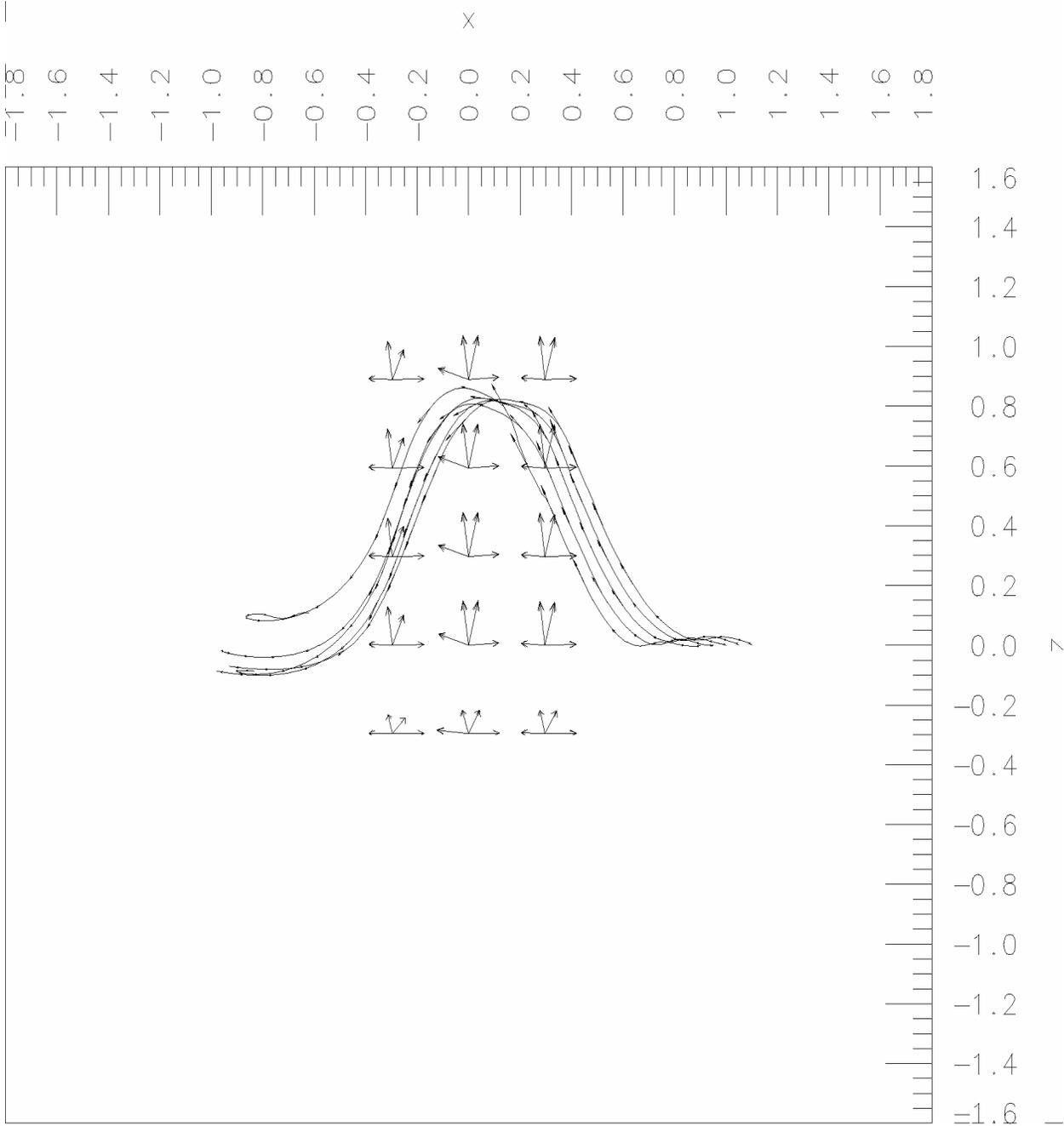}
\caption{Creation  of the poloidal magnetic field from
the toroidal by the rising and unwinding  jet produced by
star-disk collisions in a differentially rotating 
plasma. Side view from $r$-direction. Arrows indicate
velocities of  the flow in the frame corotating with the
base of the plume. The  picture shows the rising bundle of
field lines.
\label{f_num6a}  }
\end{figure}
\clearpage

\begin{figure}
\epsscale{0.85}
\plotone{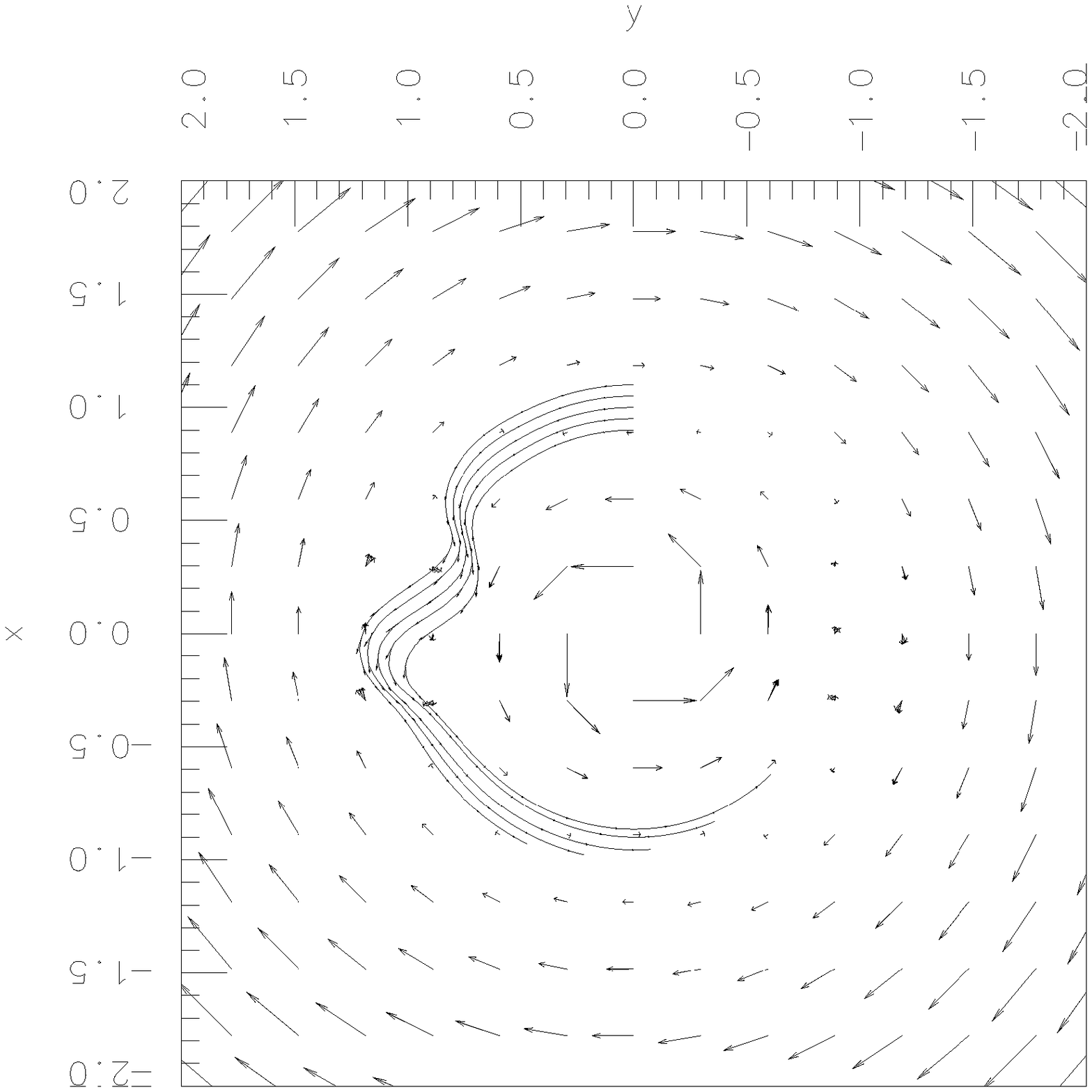}
\caption{Same  field configuration as in
Fig.~\protect\ref{f_num6a} viewed from the  top
$z$-direction. Arrows indicate the velocities of the flow
in the  frame corotating with the base of the plume. The
picture shows the  twisting of the bundle of field lines.
\label{f_num6b}  }
\end{figure}
\clearpage

\begin{figure}
\epsscale{1.0}
\plotone{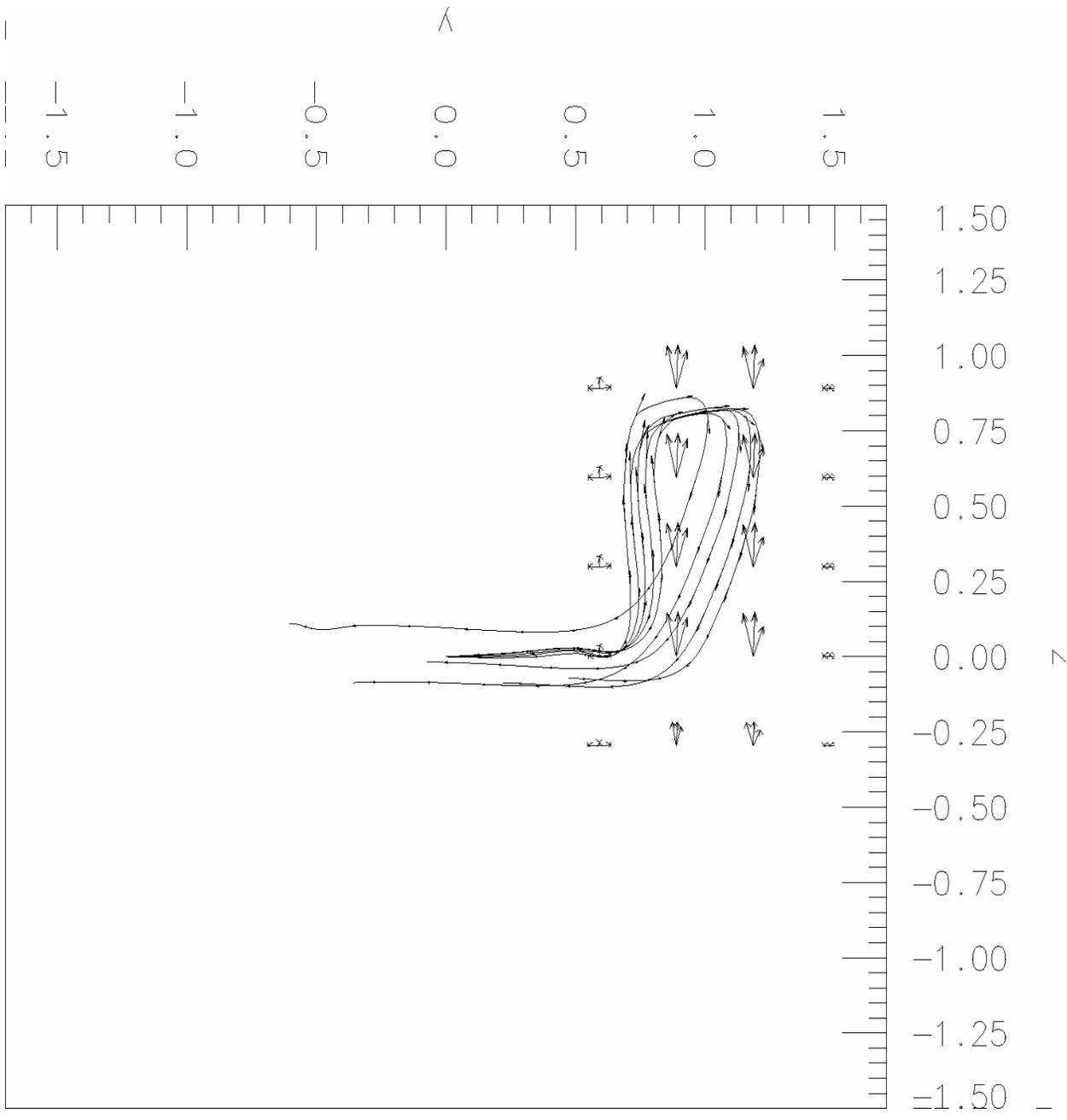}
\caption{Same  field configuration as in
Fig.~\protect\ref{f_num6a} viewed from the  side
$\phi$-direction. Arrows indicate velocities of the flow
in the  frame corotating with the base of the plume. The
picture shows the  formation of the loop of poloidal
field lines by the rising  plume.
\label{f_num6c}  }
\end{figure}
\clearpage

\begin{figure}
\epsscale{1.0}
\plotone{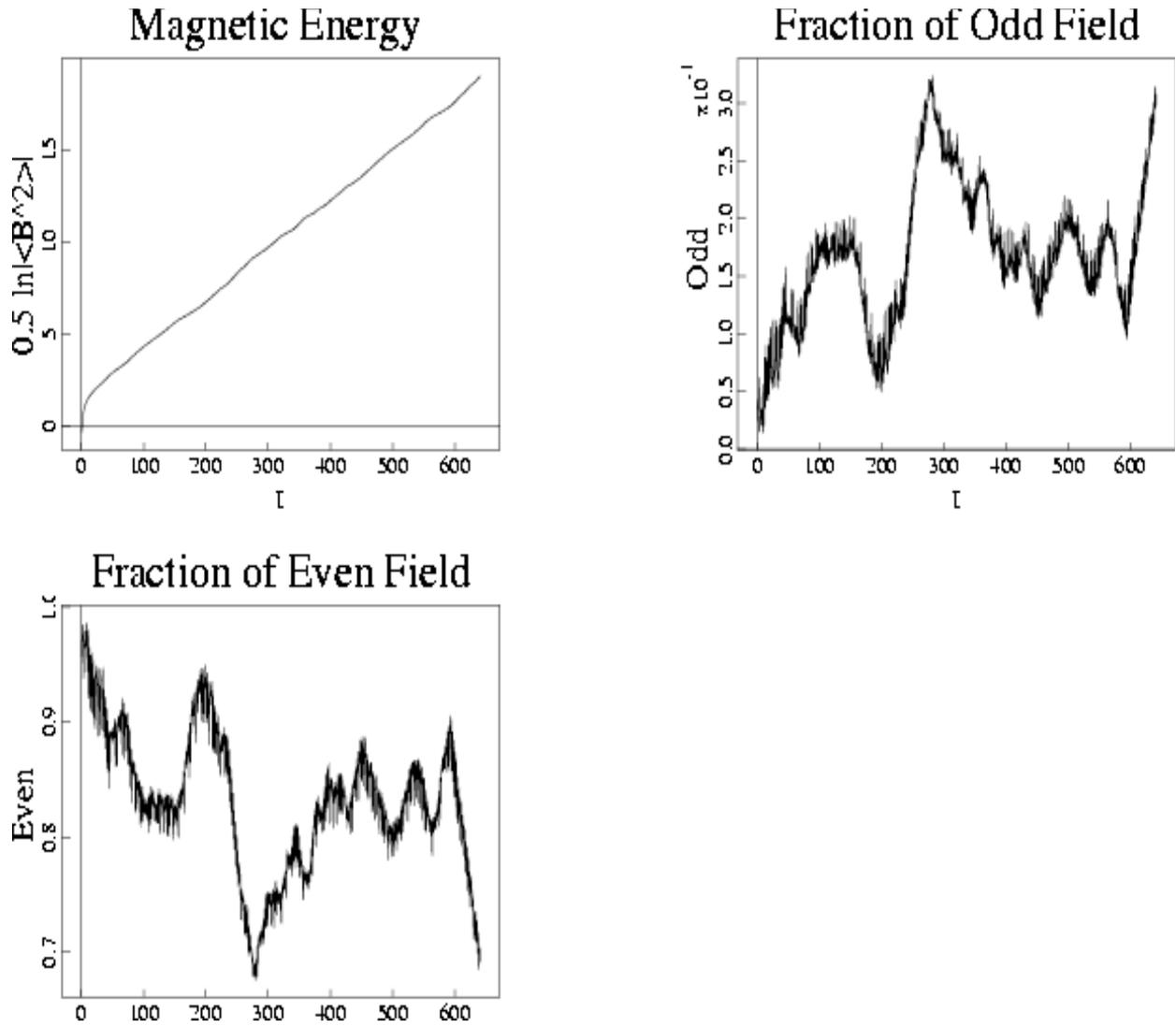}
\caption{Exponential  growth of the dynamo magnetic
field. Half of the logarithm of the 
$B^2$ averaged over all computational domain is plotted
versus  time in the top-left plot. Time evolution of the
fractions of the  energy of the odd and even components
of the magnetic field is in the  top-right and
bottom-left plots. The sum of the fractions of odd and 
even components is always equal to $1$.
\label{f_num7}  }
\end{figure}
\clearpage

\begin{figure}
\epsscale{1.0}
\plotone{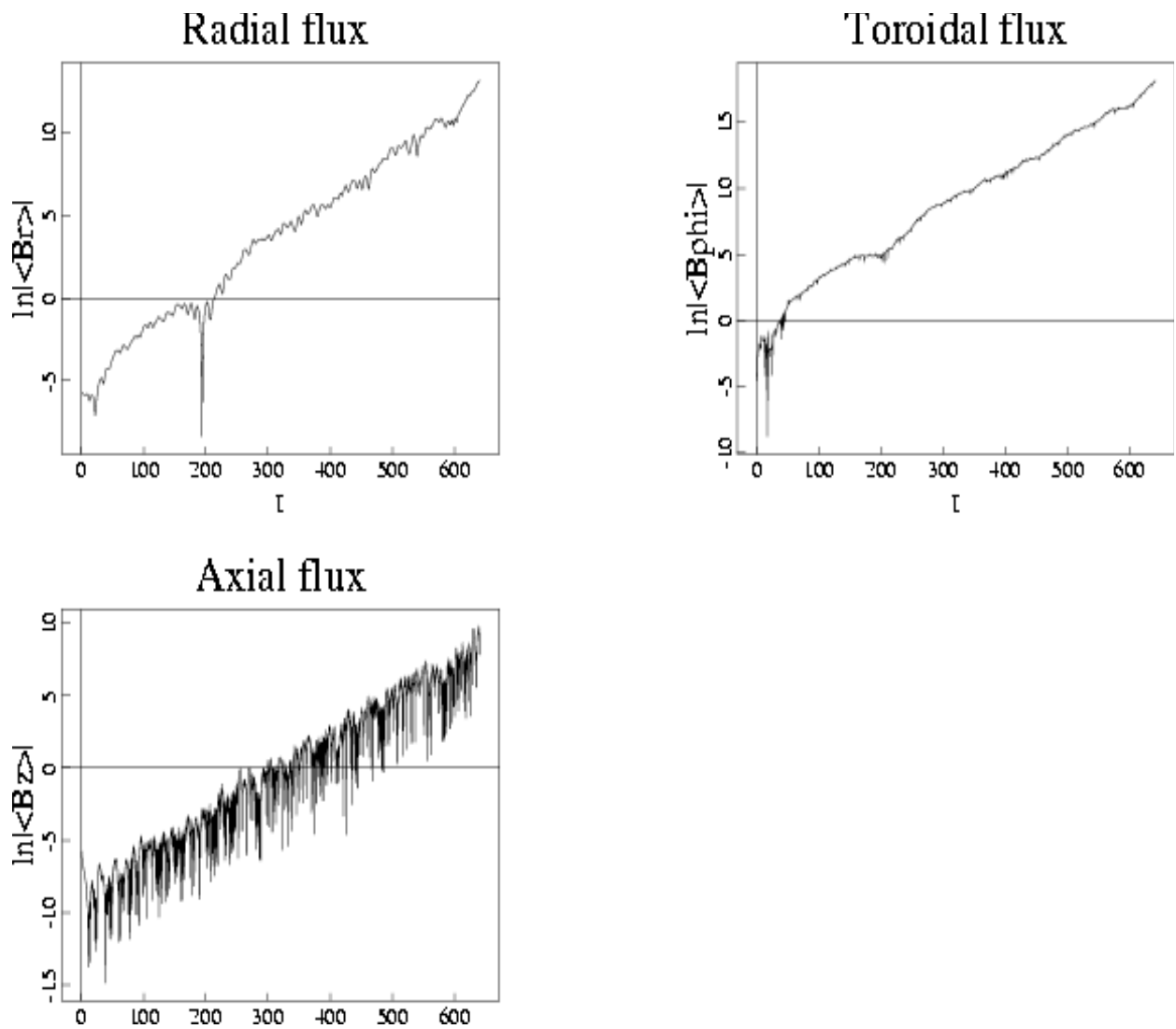}
\caption{Time  evolution of logarithms of the absolute
values of the components of  magnetic field averaged over
the surfaces described in the text.  Exponential growth
of all three components is evident.
\label{f_num8}  }
\end{figure}
\clearpage

\begin{figure}
\epsscale{1.0}
\plotone{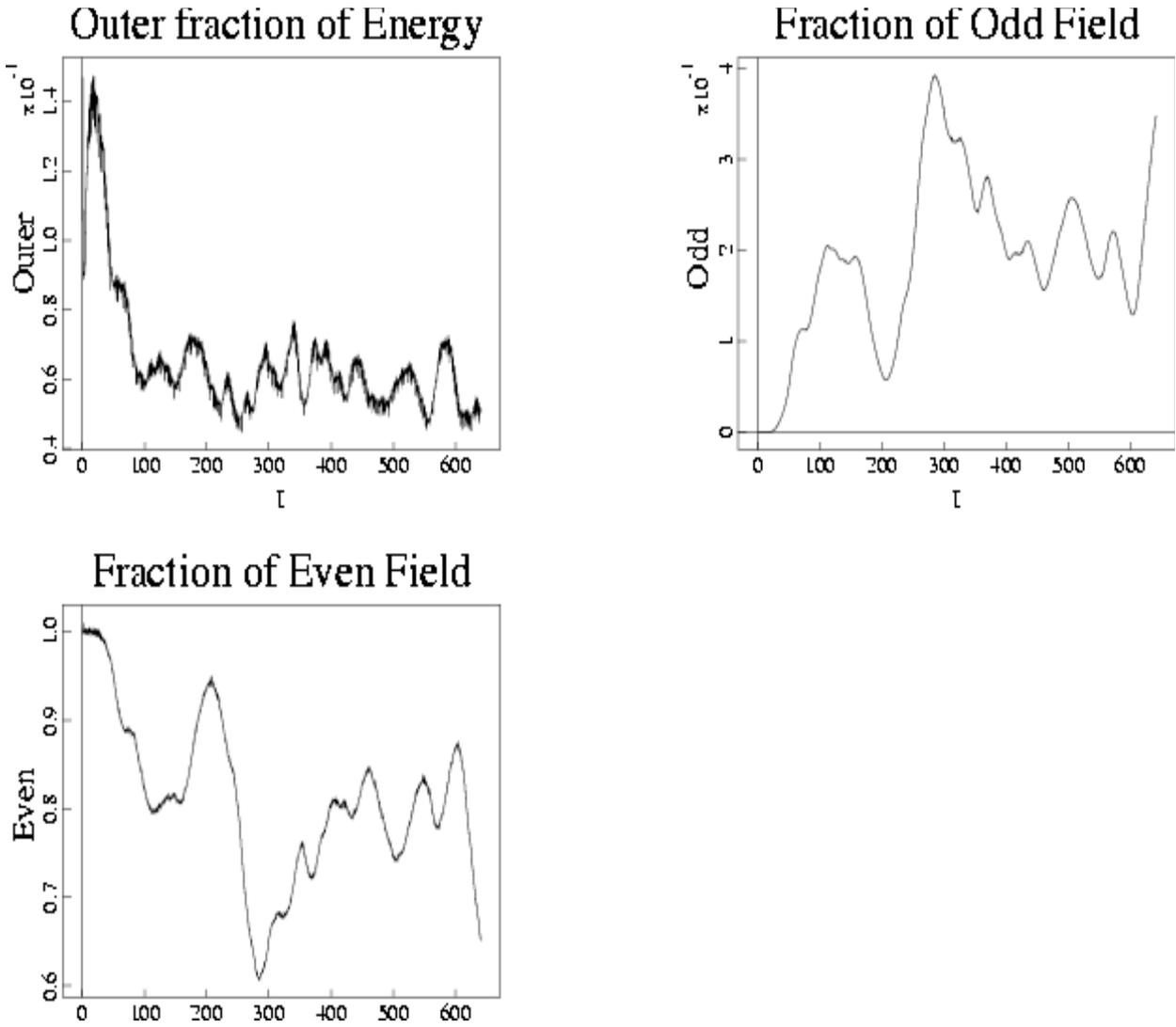}
\caption{Time  evolution of the ratio of the energy of
the magnetic field in the  outer domain to the total
energy of the magnetic field in the  computational domain
is on the top-left plot. Time evolution of the  fractions
of energy of the odd and even components of the magnetic 
field in the outer domain is on the top-right and
bottom-left plots.  The sum of the fractions of odd and
even components is always equal  to $1$.
\label{f_num9}  }
\end{figure}
\clearpage

\begin{figure}
\epsscale{1.0}
\plotone{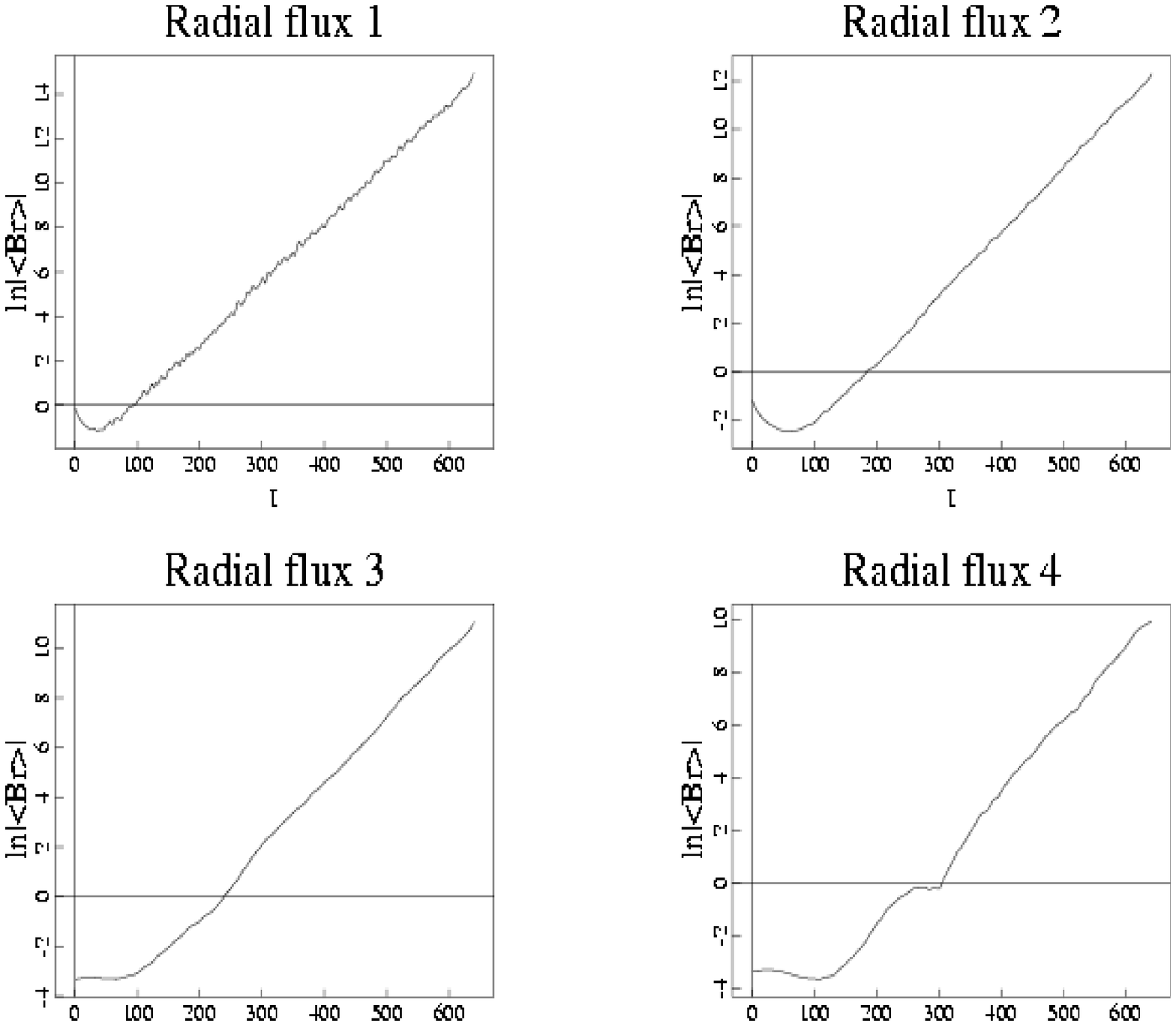}
\caption{Time  evolution of logarithms of the absolute
values of the radial  component of the magnetic field
averaged over the four surfaces  described in the text.
All four surfaces are located in the outer  region
described in the text.
\label{f_num10}  }
\end{figure}
\clearpage

\begin{figure}
\epsscale{1.0}
\plotone{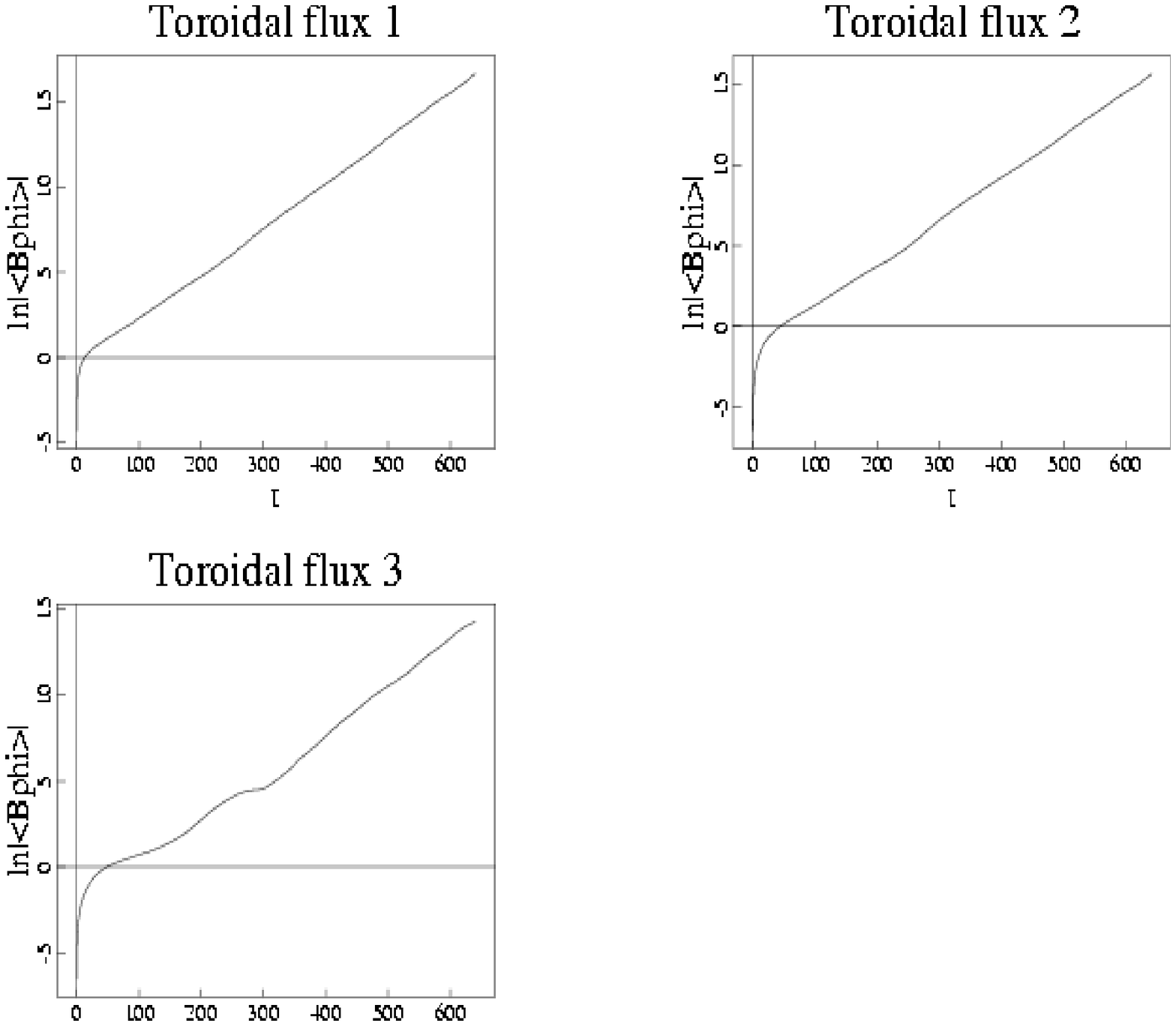}
\caption{Time  evolution of logarithms of the absolute
values of the toroidal  component of the magnetic field
averaged over the three surfaces  described in the text.
All three surfaces are located in the outer  region
described in the text.
\label{f_num11}  }
\end{figure}
\clearpage

\begin{figure}
\epsscale{0.5}
\plotone{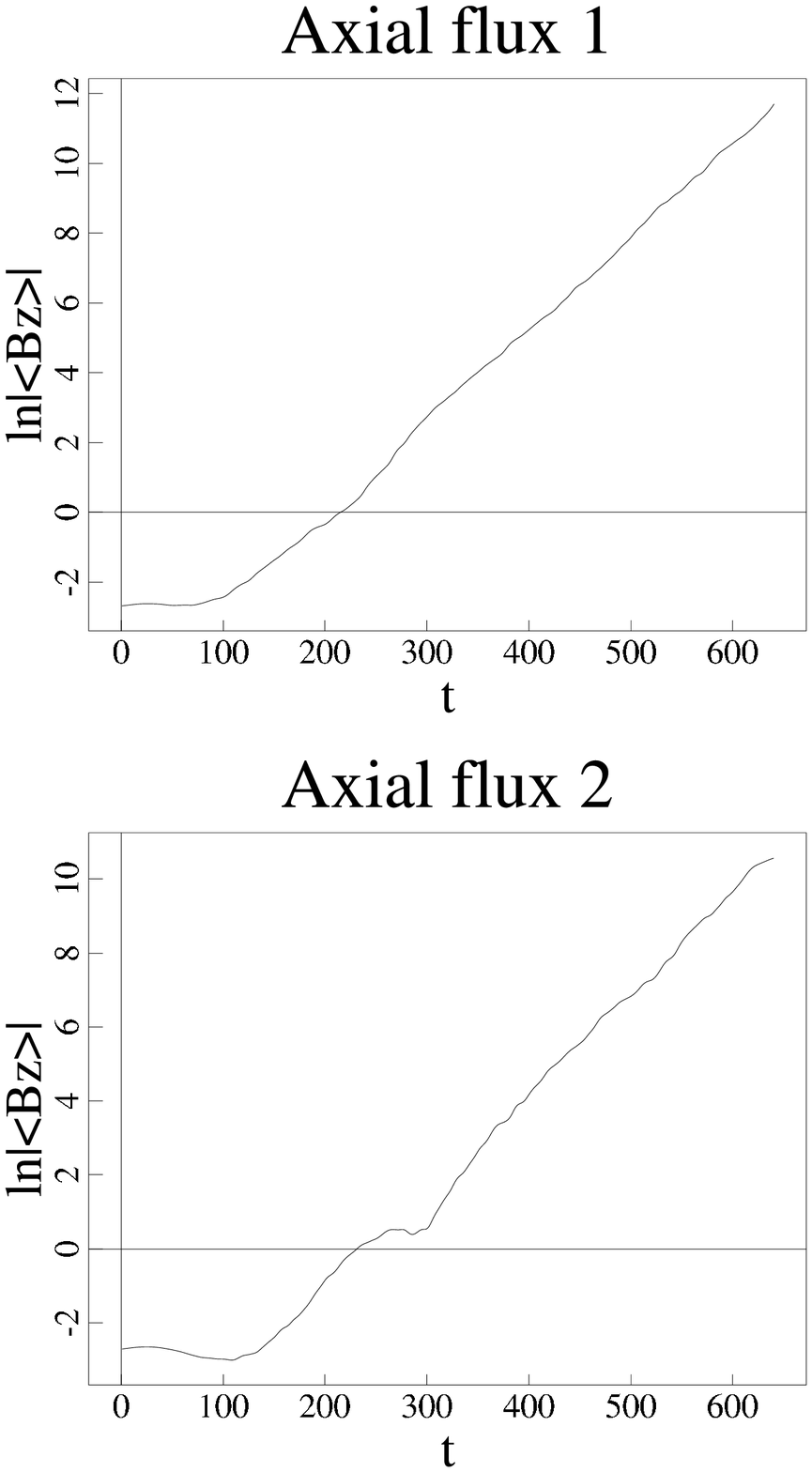}
\caption{Time  evolution of logarithms of the absolute
values of the axial component  of the magnetic field
averaged over the two surfaces described in the  text.
Both surfaces are located in the outer region described
in the  text.
\label{f_num12}  }
\end{figure}
\clearpage

\begin{figure}
\epsscale{0.6}
\plotone{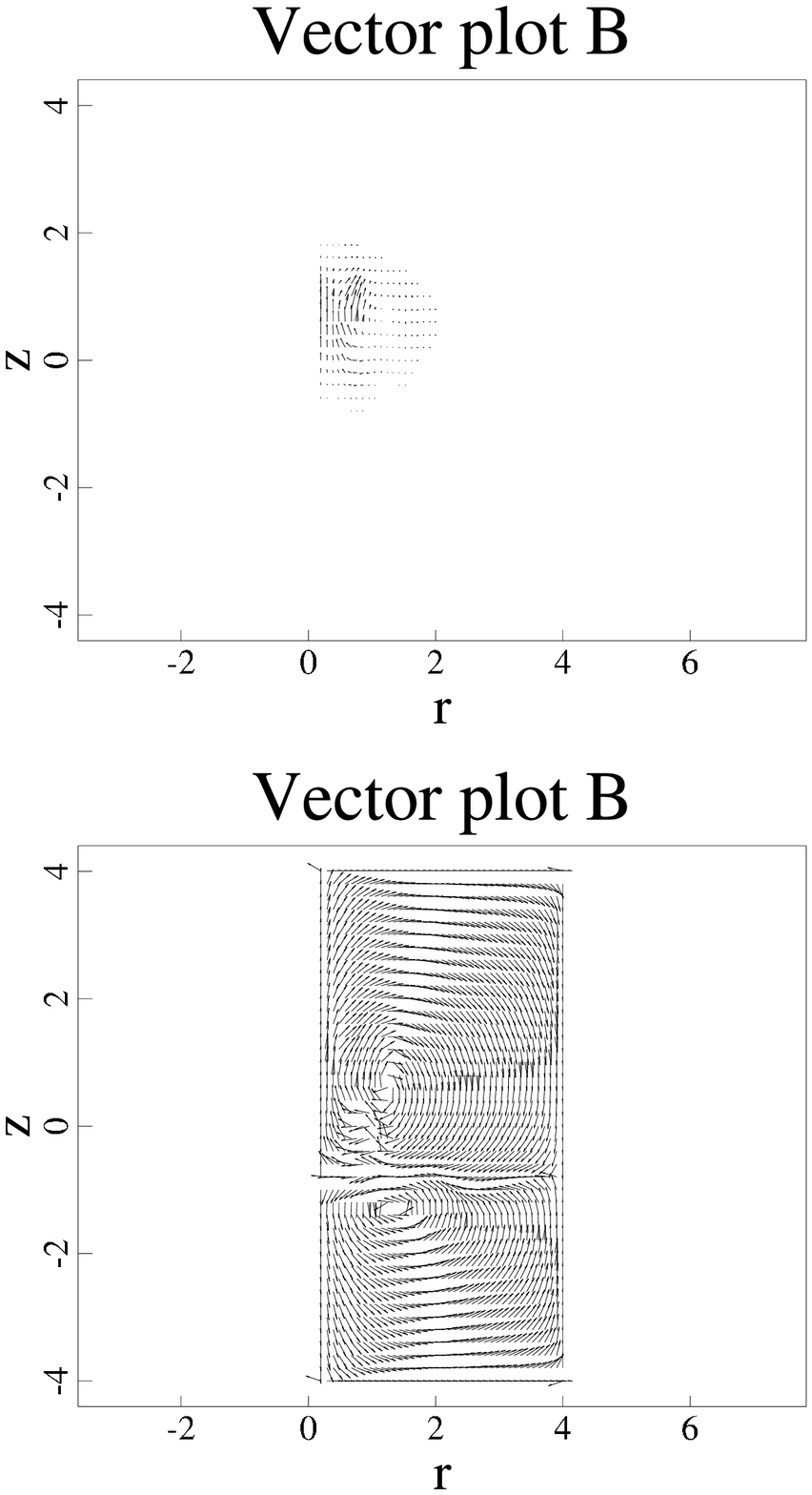}
\caption{Vector  plots of the poloidal magnetic field of
the growing dynamo at the  time $t=640$ in the plane
$\phi=0$. The length of arrows on the top  plot is
proportional to the magnitude of the poloidal magnetic
field.  Arrows on the bottom plot have unit length and
are directed along  the poloidal field.
\label{f_num13}  }
\end{figure}
\clearpage

\begin{figure}
\epsscale{0.8}
\plotone{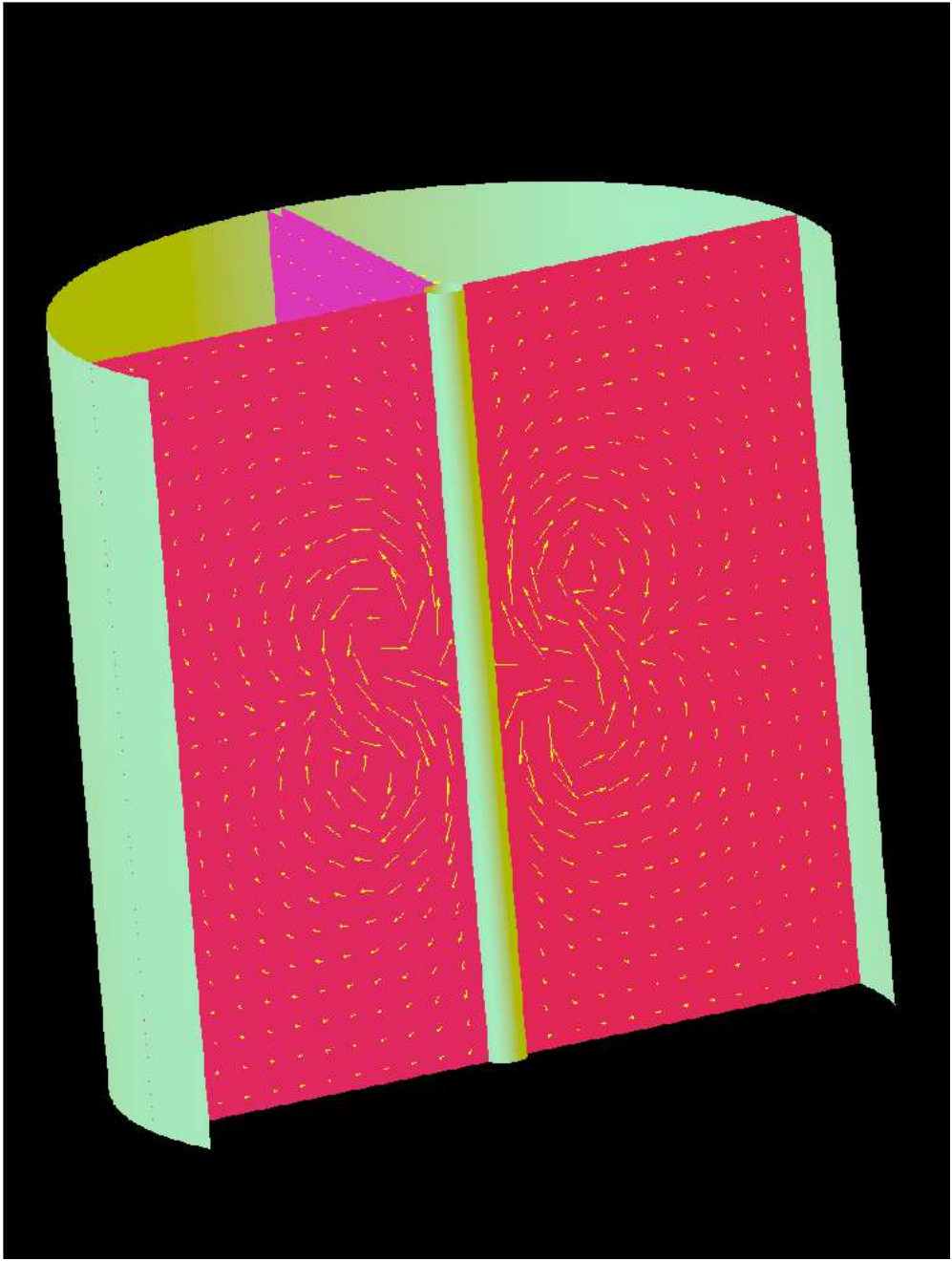}
\caption{Three  dimensional plot of the poloidal
magnetic field of the growing dynamo  at the time
$t=320$. The length arrows scales as the $1/3$ power of 
the magnitude of the poloidal magnetic field. The
magnetic field on  the planes $\phi=\pi/2$ and
$\phi=3\pi/2$ is shown. The boundaries of  the
cylindrical region are $z=\pm 4$, $R_1=0.2$, 
$R_2=4$.
\label{f_num_col}  }
\end{figure}
\clearpage


\begin{thebibliography}{}

\bibitem[Abramowitz \& Stegun(1972)]{abramowitz72}
Abramowitz,~M., \& Stegun,~I.~A. 1972, Handbook of
Mathematical Functions. (New York: Dover)

\bibitem[Bayliss et~al.(2006)]{bayliss06}
Bayliss,~R.A.,   Nornberg,~M.D., Terry,~P.W., \&
Forest,~C.B. 2006, Phys. Rev. E, submitted
(arXiv:physics/0602126) 
\bibitem[Beckley et~al.(2003)]{beckley03} Beckley,~H.F.,
Colgate,~S.A., Romero,~V.D., \& Ferrel,~R. 2003, \apj,
599, 702

\bibitem[Biskamp(1993)]{biskamp93} Biskamp,~D. 1993,
Nonlinear Magnetohydrodynamics. (Cambridge: Cambridge
Univ. Press) 
\bibitem[Blackman \& Brandenburg(2003)]{blackman03}
Blackman,~E.G., \& Brandenburg,~A. 2003, \apj, 584, L99

\bibitem[Blandford \& Payne(1982)]{blandford82}
Blandford,~R.D.,
\& Payne,~D.G. 1982, \mnras, 199, 883

\bibitem[Boldyrev \& Cattaneo(2004)]{boldyrev04}
Boldyrev,~S.A., \& Cattaneo, F. 2004, \prl, 92, 144501

\bibitem[Boldyrev(2006)]{boldyrev06} Boldyrev,~S.A.
2006, \prl, 96, 115002

\bibitem[Bondi \& Hoyle(1944)]{bondi44} Bondi,~H., 
\& Hoyle,~F. 1944, \mnras, 104, 273

\bibitem[Bondi, Hoyle \& Lyttleton(1947)]{bondi47}
Bondi,~H., Hoyle,~F., \& Lyttleton,~R.A. 1947, \mnras,
107, 184

\bibitem[Bondi(1952)]{bondi52} Bondi,~H. 1952, \mnras,
112, 195

\bibitem[Bourgoin et~al.(2002)]{bourgain02}
Bourgoin,~M., Marie,~L., Petrelis,~F., Gasquet,~C.,
Guigon,~A., Luciani,~J.-P., Moulin,~M., Namer,~F.,
et~al.   2002, Phys. Fluids., 14, 3046

\bibitem[Bourgoin et~al.(2004)]{bourgain04}
Bourgoin,~M., Odier,~P., Pinton,~J.-F., \& Ricard,~Y. 2004,
Phys. Fluids., 16, 2529

\bibitem[Busse(1991)]{busse91} Busse,~F.H. 1991, in
Advances in Solar System Magnetohydrodynamics, eds.
Priest~E.R., Wood~A.W. (Cambridge: Cambridge Univ.
Press), p.~51 

\bibitem[Chakrabarti, Rosner, \&
Vainshtein(1994)]{chakrabarti94} Chakrabarti,~S.K.,
Rosner,~R., \& Vainshtein,~S.I. 1994, \nat, 368, 434 

\bibitem[Childress et~al.(1990)]{childress90}
Childress,~S., Collet,~P., Frish,~U., Gilbert,~A.D.,
Moffatt,~H.K., \& Zaslavsky,~G.M. 1990, Geophys.
Astrophys. Fluid Dyn., 52, 263 

\bibitem[Colgate \& Li(1997)]{colgate97} Colgate,~S.A.,
\& Li,~H. 1997, in Relativistic Jets in AGNs, ed.
Ostrowski~M. (Crakow: Poland), p.~170

\bibitem[Colgate \& Li(1999)]{colgate99} Colgate,~S.A.,
\& Li,~H. 1999, \apss, 264, 357

\bibitem[Colgate, Li \& Pariev(2001)]{colgate01}
Colgate,~S.A., Li,~H., \& Pariev,~V.I. 2001, Physics of
Plasmas, 8, 2425 

\bibitem[Colgate et~al.(2003)]{colgate03} Colgate,~S.A.,
Cen,~R., Li,~H., Currier,~N., \&  Warren,~M.S. 2003,
\apj, 598, L7

\bibitem[Cowling(1981)]{cowling81} Cowling,~T.G. 1981,
\araa, 19, 115 

\bibitem[Dudley \& James(1989)]{dudley89} Dudley,~M.L.,
\& James,~R.W. 1989 Proc. R. Soc. London A 425,
407

\bibitem[Ferri\`ere(1993a)]{ferriere93a} Ferri\`ere,~K.,
1993a, \apj, 404, 162

\bibitem[Ferri\`ere(1993b)]{ferriere93b} Ferri\`ere,~K.,
1993b, \apj, 409, 248

\bibitem[Ferri\`ere(1998)]{ferriere98} Ferri\`ere,~K.,
1998, \aap, 335, 488

\bibitem[Ferri\`ere \& Schmitt(2000)]{ferriere00}
Ferri\`ere,~K.,
\& Schmitt,~D. 2000, \aap, 358, 125

\bibitem[Finn(1992)]{finn92} Finn,~J.M. 1992, in
Electromechanical coupling of the solar atmosphere,
Proceedings of the OSL Workshop, (Capri, Italy), p.~79

\bibitem[Finn et~al.(1991)]{finn91} Finn,~J.M., Ott,~E.,
Hanson,~J.D., \& Kan,~I. 1991, Physics of Fluids, B3, 1250

\bibitem[Fletcher(1992)]{fletcher92} Fletcher,~C.A.J.
1992, Computational Techniques for Fluid Dynamics.
(Heidelberg: Springer--Verlag)

\bibitem[Frolov \& Novikov(1998)]{frolov98}
Frolov,~V.P., \& Novikov,~I.D. 1998, Black Hole Physics:
Basic Concepts and New Developments.  (Dordrecht: Kluwer)

\bibitem[Gailitis \& Freiberg(1976)]{gailitis76}
Gailitis,~A., \&  Freiberg,~Ya. 1976,
Magnetohydrodynamics, 12, 127 

\bibitem[Gailitis et~al.(2000)]{gailitis00}
Gailitis,~A., Lielausis,~O., Dement'ev,~S., et~al. 2000,
\prl, 84, 4365 

\bibitem[Gailitis et~al.(2001)]{gailitis01}
Gailitis,~A., Lielausis,~O., Platacis,~E., et~al. 2001,
\prl, 86, 3024 

\bibitem[Goldreich \& Sridhar(1995)]{goldreich95}
Goldreich,~P.,
\& Sridhar,~S. 1995, \apj, 438, 763

\bibitem[Hoyle(1949)]{hoyle49} Hoyle,~F. 1949, 
Some Recent Researches in Solar Physics.
(Cambridge: University Press)

\bibitem[Iroshnikov(1963)]{iroshnikov63}
Iroshnikov,~P.S. 1963, \azh, 40, 742

\bibitem[Khanna \& Camenzind(1996a)]{khanna96a} 
Khanna,~R.,
\& Camenzind,~M., 1996a, \aap, 307, 665

\bibitem[Khanna \& Camenzind(1996b)]{khanna96b} 
Khanna,~R.,
\& Camenzind,~M., 1996b, \aap, 313, 1028

\bibitem[Kraichnan(1965)]{kraichnan65} Kraichnan,~R.H.
1965, Phys. Fluids, 8, 1385

\bibitem[Krause \& R\"adler(1980)]{krause80} Krause,~F.,
\& R\"adler,~K.H. 1980, Mean-Field Magnetohydrodynamics
and Dynamo Theory. (Oxford: Pergamon Press)

\bibitem[Kronberg et~al.(2001)]{kronberg01}
Kronberg,~P.P., Dufton,~Q.W., Li,~H., \& Colgate,~S.A.
2001, \apj, 560, 178 

\bibitem[Kulsrud(1999)]{kulsrud99} Kulsrud,~R.M., 1999,
\araa, 37, 37

\bibitem[Lau \& Finn(1993)]{lau93} Lau,~Y.-T., \&
Finn,~J.M. 1993, Physics of Fluids, B5, 365

\bibitem[Laval et~al.(2006)]{laval06} Laval,~J-P.,
Blaineau,~P., Leprovost,~N., Dubrulle,~B., \&
Daviaud,~F.  2006, \prl, 96, 204503

\bibitem[Li et~al.(2000)]{li00} Li,~H., Finn,~J.M.,
Lovelace,~R.V.E.,
\& Colgate,~S.A. 2000, \apj, 533, 1023

\bibitem[Li et~al.(2001a)]{li01a} Li,~H.,
Lovelace,~R.V.E., Finn,~J.M., \& Colgate,~S.A. 2001a,
\apj, 561, 915 

\bibitem[Li et~al.(2001b)]{li01b} Li,~H., Colgate,~S.A.,
Wendroff,~B.,
\& Liska,~R. 2001b, \apj, 551, 874

\bibitem[Lovelace et~al.(1999)]{lovelace99}
Lovelace,~R.V.E., Li,~H., Colgate,~S.A., \& Nelson,~A.F.
1999, \apj, 513, 805
 
\bibitem[Mari\'e et~al.(2003)]{marie03} Mari\'e,~L.,
Burguete,~J.,  Daviaud,~F., \& L\'e{}orat,~J. 2003,
European Physical Journal B,  33, 469

\bibitem[McCrea(1953)]{mccrea53} McCrea,~W.H.
1953, \mnras, 113, 162

\bibitem[Mestel(1999)]{mestel99} Mestel,~L. 1999,
Stellar Magnetism. (Oxford: Clarendon)

\bibitem[Moffatt(1978)]{moffatt78} Moffatt,~H.K. 1978,
Magnetic Field Generation in Electrically Conducting
Fluids. (Cambridge: Cambridge University Press)

\bibitem[Molchanov, Ruzmaikin, \&
Sokoloff(1983)]{molchanov83} Molchanov,~S.A.,
Ruzmaikin,~A.A., \& Sokoloff,~ D.D. 1983,
Magnetohydrodynamics, 19,  402 

\bibitem[ Nornberg et~al.(2006)]{nornberg06}  Nornberg, M.D.,
Spence, E.J.,  Kendrick R.D., \& Forest C.B. 2006,  Phys.
Plasmas, 13, 055901

\bibitem[O'Connell et~al.(2005)]{oconnell05} 
O'Connell,~R., Kendrick,~R.D.,  Nornberg, M.D.,  Spence,
E.J.,  Bayliss, R.A., \& Forest, C.B. 2005, in Dynamo and
Dynamics, a Mathematical Challenge, v.26 of NATO Science
Series,  eds. Chossat,~P., Ambruster,~D., Oprea,~I.

\bibitem[ Ponty et~al.(2005)]{ponty05} Ponty, Y.,
Mininni, P.D., Montgomery, D.C., Pinton, J.-F.,
Politano, H., \&  Pouquet, A. 2005, \prl, 94, 164502

\bibitem[Pariev \& Colgate(2006)]{pariev06}
Pariev,~V.I., \& Colgate,~S.A. 2006, \apj, in press
(paper~I) 

\bibitem[Parker(1955)]{parker55} Parker,~E.N. 1955, \apj,
121, 29

\bibitem[Parker(1979)]{parker79} Parker,~E.N. 1979,
Cosmical Magnetic Fields, their Origin and their
Activity. (Oxford: Claredon) 

\bibitem[Peffley, Cawthorne \& Lathrop(2000)]{peffley00}
Peffley,~N.L., Cawthorne,~A.B., \& Lathrop,~D.P. 2000,
Phys. Rev. E,  61, 5287

\bibitem[P\'e{}tr\'e{}lis et~al.(2003)]{petrelis03}
P\'e{}tr\'e{}lis,~F., Bourgoin,~M., Mari\'e,~L.,
Burguete,~J., Chiffaudel,~A., Daviaud,~F., Fauve,~S.,
Odier,~P., et~al. 2003,  \prl,  90, 174501.

\bibitem[Ponomarenko(1973)] {ponomarenko73}  
Ponomarenko,~Yu.B. 1973, J. Appl. Mech. Tech. Phys., 14,
775 
\bibitem[Priest(1982)]{priest82} Priest,~E.R. 1982, Solar
Magneto-hydrodynamics. (Boston: Kluwer, Inc.)

\bibitem[Reyes--Ruiz \& Stepinski (1999)]{reyes99}
Reyes-Ruiz,~M. \& Stepinski,~T.F., 1999, \aap, 342, 892

\bibitem[Roberts \& Soward(1992)]{roberts92}
Roberts,~P.H.,
\& Soward,~A.M. 1992, Ann. Rev. of Fluid Mechanics, 24,
459

\bibitem[Ruzmaikin, Sokoloff \&
Shukurov(1988)]{ruzmaikin88} Ruzmaikin,~A.A.,
Sokoloff,~D.D., \& Shukurov,~A.M. 1988, Magnetic fields
in galaxies. (Moscow: Nauka) 

\bibitem[Sakharov(1982)]{sakharov82} Sakharov,~A.D.
1982, Selected Scientific Works. Divertissment 11. (New
York: Marcel Dekker) 

\bibitem[Shakura(1972)]{shakura72} Shakura,~N.I. 1972,
\azh, 49, 921 

\bibitem[Shakura \& Sunyaev(1973)]{shakura73}
Shakura,~N.I., \& Sunyaev,~R.A. 1973, \aap, 24, 337

\bibitem[Sovinec, Finn \&
del-Castillo-Negrete(2001)]{sovinec01} Sovinec,~C.R.,
Finn,~J.M., \& del-Castillo-Negrete,~D. 2001, Phys.
Plasmas, 8, 475 

\bibitem[Sisan et~al.(2004)]{sisam04} Sisan,~D.R.,
Mujica,~N., Tillotson,~W.A.,  Huang,~Y.-M.,
Dorland,~W.,  Hassam,~A.B., Antonsen~T.M., \& 
Lathrop,~D.P. 2004, \prl, 93, 114502

\bibitem[Spence et~al.(2006)]{spence06}  Spence, ~E.J.,
Nornberg, M.D., Jacobson, C.M., Kendrick R.D., \&
Forest,~C.B. 2006, Phys. Rev. Lett., 96, 055002

\bibitem[Steenbeck, Krause \&
R\"adler(1966)]{steenbeck66} Steenbeck,~M., Krause,~F.,
\& R\"adler,~K.H. 1966, Z.~Naturforsch., 21a, 369

\bibitem[Stepinski \& Levy(1988)]{stepinski88}
Stepinski,~T.F.,
\& Levy,~E.H. 1988, \apj, 331, 416

\bibitem[Stieglitz \& M\"uller(2001)]{stieglitz01}
Stieglitz,~R., \& M\"uller,~U. 2001, Physics of Fluids,
13, 561 

\bibitem[Stix(1975)]{stix75} Stix,~M. 1975, \aap, 42, 85

\bibitem[Sweet et~al.(2001)]{sweet01} Sweet,~D.,
Ott,~E., Antonsen,~T.M.,  Lathrop,~D.P., \& Finn,~J.M. 
2001, Physics of Plasmas, 8, 1944 

\bibitem[Vainshtein \& Zeldovich(1972)]{vainshtein72}
Vainshtein,~S.I., \& Zeldovich,~Ya.B. 1972, Soviet
Physics Uspekhi, 15, 159

\bibitem[Zeldovich, Ruzmaikin, \&
Sokoloff(1983)]{zeldovich83} Zeldovich,~Ya.B.,
Ruzmaikin,~A.A., \& Sokoloff,~D.D. 1983, Magnetic Fields
in Astrophysics. (New York: Gordon and Breach Science
Publishers) 

\end{thebibliography}
\end{document}